# Theoretical morphology of a cichlid according to the approach of Systemic Morphometry

Systemic Morphometry and Theoretical Morphology of a fish

Juan Rivera Cázares[1], Xavier Valencia Díaz[1], Christian Lambarri Martínez[1, 2]

[1]Colección Nacional de Peces, Instituto de Biología, Universidad Nacional Autónoma de México. Circuito Centro Cultural, Ciudad Universitaria, Coyoacán, 04510, Ciudad de México, CDMX.

[2] Posgrado en Ciencias Biológicas – UNAM, Unidad de Posgrado Edificio D primer piso, Ciudad Universitaria, CDMX, México.

*Corresponding author:

Email: juan.rivera@st.ib.unam.mx

# Abstract

We analyzed the body structure of the Blackstripe Cichlid *Vieja fenestrata* (Günther, 1860), a species with highly phenotypic variability, by the Systemics Morphometrics Methodology, previously proposed by one of the authors. From this perspective and considering the properties of its bauplan, we describe the expected morphometrics variability of this species. The Infinitesimal Change Rates (IChR) were obtained deriving the allometric equations that relate pairs of morphometric variables, and they demonstrated that the species' growth is continuous throughout its ontogeny. For some of the morphometric variables, relative growth trajectories were traced and their relationship with the IChR showed. Also, the observed and theoretical Systemic Phenotypical Spaces (SPS) were described by using three dimensional graphs and Mahalanobis Quadratic Distances (MQD). This was an alternate approach that allowed the analysis of the phenotypical spaces' properties in a wider, more objective, and analytical manner.

We conclude that the morphometric variability observed in *V. fenestrata* agrees with the variability expected in the times and places sampled, although there are still some issues to be explained. We propose to incorporate the structural variance into the classical phenotypic variance equation, and consider the equality: $\sigma^2_{phenotypic} = SPS_{theoretical}$, (the phenotypic variance is equal to the theoretical Systemic Phenotypic Space), as a point of convergence between Quantitative Genetics and Systemic Morphometry.

# Introduction

In the last section of their publication, [1] proposed the development of Systemic Morphometry as an alternative methodology for the study of the body structure, or body plan, of living beings. Among the differences between this approach and the previous ones, is the clearly systemic character, together with its focus on the study of the body structure of the taxon, which is not necessarily the same as the shape, since the latter can be modified at any time throughout the ontogeny of the organism due to growth (allomeric or isometric), and the presence of other structures such as fat, scales, fur, etc., while the body structure, architecture or spatial arrangement of the elements that constitute the organism, remain constant from the end of embryogenesis, through the development of juvenile phases and until death, as in the case of vertebrates.

[2] referred [3], who proposed that Theoretical Morphology includes a mathematical simulation of the body plan and the construction of the theoretical and observed phenotypical space (of a determined taxon), and that both of them can be compared. Hereby we develop both aspects for the Blackstripe Cichlid *Vieja fenestrata* (Günther, 1860) [4], a secondary freshwater fish species [5] whose marine origins confer it wide salinity tolerances. By considering the corporal structure of an organism as a system, it is possible to describe relationships among its constituent elements through a system of simultaneous equations (linear, allomeric, etc.) with descriptive and inferential capacities [1], named the Systemic Morphometrics Model (SMM).

The quantitative analysis of the body plan from the systemic perspective and its quantitative expression, the systemic morphometric model, allows us to analyze the changes that occur in a taxon, both in space and time. On the other hand, the characterization of the phenotypic space of a taxon makes it possible to compare it against its observed phenotypic diversity and then propose hypotheses to explain the similarities and differences between them. The systemic approach to the study of theoretical morphology has been successfully applied in groups of vertebrates such as amphibians and reptiles [6].

Morphometrical studies on fishes of the family Cichlidae have focused on traditional and geometric morphometrics, to discriminate differences and morphological tendencies on the shape of species and populations of the *Amphilophus citrinellus* species complex [7]. Also, [8] found differences on the body structure, notably on the cephalic region, of six species of *Vieja* Fernández-Yépez, 1969 [9] from Southeastern Mexico.

The need to understand biological diversity in general, and that of vertebrates in particular, has led to the development of various tools, including the mathematical modeling of the body plan, (morphometric model), of a certain taxon and the simulation of its total expected morphological variability, obtaining a representation of its theoretical phenotypic space.

Both the morphometric model and the phenotypic space are useful tools to describe and infer the body structure properties of a taxon and to study its expected and observed phenotypic diversity, thus being valuable tools in evolutionary, taxonomic, ecological, physiological, and other biological studies.

This work is part of a project that focuses on the study of the theoretical morphology of the five groups of vertebrates. In particular, our objective was to apply a systemic approach to quantitatively analyze the body plan of *V. fenestrata*, obtaining a morphometric model through which we could explore the structural properties of the species and simulate its expected total phenotypic variation, that is, its theoretical phenotypic space, as well as studying some properties of both spaces, the theoretical and the observed.

The black striped cichlid *V. fenestrata* has gone through a long taxonomic history and genus placement. It was first described in the genus *Chromis* and later placed in *Cichlasoma*, until [10] restricted the latter to South America. Then, using nuclear and mitochondrial DNA phylogenetic analysis, [11] assigned it to the genus *Paraneetroplus* [12]; and finally, it was assigned to the genus *Vieja* by [13,14].

[15] mentioned that *V. fenestrata* was often confounded with *Cichlasoma gadovii* Regan, 1905 (now considered its synonym) and with *Cichlasoma sexfasciatum* Regan, 1905 (recognized then as a valid species and a synonym nowadays), although the populations in Alvarado and Boca del Río, in Veracruz, México, were thought to be *V. fenestrata*, even though they inhabited high salinity waters.

The Blackstripe Cichlid is fundamentally herbivorous but occasionally consumes invertebrates such as insects and crustaceans. Its shape is that of a typical fish but characterizes by a tall and moderately long body and a short to medium caudal peduncle; its total height, over the pelvic fins, fits about 2.1 times in the standard length [15]. Nuptial males and mature females measure up to 300 mm and 255 mm, respectively. Its mouth is big and horizontal, with a more oblique positioning in adult individuals. It has seven scales, (eight in individuals of the Coatzacoalcos, México), between the origin of the spinous dorsal fin and the lateral line; and has between 29 and 31 vertebrae. Its coloration is usually a brownish background with silvery edges on the body scales, reddish head, and yellow dorsal and caudal fin bases; but there are populations with pink, white and black coloration fish in the Catemaco Lake, Veracruz, México. It has six well developed vertical bars on the sides, which form a wide longitudinal stripe (sometimes incomplete) that extends into a prominent black caudal blotch.

The species *V. fenestrata* is endemic to Mexico, particularly on the Atlantic Slope in Veracruz: from the Actopan River, in the northern Chachalacas basin, with populations in Alvarado and Catemaco, in the Papaloapan River, to the south in the low Coatzacoalcos River. Like many other cichlid species, it has broad phenotypical plasticity and overlap of features, which makes its discrimination against other sister species difficult.

Even though cichlids show diverse reproductive strategies, this species is sexually dimorphic, and males and females distinguish one from each other until they are adults: males display a frontal hump and a dark turquoise dorsal coloration (including the dorsal fin). At the same time, females develop their ovaries so significantly, that they occupy a large percentage of their volume.

# Materials and methods

This work followed the methods proposed by [1]. For the construction of the SMM (Fig. 1), we defined the set of variables that characterize the external morphology of *V. fenestrata* (Fig. 2), which were measured in organisms preserved from the National Fish Collection of the Biology Institute of the *Universidad Nacional Autónoma de México* (CNPE-IBUNAM). The measurements were taken by the same researcher, using an electronic vernier with a precision of 0.01 mm.

**Figure 1. Flux diagram of the method proposed by [2].**

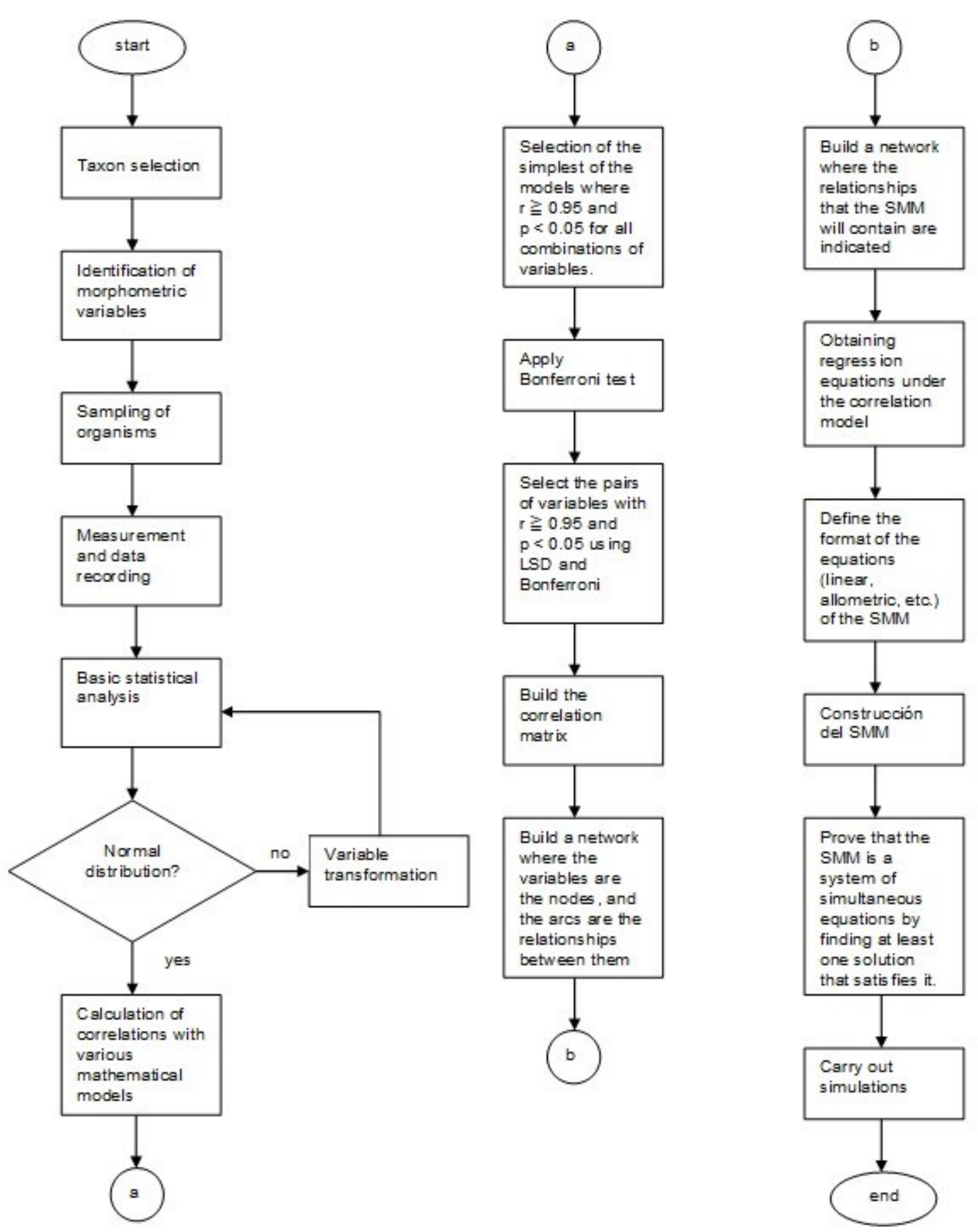

The stages for building a systemic morphometric model are shown, from the identification of the taxon to be studied to the verification of the functionality of said model.

**Fig. 2. External body structure of Vieja fenestrata**

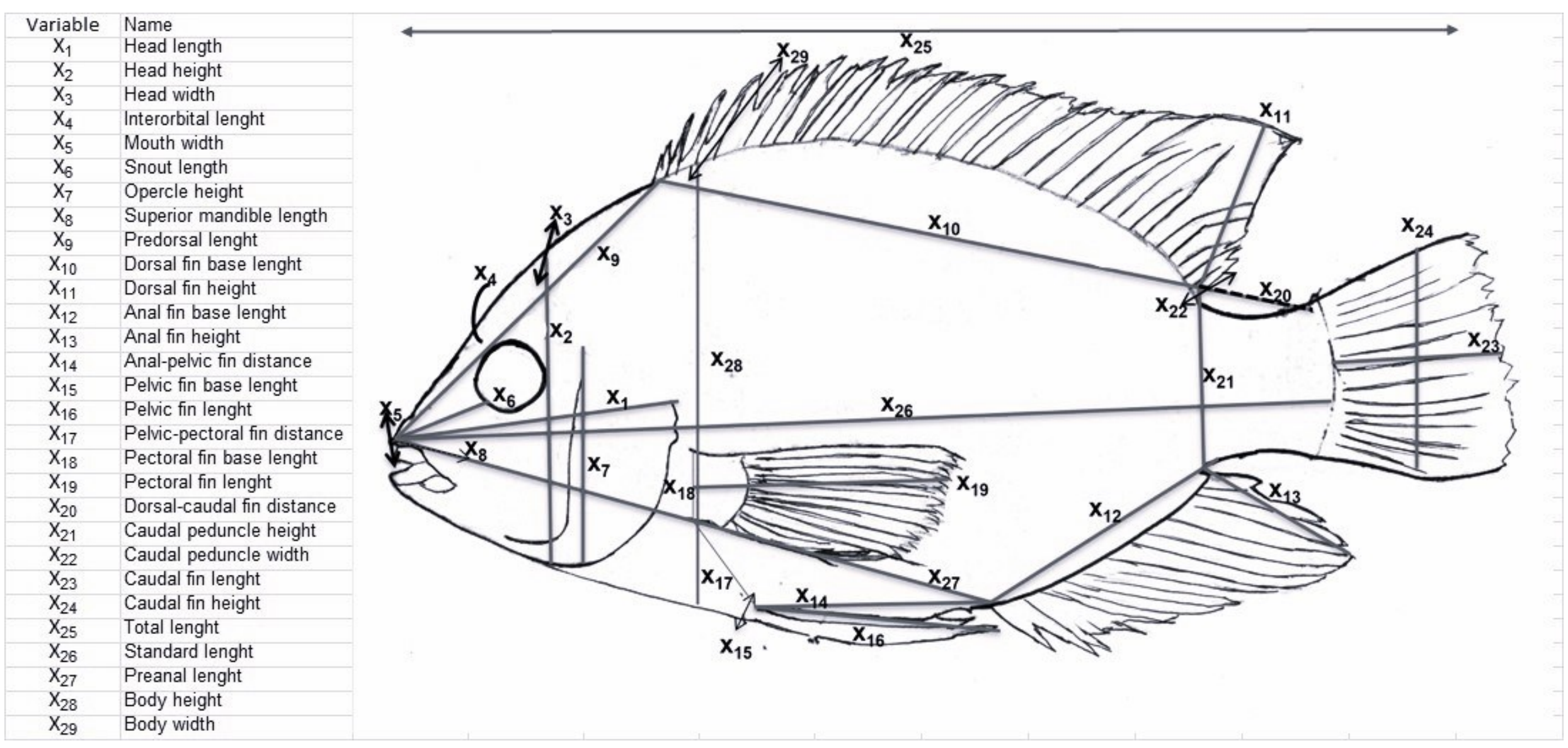

This image shows the morphometric variables that are considered descriptors of the body structure of *Vieja fenestrata*.

We measured 48 individuals (n = 48), of which 32 were collected in Laguna Escondida and 16 in Laguna Alvarado, Veracruz. In each individual we measured the 29 morphometric variables used to describe the external morphology of this species, resulting in 1,392 measurements. The catalog numbers and collection dates are: CNPE-IBUNAM 496 (07/13/1967), CNPE-IBUNAM 501 (01/15/1975), CNPE-IBUNAM 857 (05/24/1981), CNPE-IBUNAM 2824 (09/04/1986), CNPE-IBUNAM 4989 (07/09/1986) and CNPE-IBUNAM 10128 (03/25/1990).

A univariate statistical analysis of all data was performed for each of the 29 morphometric variables. The measures of central tendency, variability and shape are shown in Table 1. Skewness values show that the data is not normally distributed.

**Table 1. Central tendencies, variability and shape**

Table 1

| | $X_1$ | $X_2$ | $X_3$ | $X_4$ | $X_5$ | $X_6$ | $X_7$ | $X_8$ | $X_9$ | $X_{10}$ | $X_{11}$ | $X_{12}$ | $X_{13}$ | $X_{14}$ |
|---|---|---|---|---|---|---|---|---|---|---|---|---|---|---|
| Count | 48 | 48 | 48 | 48 | 48 | 48 | 48 | 48 | 48 | 48 | 48 | 48 | 48 | 48 |
| Average | 31.61 | 33.44 | 16.26 | 12.46 | 9.22 | 12.96 | 21.75 | 10.02 | 37.65 | 54.27 | 18.99 | 24.08 | 17.77 | 24.47 |
| Standar deviation | 11.67 | 16.10 | 6.52 | 5.83 | 4.78 | 5.92 | 9.26 | 4.41 | 15.11 | 24.40 | 10.36 | 10.85 | 10.41 | 12.07 |
| Coeff. of variation | 36.9% | 48.1% | 40.1% | 46.8% | 51.8% | 45.7% | 42.6% | 44.0% | 40.1% | 45.0% | 54.6% | 45.0% | 58.6% | 49.3% |
| Minimum | 14.67 | 13.1 | 7.08 | 4.79 | 2.92 | 4.57 | 9.07 | 3.91 | 18.06 | 21.1 | 7.13 | 9.57 | 6.49 | 9.75 |
| Maximum | 60.89 | 79.2 | 33.2 | 27.93 | 24.13 | 31.29 | 46.97 | 20.24 | 79.02 | 115.47 | 46.08 | 51.01 | 43.28 | 57.49 |
| Range | 46.22 | 66.1 | 26.12 | 23.14 | 21.21 | 26.72 | 37.9 | 16.33 | 60.96 | 94.37 | 38.95 | 41.44 | 36.79 | 47.74 |
| Stnd. skewness | 2.26 | 3.10 | 2.35 | 2.75 | 3.15 | 2.93 | 2.79 | 1.72 | 2.57 | 2.38 | 2.77 | 2.52 | 3.02 | 2.65 |
| Stnd. kurtosis | -0.48 | 0.60 | -0.36 | 0.17 | 1.06 | 0.88 | 0.20 | -0.95 | -0.16 | -0.59 | -0.28 | -0.45 | -0.08 | -0.08 |

| | $X_{15}$ | $X_{16}$ | $X_{17}$ | $X_{18}$ | $X_{19}$ | $X_{20}$ | $X_{21}$ | $X_{22}$ | $X_{23}$ | $X_{24}$ | $X_{25}$ | $X_{26}$ | $X_{27}$ | $X_{28}$ | $X_{29}$ |
|---|---|---|---|---|---|---|---|---|---|---|---|---|---|---|---|
| Count | 48 | 48 | 48 | 48 | 48 | 48 | 48 | 48 | 48 | 48 | 48 | 48 | 48 | 48 | 48 |
| Average | 7.72 | 25.11 | 8.62 | 6.99 | 24.29 | 8.88 | 15.13 | 4.48 | 27.24 | 17.67 | 116.95 | 89.64 | 57.53 | 42.82 | 16.11 |
| Standar deviation | 3.45 | 10.99 | 3.81 | 2.77 | 9.40 | 3.66 | 6.61 | 2.14 | 11.79 | 10.15 | 49.52 | 37.42 | 22.98 | 19.91 | 6.95 |
| Coeff. of variation | 44.7% | 43.8% | 44.3% | 39.6% | 38.7% | 41.3% | 43.7% | 47.8% | 43.3% | 57.5% | 42.3% | 41.7% | 40.0% | 46.5% | 43.1% |
| Minimum | 3.03 | 11.09 | 3.88 | 3.02 | 11.6 | 3.44 | 6.15 | 1.45 | 12.22 | 5.62 | 47.71 | 35.5 | 23.64 | 17.41 | 6.09 |
| Maximum | 16.5 | 53.31 | 19.09 | 13.94 | 45.41 | 19.23 | 31.88 | 8.8 | 55.13 | 39.09 | 245 | 191 | 117.32 | 92.98 | 33 |
| Range | 13.47 | 42.22 | 15.21 | 10.92 | 33.81 | 15.79 | 25.73 | 7.35 | 42.91 | 33.47 | 197.29 | 155.5 | 93.68 | 75.57 | 26.91 |
| Stnd. skewness | 2.44 | 2.95 | 2.46 | 2.27 | 2.28 | 2.05 | 2.36 | 2.01 | 2.33 | 2.24 | 2.41 | 2.46 | 2.34 | 2.50 | 2.29 |
| Stnd. kurtosis | -0.50 | 0.32 | -0.36 | -0.53 | -0.81 | -0.33 | -0.64 | -1.09 | -0.73 | -1.21 | -0.51 | -0.27 | -0.53 | -0.50 | -0.51 |

Measures of central tendencies, variability and shape for each morphometric variable used to describe the external morphology of *V. fenestrata* (mm).

Considering the possibility that the morphometric variables are allometrically related to each other, we calculate the natural logarithm of the entire data set and again perform a univariate statistical analysis. Table 2 shows that the transformed data is normally distributed, according to the skewness and kurtosis values, which is needed to apply the pairwise correlation model of the variables and validate statistical inferences, one of the relevant properties of the SMM.

**Table 2. Central tendency of variability and shape**

Table 2

| | $X_1$ | $X_2$ | $X_3$ | $X_4$ | $X_5$ | $X_6$ | $X_7$ | $X_8$ | $X_9$ | $X_{10}$ | $X_{11}$ | $X_{12}$ | $X_{13}$ | $X_{14}$ |
|---|---|---|---|---|---|---|---|---|---|---|---|---|---|---|
| Count | 48 | 48 | 48 | 48 | 48 | 48 | 48 | 48 | 48 | 48 | 48 | 48 | 48 | 48 |
| Average | 3.39 | 3.41 | 2.71 | 2.42 | 2.10 | 2.47 | 3.00 | 2.21 | 3.50 | 3.90 | 2.81 | 3.09 | 2.73 | 3.09 |
| Standar deviation | 0.35 | 0.44 | 0.38 | 0.44 | 0.49 | 0.43 | 0.40 | 0.45 | 0.38 | 0.43 | 0.51 | 0.43 | 0.54 | 0.47 |
| Coeff. of variation | 10.4% | 13.0% | 14.2% | 18.3% | 23.3% | 17.6% | 13.3% | 20.2% | 10.7% | 11.0% | 18.2% | 13.0% | 19.7% | 15.2% |
| Minimum | 2.69 | 2.57 | 1.96 | 1.57 | 1.07 | 1.52 | 2.2 | 1.36 | 2.89 | 3.05 | 1.96 | 2.26 | 1.87 | 2.26 |
| Maximum | 4.11 | 4.37 | 3.5 | 3.33 | 3.18 | 3.44 | 3.85 | 3.01 | 4.37 | 4.75 | 3.83 | 3.93 | 3.77 | 4.05 |
| Range | 1.42 | 1.8 | 1.54 | 1.76 | 2.11 | 1.92 | 1.65 | 1.65 | 1.48 | 1.7 | 1.87 | 1.67 | 1.9 | 1.77 |
| Stnd. skewness | 0.80 | 1.18 | 0.79 | 0.87 | 0.76 | 0.68 | 1.08 | -0.03 | 1.09 | 0.82 | 1.18 | 0.97 | 1.40 | 0.86 |
| Stnd. kurtosis | -1.21 | -1.09 | -1.29 | -1.22 | -1.01 | -0.90 | -1.10 | -1.44 | -1.28 | -1.34 | -1.49 | -1.31 | -1.42 | -1.42 |

| | $X_{15}$ | $X_{16}$ | $X_{17}$ | $X_{18}$ | $X_{19}$ | $X_{20}$ | $X_{21}$ | $X_{22}$ | $X_{23}$ | $X_{24}$ | $X_{25}$ | $X_{26}$ | $X_{27}$ | $X_{28}$ | $X_{29}$ |
|---|---|---|---|---|---|---|---|---|---|---|---|---|---|---|---|
| Count | 48 | 48 | 48 | 48 | 48 | 48 | 48 | 48 | 48 | 48 | 48 | 48 | 48 | 48 | 48 |
| Average | 1.95 | 3.14 | 2.06 | 1.87 | 3.12 | 2.10 | 2.63 | 1.39 | 3.22 | 2.72 | 4.68 | 4.42 | 3.98 | 3.66 | 2.69 |
| Standar deviation | 0.43 | 0.41 | 0.42 | 0.38 | 0.37 | 0.41 | 0.42 | 0.47 | 0.41 | 0.55 | 0.40 | 0.40 | 0.38 | 0.44 | 0.42 |
| Coeff. of variation | 21.8% | 12.9% | 20.4% | 20.3% | 11.8% | 19.4% | 15.9% | 34.1% | 12.8% | 20.4% | 8.6% | 9.0% | 9.6% | 12.0% | 15.6% |
| Minimum | 1.11 | 2.41 | 1.36 | 1.11 | 2.45 | 1.24 | 1.82 | 0.37 | 2.5 | 1.73 | 3.87 | 3.57 | 3.16 | 2.86 | 1.81 |
| Maximum | 2.8 | 3.98 | 2.95 | 2.63 | 3.82 | 2.96 | 3.46 | 2.17 | 4.01 | 3.67 | 5.5 | 5.25 | 4.76 | 4.53 | 3.5 |
| Range | 1.69 | 1.57 | 1.59 | 1.52 | 1.37 | 1.72 | 1.64 | 1.8 | 1.51 | 1.94 | 1.63 | 1.68 | 1.6 | 1.67 | 1.69 |
| Stnd. skewness | -0.92 | 1.27 | 0.93 | 0.74 | 1.03 | 0.26 | 0.50 | 0.25 | 0.95 | 0.98 | 0.88 | 0.77 | 0.79 | 0.97 | 0.46 |
| Stnd. kurtosis | -1.39 | -1.11 | -1.42 | -1.26 | -1.40 | -1.27 | -1.41 | -1.28 | -1.46 | -1.80 | -1.28 | -1.15 | -1.18 | -1.43 | -1.06 |

Measurements of central tendency of variability and shape for each transformed morphometric variable, through its natural logarithm.

Correlation analysis was performed with the 406 pairs of natural logarithms of the variables for 12 different mathematical models. The linear model was chosen as it satisfies three of the conditions stipulated by the methodology proposed by [1] for the construction of systemic morphometric models: a) it is a model for which the majority of the linear regressions have an r-value of at least 0.95, with $p < 0.05$, in accordance with the Least Significant Difference method (LSD), b) it is the simplest model with the aforementioned property and c) it is valid for all combinations of values we are interested in this study.

In Table 3, the linear correlation coefficient values for each pair of transformed variables are presented. Those with $r \geq 0.95$ was selected from these, in order to give the model sufficient inference capabilities, which is achieved when $r > 0.95$, so that $r^2$ is greater than 0.9, implying the fraction of the total variability of the values out of the model's reach is $(1 - r^2)$, which is less than 0.1 [16].

**Table 3. Matrix of correlations between pairs of morphometric variables.**

Table 3

morphometric variables

| | $X_1$ | $X_2$ | $X_3$ | $X_4$ | $X_5$ | $X_6$ | $X_7$ | $X_8$ | $X_9$ | $X_{10}$ | $X_{11}$ | $X_{12}$ | $X_{13}$ | $X_{14}$ | $X_{15}$ | $X_{16}$ | $X_{17}$ | $X_{18}$ | $X_{19}$ | $X_{20}$ | $X_{21}$ | $X_{22}$ | $X_{23}$ | $X_{24}$ | $X_{25}$ | $X_{26}$ | $X_{27}$ | $X_{28}$ | $X_{29}$ |
|---|---|---|---|---|---|---|---|---|---|---|---|---|---|---|---|---|---|---|---|---|---|---|---|---|---|---|---|---|---|
| $X_1$ | | | | | | | | | | | | | | | | | | | | | | | | | | | | | |
| $X_2$ | 0.99 | | | | | | | | | | | | | | | | | | | | | | | | | | | | |
| $X_3$ | 0.99 | 0.98 | | | | | | | | | | | | | | | | | | | | | | | | | | | |
| $X_4$ | 0.99 | 0.99 | 0.99 | | | | | | | | | | | | | | | | | | | | | | | | | | |
| $X_5$ | 0.99 | 0.98 | 0.98 | 0.98 | | | | | | | | | | | | | | | | | | | | | | | | | |
| $X_6$ | 0.99 | 0.98 | 0.97 | 0.96 | 0.97 | | | | | | | | | | | | | | | | | | | | | | | | |
| $X_7$ | 0.98 | 0.98 | 0.98 | 0.98 | 0.98 | 0.96 | | | | | | | | | | | | | | | | | | | | | | | |
| $X_8$ | 0.94 | 0.94 | 0.95 | 0.96 | 0.95 | 0.90 | 0.94 | | | | | | | | | | | | | | | | | | | | | | |
| $X_9$ | 0.99 | 0.99 | 0.99 | 0.99 | 0.99 | 0.98 | 0.99 | 0.95 | | | | | | | | | | | | | | | | | | | | | |
| $X_{10}$ | 0.99 | 0.99 | 0.99 | 0.99 | 0.99 | 0.97 | 0.99 | 0.96 | 0.99 | | | | | | | | | | | | | | | | | | | | |
| $X_{11}$ | 0.97 | 0.98 | 0.98 | 0.98 | 0.97 | 0.95 | 0.98 | 0.95 | 0.98 | 0.99 | | | | | | | | | | | | | | | | | | | |
| $X_{12}$ | 0.99 | 0.99 | 0.99 | 0.99 | 0.98 | 0.97 | 0.99 | 0.95 | 0.99 | 1.00 | 0.98 | | | | | | | | | | | | | | | | | | |
| $X_{13}$ | 0.97 | 0.98 | 0.98 | 0.98 | 0.97 | 0.95 | 0.98 | 0.94 | 0.98 | 0.98 | 0.99 | 0.98 | | | | | | | | | | | | | | | | | |
| $X_{14}$ | 0.96 | 0.97 | 0.98 | 0.98 | 0.96 | 0.94 | 0.97 | 0.94 | 0.97 | 0.98 | 0.97 | 0.97 | 0.97 | | | | | | | | | | | | | | | | |
| $X_{15}$ | 0.98 | 0.99 | 0.99 | 0.99 | 0.98 | 0.96 | 0.98 | 0.95 | 0.99 | 0.99 | 0.99 | 0.99 | 0.98 | 0.98 | | | | | | | | | | | | | | | |
| $X_{16}$ | 0.98 | 0.98 | 0.98 | 0.98 | 0.97 | 0.96 | 0.99 | 0.94 | 0.99 | 0.99 | 0.98 | 0.98 | 0.98 | 0.96 | 0.99 | | | | | | | | | | | | | | |
| $X_{17}$ | 0.95 | 0.96 | 0.96 | 0.96 | 0.95 | 0.93 | 0.95 | 0.95 | 0.97 | 0.97 | 0.97 | 0.96 | 0.96 | 0.95 | 0.97 | 0.95 | | | | | | | | | | | | | |
| $X_{18}$ | 0.98 | 0.98 | 0.98 | 0.99 | 0.97 | 0.96 | 0.98 | 0.96 | 0.98 | 0.99 | 0.98 | 0.98 | 0.97 | 0.97 | 0.98 | 0.97 | 0.96 | | | | | | | | | | | | |
| $X_{19}$ | 0.98 | 0.98 | 0.98 | 0.98 | 0.98 | 0.97 | 0.98 | 0.94 | 0.99 | 0.99 | 0.98 | 0.98 | 0.98 | 0.96 | 0.98 | 0.98 | 0.96 | 0.97 | | | | | | | | | | | |
| $X_{20}$ | 0.95 | 0.93 | 0.96 | 0.95 | 0.95 | 0.92 | 0.95 | 0.95 | 0.96 | 0.96 | 0.95 | 0.95 | 0.94 | 0.94 | 0.95 | 0.94 | 0.95 | 0.95 | 0.95 | | | | | | | | | | |
| $X_{21}$ | 0.98 | 0.98 | 0.99 | 0.99 | 0.98 | 0.96 | 0.98 | 0.96 | 0.99 | 0.99 | 0.99 | 0.99 | 0.98 | 0.97 | 0.99 | 0.98 | 0.98 | 0.98 | 0.98 | 0.96 | | | | | | | | | |
| $X_{22}$ | 0.97 | 0.96 | 0.97 | 0.96 | 0.97 | 0.95 | 0.96 | 0.92 | 0.97 | 0.97 | 0.96 | 0.97 | 0.95 | 0.94 | 0.97 | 0.96 | 0.94 | 0.95 | 0.97 | 0.95 | 0.98 | | | | | | | | |
| $X_{23}$ | 0.98 | 0.98 | 0.98 | 0.98 | 0.98 | 0.96 | 0.98 | 0.95 | 0.99 | 0.99 | 0.98 | 0.98 | 0.98 | 0.97 | 0.98 | 0.98 | 0.97 | 0.98 | 0.99 | 0.96 | 0.99 | 0.97 | | | | | | | |
| $X_{24}$ | 0.95 | 0.96 | 0.96 | 0.96 | 0.94 | 0.93 | 0.95 | 0.92 | 0.95 | 0.97 | 0.97 | 0.96 | 0.96 | 0.94 | 0.96 | 0.95 | 0.95 | 0.95 | 0.95 | 0.92 | 0.97 | 0.93 | 0.96 | | | | | | |
| $X_{25}$ | 1.00 | 0.99 | 0.99 | 0.99 | 0.99 | 0.98 | 0.99 | 0.95 | 0.99 | 1.00 | 0.98 | 0.99 | 0.98 | 0.98 | 0.99 | 0.99 | 0.97 | 0.98 | 0.99 | 0.96 | 0.99 | 0.97 | 0.99 | 0.96 | | | | | |
| $X_{26}$ | 1.00 | 0.99 | 0.99 | 0.99 | 0.99 | 0.98 | 0.99 | 0.95 | 0.99 | 1.00 | 0.98 | 0.99 | 0.98 | 0.97 | 0.99 | 0.98 | 0.96 | 0.98 | 0.98 | 0.96 | 0.99 | 0.97 | 0.99 | 0.96 | 1.00 | | | | |
| $X_{27}$ | 0.99 | 0.99 | 0.99 | 0.99 | 0.99 | 0.97 | 0.98 | 0.96 | 0.99 | 0.99 | 0.98 | 0.99 | 0.98 | 0.98 | 0.99 | 0.98 | 0.96 | 0.98 | 0.98 | 0.96 | 0.99 | 0.96 | 0.98 | 0.96 | 1.00 | 1.00 | | | |
| $X_{28}$ | 0.98 | 0.99 | 0.99 | 0.99 | 0.98 | 0.95 | 0.98 | 0.96 | 0.99 | 0.99 | 0.99 | 0.99 | 0.98 | 0.98 | 0.99 | 0.98 | 0.98 | 0.98 | 0.98 | 0.96 | 0.99 | 0.96 | 0.98 | 0.96 | 0.99 | 0.99 | 0.98 | | |
| $X_{29}$ | 0.98 | 0.97 | 0.98 | 0.98 | 0.97 | 0.96 | 0.97 | 0.94 | 0.98 | 0.98 | 0.97 | 0.98 | 0.97 | 0.96 | 0.98 | 0.97 | 0.96 | 0.97 | 0.98 | 0.95 | 0.98 | 0.97 | 0.97 | 0.94 | 0.98 | 0.98 | 0.98 | 0.97 | |
| | $X_1$ | $X_2$ | $X_3$ | $X_4$ | $X_5$ | $X_6$ | $X_7$ | $X_8$ | $X_9$ | $X_{10}$ | $X_{11}$ | $X_{12}$ | $X_{13}$ | $X_{14}$ | $X_{15}$ | $X_{16}$ | $X_{17}$ | $X_{18}$ | $X_{19}$ | $X_{20}$ | $X_{21}$ | $X_{22}$ | $X_{23}$ | $X_{24}$ | $X_{25}$ | $X_{26}$ | $X_{27}$ | $X_{28}$ | $X_{29}$ |

Afterwards, linear regression equations were obtained, through the correlation model, for all pairs of transformed variables satisfying the conditions mentioned above; these equations are of the form $X_i = mX_j + b$, where $X_i$, $X_j$ are the transformed morphometric variables; m and b are constants.

Bonferroni's multiple comparison procedure was performed to adjust the significance level α, following [17], and afterwards the T test statistic was used to evaluate the significance of the correlation coefficients [16,18]. Statistically significant correlation in the LSD sense was preserved after using Bonferroni's correction method. Throughout the construction of the systemic morphometric model, it is convenient to use those relations between morphometric variables which have the highest possible correlation value. In Figure 3A, we present a graph showing the relations between transformed morphometric variables; 8 of which present an r value of 1 and 96 have r = 0:99. We find high correlation coefficient relations between 22 of the 29 morphometric variables considered; the other 7, $X_8$, $X_{14}$, $X_{17}$, $X_{20}$, $X_{22}$, $X_{24}$, $X_{29}$, exhibit a correlation coefficient in the interval $0.96 < r < 0.99$. In the graph in Figure 3B we present the relations between the variables we considered to construct the systemic morphometric model.

**Fig. 3. (A): Network of relationship among morphometric variables and (B): Relationships used to build the systemic morphometric model**

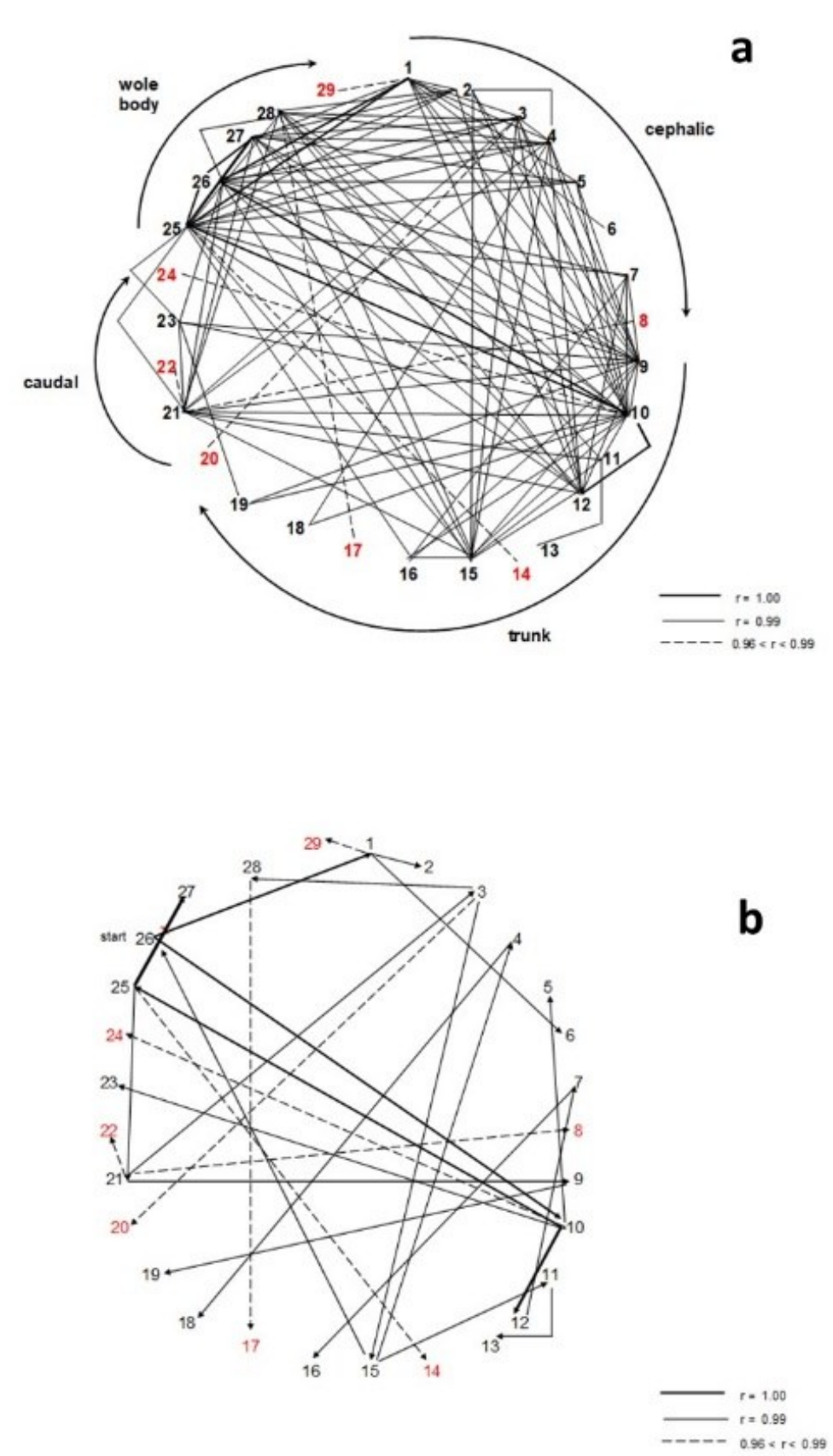

3A. The graph of the relationships established between the morphometric variables corresponds to a network, and it shows the number and intensity of relationships that each variable establishes with the rest of the set. The variables are grouped for the different modules of the body structure of Vieja fenestrata and the entire body.
3B. From the set of variables shown in Fig. 3A, a subset was selected that describes the complete external body structure of Vieja fenestrata. The systemic morphometric model was built based on the relationships between this subset of variables.

# Results

## Systemic Morphometric Model

In the linear regression equations obtained through the aforementioned methodology presented in Table 4, we must keep in mind the linearity condition that was reached when transforming the variables through calculating their natural logarithm. It can be shown that these equations make up a linear system of simultaneous equations, as there is a solution set for it. Linear equations mentioned in the earlier paragraph may be rewritten considering their slope corresponds to the exponent, and their intercept to the coefficient, of an allometric equation of the non-transformed variables of the form:

$$X_i = bX_j^{\alpha}$$

Where $X_i$, $X_j$ are the morphometric variables and b, α are constants. In Table 5, the simultaneous allometric system of equations is presented, which we will henceforth refer to as the SMM.

**Table 4. Linearized allometric equations system**

Table 4

| equation | | $r^2$ | p |
|---|---|---|---|
| Eq.1 | $\ln X_1 = -0.5075 + 0.8827 * \ln X_{26}$ | 0.991 | 0.000 |
| Eq.2 | $\ln X_{10} = -0.8484 + 1.0755 * \ln X_{26}$ | 0.992 | 0.000 |
| Eq.3 | $\ln X_{25} = 1.0411 + 0.9329 * \ln X_{10}$ | 0.993 | 0.000 |
| Eq.4 | $\ln X_{27} = -0.4401 + 0.9443 * \ln X_{25}$ | 0.992 | 0.000 |
| Eq.5 | $\ln X_{12} = -0.7543 + 0.9855 * \ln X_{10}$ | 0.991 | 0.000 |
| Eq.6 | $\ln X_2 = -0.8041 + 1.2426 * \ln X_1$ | 0.976 | 0.000 |
| Eq.7 | $\ln X_{21} = -2.1783 + 1.0272 * \ln X_{25}$ | 0.983 | 0.000 |
| Eq.8 | $\ln X_3 = 0.3230 + 0.9095 * \ln X_{21}$ | 0.976 | 0.000 |
| Eq.9 | $\ln X_{15} = -1.0161 + 1.0938 * \ln X_3$ | 0.978 | 0.000 |
| Eq.10 | $\ln X_4 = 0.4098 + 1.0313 * \ln X_{15}$ | 0.980 | 0.000 |
| Eq.11 | $\ln X_{26} = 2.6036 + 0.9282 * \ln X_{15}$ | 0.980 | 0.000 |
| Eq.12 | $\ln X_7 = 0.1414 + 0.9244 * \ln X_{12}$ | 0.975 | 0.000 |
| Eq.13 | $\ln X_6 = -1.6425 + 1.2122 * \ln X_1$ | 0.977 | 0.000 |
| Eq.14 | $\ln X_5 = -2.2590 + 1.1180 * \ln X_{10}$ | 0.970 | 0.000 |
| Eq.15 | $\ln X_9 = 1.1914 + 0.8994 * \ln X_{21}$ | 0.980 | 0.000 |
| Eq.16 | $\ln X_{11} = 0.5015 + 1.1829 * \ln X_{15}$ | 0.972 | 0.000 |
| Eq.17 | $\ln X_{28} = 0.5959 + 1.1285 * \ln X_3$ | 0.971 | 0.000 |
| Eq.18 | $\ln X_{23} = -0.4882 + 0.9502 * \ln X_{10}$ | 0.980 | 0.000 |
| Eq.19 | $\ln X_{13} = -0.1986 + 1.0414 * \ln X_{11}$ | 0.977 | 0.000 |
| Eq.20 | $\ln X_{16} = 0.1353 + 1.0023 * \ln X_7$ | 0.972 | 0.000 |
| Eq.21 | $\ln X_{18} = -0.1811 + 0.8469 * \ln X_4$ | 0.971 | 0.000 |
| Eq.22 | $\ln X_{19} = -0.2873 + 0.9589 * \ln X_9$ | 0.976 | 0.000 |
| Eq.23 | $\ln X_8 = -0.4714 + 1.0195 * \ln X_{21}$ | 0.915 | 0.000 |
| Eq.24 | $\ln X_{14} = -2.2263 + 1.1353 * \ln X_{25}$ | 0.951 | 0.000 |
| Eq.25 | $\ln X_{17} = -1.3679 + 0.9381 * \ln X_{28}$ | 0.960 | 0.000 |
| Eq.26 | $\ln X_{20} = -0.6627 + 1.0187 * \ln X_3$ | 0.924 | 0.000 |
| Eq.27 | $\ln X_{22} = -1.5205 + 1.1072 * \ln X_{21}$ | 0.952 | 0.000 |
| Eq.28 | $\ln X_{24} = -2.1163 + 1.2392 * \ln X_{10}$ | 0.932 | 0.000 |
| Eq.29 | $\ln X_{29} = -1.2428 + 1.1610 * \ln X_1$ | 0.960 | 0.000 |

**Table 5. Systemic Morphometrics Model (SMM) for the external morphology of *V. fenestrata*.**

**Table 5**

| Equation | |
|---|---|
| Eq.1 | $X_1 = 0.6020 * X_{26}^{0.8827}$ |
| Eq.2 | $X_{10} = 0.4281 * X_{26}^{1.0755}$ |
| Eq.3 | $X_{25} = 2.8323 * X_{10}^{0.9329}$ |
| Eq.4 | $X_{27} = 0.6440 * X_{25}^{0.9443}$ |
| Eq.5 | $X_{12} = 0.4703 * X_{10}^{0.9855}$ |
| Eq.6 | $X_2 = 0.4475 * X_1^{1.2426}$ |
| Eq.7 | $X_{21} = 0.1132 * X_{25}^{1.0272}$ |
| Eq.8 | $X_3 = 1.3813 * X_{21}^{0.9095}$ |
| Eq.9 | $X_{15} = 0.3620 * X_3^{1.0938}$ |
| Eq.10 | $X_4 = 1.5065 * X_{15}^{1.0313}$ |
| Eq.11 | $X_{26} = 13.5123 * X_{15}^{0.9282}$ |
| Eq.12 | $X_7 = 1.1519 * X_{12}^{0.9244}$ |
| Eq.13 | $X_6 = 0.1935 * X_1^{1.2122}$ |
| Eq.14 | $X_5 = 0.1045 * X_{10}^{1.1180}$ |
| Eq.15 | $X_9 = 3.2917 * X_{21}^{0.8994}$ |
| Eq.16 | $X_{11} = 1.6512 * X_{15}^{1.1829}$ |
| Eq.17 | $X_{28} = 1.8147 * X_3^{1.1285}$ |
| Eq.18 | $X_{23} = 0.6137 * X_{10}^{0.9502}$ |
| Eq.19 | $X_{13} = 0.8199 * X_{11}^{1.0414}$ |
| Eq.20 | $X_{16} = 1.1449 * X_7^{1.0023}$ |
| Eq.21 | $X_{18} = 0.8344 * X_4^{0.8469}$ |
| Eq.22 | $X_{19} = 0.7503 * X_9^{0.9589}$ |
| Eq.23 | $X_8 = 0.6241 * X_{21}^{1.0195}$ |
| Eq.24 | $X_{14} = 0.1079 * X_{25}^{1.1353}$ |
| Eq.25 | $X_{17} = 0.2546 * X_{28}^{0.9381}$ |
| Eq.26 | $X_{20} = 0.5155 * X_3^{1.0187}$ |
| Eq.27 | $X_{22} = 0.2186 * X_{21}^{1.1072}$ |
| Eq.28 | $X_{24} = 0.1205 * X_{10}^{1.2392}$ |
| Eq.29 | $X_{29} = 0.2886 * X_1^{1.1610}$ |

The SMM may be used as an inference instrument for the dimensions of the elements of the body plane, starting with the value of a single measurement for any of the variables in the model. We should note that in doing so, numerical uncertainty is propagated as we calculate the values of the other variables successively, which could potentially affect the inference ability of the model. As a simple test, we solved the SMM for the variable $X_{26}$, standard length, and afterwards introduced corresponding values for three organisms that were not part of the sample from which the SMM was obtained. In Table 6, the differences between real and calculated values for one of these three organisms are shown, we present a graph of these differences in Figure 4. We can see the error is distributed at random, the same situation was observed for the other two organisms.

**Table 6. Differences between real and calculated values (mm), using SMM.**

**Table 6**

$X_{26}$ = **118.36** organism not included in the SMM

| variable | calculated | real | difference | dif. % | dif.% abs |
|---|---|---|---|---|---|
| $X_1$ | 40.70 | 41.13 | 0.43 | 1.0% | 0.01 |
| $X_2$ | 44.76 | 43.26 | -1.50 | -3.5% | 0.03 |
| $X_3$ | 21.10 | 20.79 | -0.31 | -1.5% | 0.02 |
| $X_4$ | 16.47 | 17.48 | 1.01 | 5.8% | 0.06 |
| $X_5$ | 12.59 | 13.50 | 0.91 | 6.7% | 0.07 |
| $X_6$ | 17.29 | 17.19 | -0.10 | -0.6% | 0.01 |
| $X_7$ | 28.46 | 26.65 | -1.81 | -6.8% | 0.07 |
| $X_8$ | 13.26 | 11.97 | -1.29 | -10.8% | 0.11 |
| $X_9$ | 48.79 | 49.70 | 0.91 | 1.8% | 0.02 |
| $X_{10}$ | 72.66 | 72.26 | -0.40 | -0.5% | 0.01 |
| $X_{11}$ | 25.66 | 25.21 | -0.45 | -1.8% | 0.02 |
| $X_{12}$ | 32.11 | 31.44 | -0.67 | -2.1% | 0.02 |
| $X_{13}$ | 24.07 | 23.96 | -0.11 | -0.5% | 0.00 |
| $X_{14}$ | 32.93 | 27.76 | -5.17 | -18.6% | 0.19 |
| $X_{15}$ | 10.17 | 9.82 | -0.35 | -3.6% | 0.04 |
| $X_{16}$ | 37.06 | 35.41 | -1.65 | -4.7% | 0.05 |
| $X_{17}$ | 11.24 | 10.50 | -0.74 | -7.0% | 0.07 |
| $X_{18}$ | 8.95 | 9.15 | 0.20 | 2.2% | 0.02 |
| $X_{19}$ | 30.66 | 33.04 | 2.38 | 7.2% | 0.07 |
| $X_{20}$ | 11.52 | 11.39 | -0.13 | -1.1% | 0.01 |
| $X_{21}$ | 20.04 | 19.81 | -0.23 | -1.2% | 0.01 |
| $X_{22}$ | 6.04 | 5.23 | -0.81 | -15.5% | 0.16 |
| $X_{23}$ | 36.02 | 37.35 | 1.33 | 3.6% | 0.04 |
| $X_{24}$ | 24.41 | 22.00 | -2.41 | -10.9% | 0.11 |
| $X_{25}$ | 154.36 | 155.00 | 0.64 | 0.4% | 0.00 |
| $X_{27}$ | 75.08 | 74.65 | -0.43 | -0.6% | 0.01 |
| $X_{28}$ | 56.67 | 51.29 | -5.38 | -10.5% | 0.10 |
| $X_{29}$ | 21.33 | 21.37 | 0.04 | 0.2% | 0.00 |

**Fig. 4. Propagation of the error generated when inferring morphometric values using the SMM**

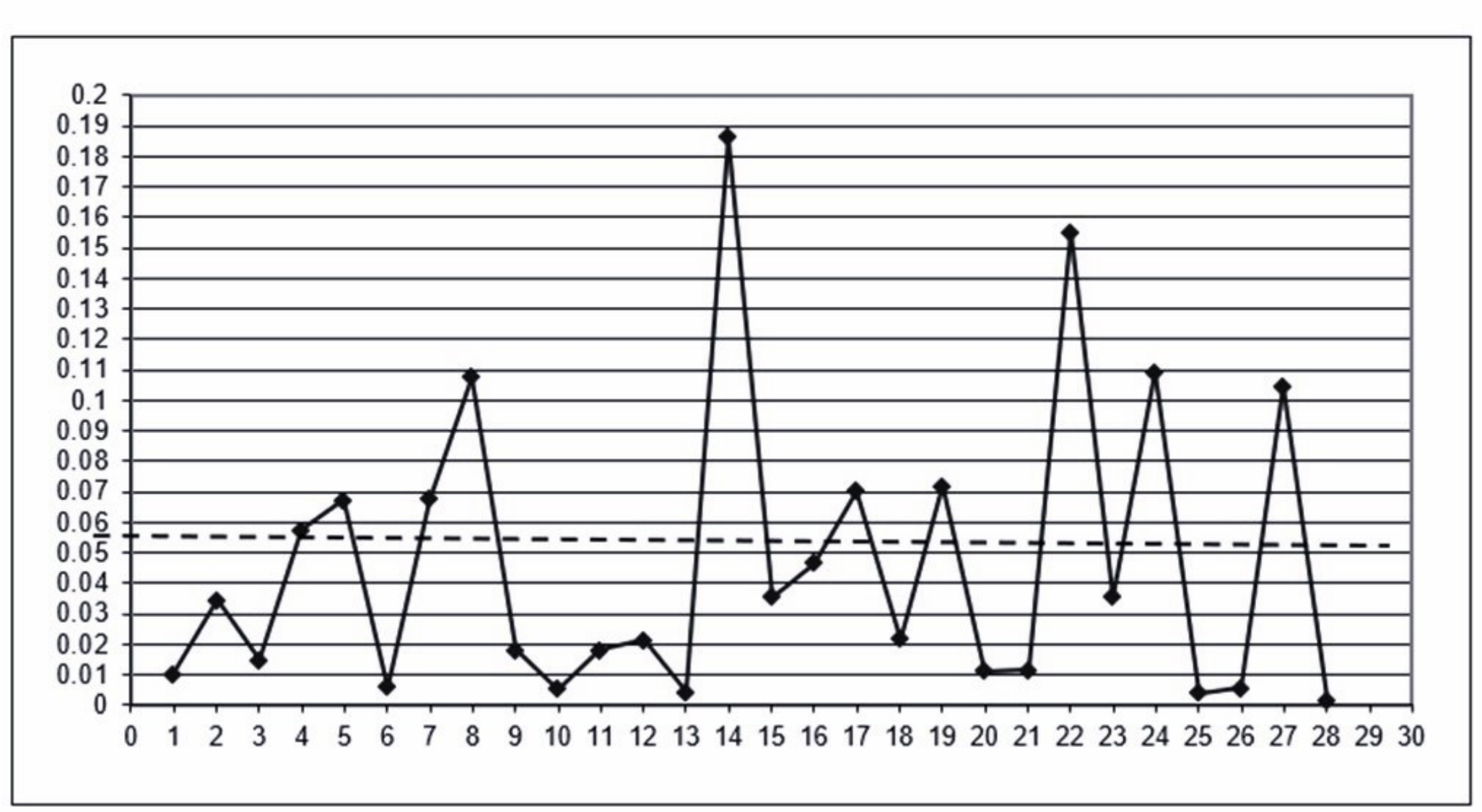

The graph shows the random behavior of the difference, or error, between the calculated and real values for the morphometric variables of an individual not included in the sample of fishes used to formulate the SMM. The model was solved using the value of the $X_{26}$ variable for that individual.

# Theoretical systemic phenotypic space of *Vieja fenestrata*

Using the SMM, it is possible to simulate the total expected phenotypic diversity, the theoretical systemic phenotypic space, for *V. fenestrata*. To achieve this, we must solve the model for one of the variables involved, in Table 7 values for 28 of the 29 variables describing the external morphology of *V. fenestrata* calculated from $X_{26}$, the standard length, in the interval [μ – 2.3σ, μ+ 3.0σ] are shown. Values for μ and σ were estimated from values in Table 1. Initially we considered the interval (μ-3.0σ, μ+3.0σ) for $X_{26}$, as it has 99.7% of the values this variable may take [18]. However, we found that for values less than μ-2.3σ, the value of this variable turns negative, which lacks biological sense.

**Table 7. Values of morphometric variables**

**Table 7**

| $X_{26}$ | 3.57 | ... | 74.67 | 82.16 | **89.64** | 97.12 | 104.61 | ... | 201.90 |
|---|---|---|---|---|---|---|---|---|---|
| | -2.30σ | ... | -0.40σ | -0.20σ | **μ** | +0.20σ | +0.40σ | ... | +3.00σ |
| $X_1$ | 1.85 | ... | 27.10 | 29.49 | **31.85** | 34.18 | 36.50 | ... | 65.21 |
| $X_2$ | 0.96 | ... | 27.01 | 29.99 | **33.00** | 36.03 | 39.09 | ... | 80.41 |
| $X_3$ | 0.79 | ... | 13.70 | 14.99 | **16.26** | 17.53 | 18.80 | ... | 34.82 |
| $X_4$ | 0.41 | ... | 10.12 | 11.20 | **12.28** | 13.37 | 14.46 | ... | 28.98 |
| $X_5$ | 0.19 | ... | 7.24 | 8.12 | **9.01** | 9.93 | 10.85 | ... | 23.93 |
| $X_6$ | 0.41 | ... | 10.56 | 11.70 | **12.84** | 13.99 | 15.15 | ... | 30.62 |
| $X_7$ | 0.92 | ... | 18.12 | 19.90 | **21.67** | 23.44 | 25.21 | ... | 48.02 |
| $X_8$ | 0.34 | ... | 8.17 | 9.04 | **9.90** | 10.77 | 11.65 | ... | 23.24 |
| $X_9$ | 1.90 | ... | 31.84 | 34.78 | **37.71** | 40.62 | 43.51 | ... | 80.05 |
| $X_{10}$ | 1.68 | ... | 44.27 | 49.06 | **53.88** | 58.74 | 63.62 | ... | 129.04 |
| $X_{11}$ | 0.37 | ... | 14.68 | 16.48 | **18.32** | 20.19 | 22.09 | ... | 49.05 |
| $X_{12}$ | 0.79 | ... | 19.71 | 21.81 | **23.92** | 26.04 | 28.17 | ... | 56.56 |
| $X_{13}$ | 0.29 | ... | 13.45 | 15.18 | **16.94** | 18.75 | 20.59 | ... | 47.25 |
| $X_{14}$ | 0.61 | ... | 19.49 | 21.73 | **24.00** | 26.29 | 28.61 | ... | 60.51 |
| $X_{15}$ | 0.28 | ... | 6.34 | 6.99 | **7.65** | 8.30 | 8.96 | ... | 17.58 |
| $X_{16}$ | 1.06 | ... | 20.88 | 22.94 | **24.99** | 27.04 | 29.08 | ... | 55.47 |
| $X_{17}$ | 0.35 | ... | 7.11 | 7.82 | **8.53** | 9.24 | 9.94 | ... | 19.09 |
| $X_{18}$ | 0.39 | ... | 5.93 | 6.45 | **6.98** | 7.50 | 8.01 | ... | 14.44 |
| $X_{19}$ | 1.39 | ... | 20.72 | 22.56 | **24.37** | 26.17 | 27.96 | ... | 50.16 |
| $X_{20}$ | 0.41 | ... | 7.42 | 8.13 | **8.83** | 9.54 | 10.24 | ... | 19.18 |
| $X_{21}$ | 0.54 | ... | 12.47 | 13.76 | **15.05** | 16.35 | 17.64 | ... | 34.75 |
| $X_{22}$ | 0.11 | ... | 3.57 | 3.98 | **4.40** | 4.82 | 5.25 | ... | 11.11 |
| $X_{23}$ | 1.01 | ... | 22.50 | 24.80 | **27.11** | 29.43 | 31.75 | ... | 62.17 |
| $X_{24}$ | 0.23 | ... | 13.21 | 15.00 | **16.85** | 18.75 | 20.70 | ... | 49.73 |
| $X_{25}$ | 4.61 | ... | 97.23 | 107.01 | **116.79** | 126.58 | 136.37 | ... | 263.77 |
| $X_{27}$ | 2.72 | ... | 48.53 | 53.12 | **57.70** | 62.25 | 66.79 | ... | 124.52 |
| $X_{28}$ | 1.40 | ... | 34.81 | 38.51 | **42.23** | 45.97 | 49.73 | ... | 99.70 |
| $X_{29}$ | 0.59 | ... | 13.31 | 14.67 | **16.05** | 17.42 | 18.80 | ... | 36.88 |

Values (mm) for 28 of the 29 morphometric variables, calculated from the standard-length variable ($X_{26}$), in the range [μ-2.3σ), (μ+3.0σ)] of the untransformed values of that variable. With σ = 37.4

In Figure 5, we represent both the SMM-simulated systemic phenotypic space, the theoretical one, and the observed systemic phenotypic space obtained from the measurement of all 29 morphometric values for the 48 fish sample. The theoretical Systemic Phenotypical Space (SPS) was constructed with 28 values for each variable, taken from 28 subintervals of length 0.2σ of the interval [μ – 2.3σ, μ+ 3.0σ], in which $X_{26}$ ranges (Table 7). In turn, the observed SPS consists of 48 values for each variable, the lab measurements of 48 individuals.

**Fig. 5. Three-dimensional theoretical and observed phenotypic space for *Vieja fenestrata***

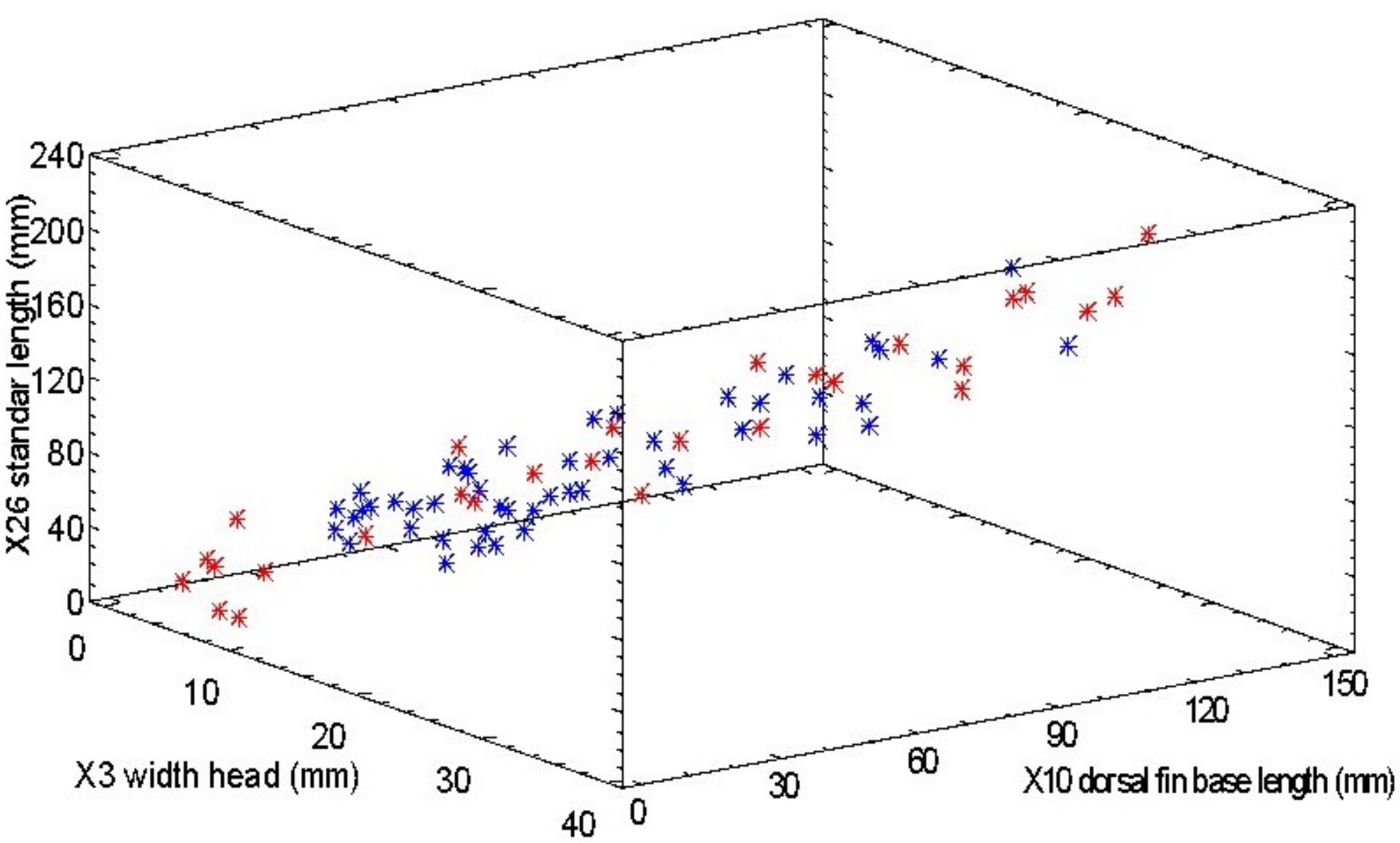

The distribution of observed and theoretical phenotypic space is compared in a three-dimensional cartesian system, the variables used were randomly selected from the set that describes the external body structure of the taxon. This is a simplified 3D version of the 29D system required to represent the phenotypic space of *Vieja fenestrata* from a systemic perspective.

Note that both SPS were built with three variables, however, the systemic approach indicates that each phenotypic space is found in a 29-dimensional system, the same number of variables used to describe the external morphology of *V. fenestrata*, but due to the impossibility of representing this multidimensional space, 2 morphometric variables, $X_3$ and $X_{10}$, were taken at random, in addition to the variable $X_{26}$, the standard length, to obtain the 3D graph of Figure 5; It is important to consider that these three variables are not independent of the other 26 contained in the MMS; on the contrary, they have an influence on the values that those used to construct the phenotypic space can take: $X_{26}$, $X_3$ and $X_{10}$, and vice versa. Accordingly, Figure 5 represents only a part of the 29-dimensional SPS observed in a three-dimensional window; The number of windows of this type that presents the 29-dimensional phenotypic space is equal to 3,654.

In Figure 5, we can see the observed SPS lies within the theoretical SPS, as it can also be established by comparing the values in Table 7 with the maximum and minimum values of each value in Table 1. It is relevant to mention that the theoretical SPS contains the organisms that are possible according to the structural properties of the *V. fenestrata* body plan expressed in the MMS, unlike the spaces constructed with other methodologies which contain all geometrically or mathematically possible phenotypes [19,20].

## Structure of the Systemic Phenotypic Space of *Vieja fenestrata*

As previously mentioned, the theoretical SPS goes beyond the limits of the observed SPS. In saying so, it's worth asserting that while a lower limit for the theoretical SPS exists (since at least one of the variables becomes less than zero when $X_{26}$ is less than $\mu - 2.3\sigma$ as previously mentioned, Tab. 7), there is no upper limit, that is, the SMM can be solved for arbitrarily large values of $X_{26}$. Notably, the SMM has solution for values of $X_{26} \geq \mu + 3\sigma$

Limits of the phenotypic space in axes $X_{26}$, $X_3$, $X_{10}$, calculated through the limit values for these variables in Table 7 and the means and standard deviations in Table 1:

$X_{26}$ (standard length): $[(\mu - 2.3\sigma), (\mu + 3.0\sigma))$, no upper bound.

$X_3$ (head width): $[(\mu - 2.4\sigma), (\mu + 2.8\sigma))$, no upper bound.

$X_{10}$ (dorsal fin's base length) $[(\mu - 2.1\sigma), (\mu + 3.1\sigma))$, no upper bound.

In an analogous manner, the limits of the phenotypic space can be determined on the axes corresponding to the other 26 morphometric variables contained in the MMS.

The SPS with dimensionality greater than three is a contribution of the systemic approach applied to Theoretical Morphology, through which the body plan can be conceptualized as a set of intertwined elements that interact and control each other; as a corollary of this situation, the SPS is constituted only by the combinations (phenotypes) of the elements of the body plan that comply with the structural rules imposed by this system. This implies that a set of individual-level properties, the body plan, determines a set of population-level properties for a given taxon, the phenotypic space. In other words, phenotypic space is an emergent property of the complex system called body structure; just as this is an emergent property of the set of elements that constitute it.

## Representation of the Systemic Phenotypic Space using the Quadratic Mahalanobis Distance

Considering the difficulty of representing the multidimensional SPS, both theoretical and observed, in a dimensionally bounded cartesian system, as was done in the previous section, an alternative way was sought to achieve such representation in a system with the smallest number of dimensions possible, maintaining the systemic character of the theoretical and observed phenotypic spaces. For this purpose, a parameter was conceptualized and sought that would concentrate all the information related to the metric variability of the elements of the body plan, its variance, and the interrelation between said elements, its covariance. We find that these properties are brought together by the so-called Mahalanobis Quadratic Distance (MQD) [21].

The MQD is a measure, in multivariate terms, of the divergence or distance between individuals, or between an individual and a centroid, which considers the variance and covariance between variables, the latter implies that it contemplates the correlation between these and is invariant to changes in scale [22–24].

The equation of the MQD (Eq. 1) between an individual and the centroid is defined by the matrix product:

$$d^2 M = (x^n - c) S^{-1} (x^n - c)^T \qquad (1)$$

Where:

$x^n$ is the row vector corresponding to the nth individual.

c is the centroid: the row vector of means of the variables considered

$S^{-1}$ is the inverse of the covariance matrix

$^T$ denotes the transpose of a matrix

Mahalanobis distance's characteristics fit very well with the approach of Systemic Morphometry which seeks to conceptualize the properties and diversity of body structures of a given taxon by considering the variability and interrelation between the elements that conform them or, in other words, in terms of the variance and covariance of said elements. We consider that, just as the standard deviation is a measure of the distance between a set of data and its arithmetic mean, in the same way the MQD represents the distance between an individual in a SPS and a hypothetical individual, the centroid, which has the average values of all the variables that were used to describe the body structure in said phenotypic space.

From the perspective of Systemic Morphometry, the MQD represents the degree of deviation of a given body structure with respect to the typical body structure (the centroid or average individual), which we can, under certain circumstances, identify as the bauplan of the taxon whose phenotypic space is being studied, (see the Discussion section).

The theoretical value of the MQD in the Systemic Morphometry approach lies in the possibility of capturing this degree of deviation in a single scalar value. This is a useful conceptualization since it allows the multidimensional SPS to be visualized in a two-dimensional space, without artificially reducing the visualizations to a maximum of 3 dimensions as presented in the previous section and also in another work to characterize the SPS of lizards of the species *Uta stansburiana* [1], where this same problem of representation was discussed.

Being able to represent the degree of deviation of a body structure with respect to the typical structure allows to describe and analyze the SPS in a more concise and analytical way, considering that it is possible to capture the entire systemic character of the morphometric variability of a taxon with a single scalar statistic that considers the variance and covariance between the variables that describe said body structure. This allows, for example, to compare the observed SPS of a given taxon vs the theoretical one, through statistics such as mean, standard deviation, and so on, in addition to the usual approach of comparing graphical visualizations of them. This is another contribution of the Systemic Morphometry approach to Theoretical Morphology.

The graphical visualization of the theoretical SPS consists of a frequency distribution of the values of the corresponding MQD of each theoretical organism, in which the frequency distribution of MQD values of the observed organisms is contained, the observed SPS.

Thus, we proceeded to calculate the MQD for both observed and theoretical organisms of *V. fenestrata* considered in this study. We shall refer to these as the Systemic Morphometric Distances (SMD) so as to bring them into the context of Systemic Morphometry.

To calculate the MQD, in general, and the SMD in particular, it is necessary to characterize each individual using a row vector of n variables. Table 8 presents the values of 28 of the 29 variables considered to describe the body structure of the sampled organisms, in this matrix each row represents an individual. The variable $X_{26}$ was not considered because it was used to solve the MMS to obtain the dimensions of the theoretical organisms, therefore, this variable cannot appear in the vector that characterizes these organisms to avoid circularity effects in the calculations. Nor can it appear in the row vectors that represent the measured organisms because the calculation of the SMD of the theoretical and observed organisms must be standardized so that their respective spaces are comparable.

**Table 8. Matrix of values for the morphometric variables of *V. fenestrata***

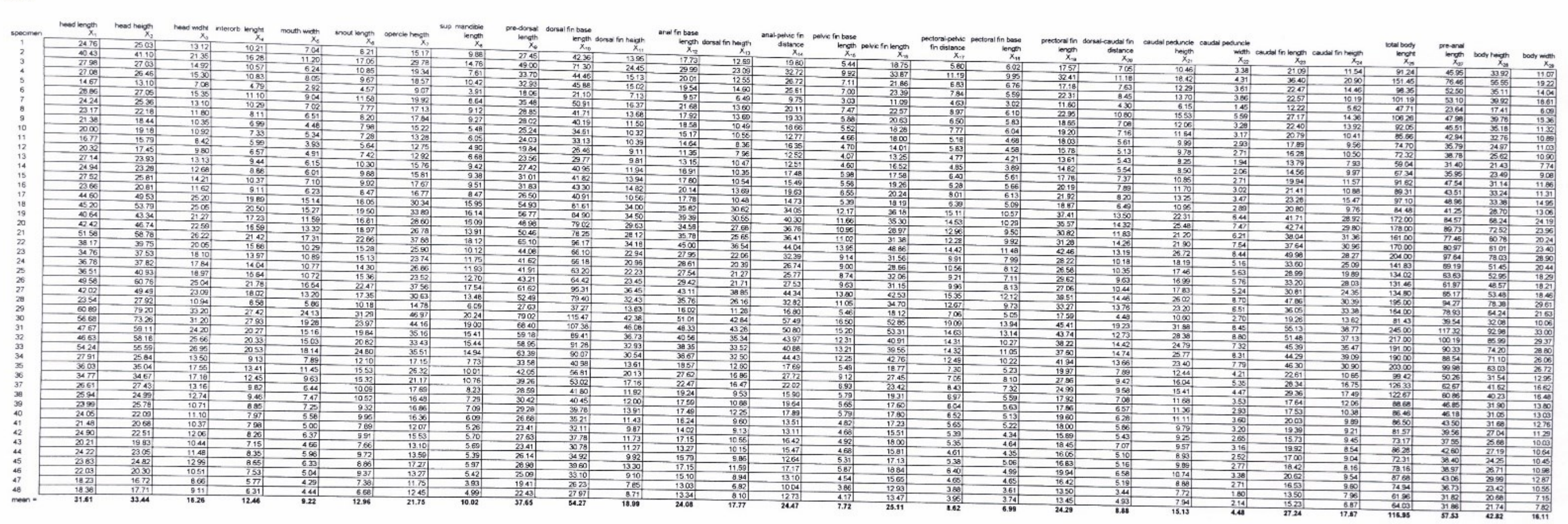

Table 8

| specimen | head length $X_1$ | head heigth $X_2$ | head widht $X_3$ | interorb. lenght $X_4$ | mouth width $X_5$ | snout length $X_6$ | opercle heigth $X_7$ | sup. mandible length $X_8$ | pre-dorsal length $X_9$ | dorsal fin base length $X_{10}$ | dorsal fin heigth $X_{11}$ | anal fin base length $X_{12}$ | dorsal fin heigth $X_{13}$ | anal-pelvic fin distance $X_{14}$ |
|---|---|---|---|---|---|---|---|---|---|---|---|---|---|---|
| 1 | 24.76 | 25.03 | 13.12 | 10.21 | 7.04 | 8.21 | 15.17 | 9.88 | 27.45 | 42.36 | 13.95 | 17.73 | 12.69 | 19.80 |
| 2 | 40.43 | 41.10 | 21.35 | 16.28 | 11.20 | 17.06 | 29.78 | 14.76 | 49.00 | 71.30 | 24.45 | 29.99 | 23.09 | 32.72 |
| 3 | 27.98 | 27.03 | 14.92 | 10.57 | 6.24 | 10.85 | 19.34 | 7.61 | 33.70 | 44.46 | 15.13 | 20.01 | 12.55 | 26.72 |
| 4 | 27.08 | 26.46 | 15.30 | 10.83 | 8.05 | 9.67 | 18.57 | 10.42 | 32.93 | 45.88 | 15.02 | 19.54 | 14.60 | 25.61 |
| 5 | 14.67 | 13.10 | 7.08 | 4.79 | 2.92 | 4.57 | 9.07 | 3.91 | 18.06 | 21.10 | 7.13 | 9.57 | 6.49 | 9.75 |
| 6 | 28.86 | 27.05 | 15.35 | 11.10 | 9.04 | 11.58 | 19.92 | 8.64 | 35.48 | 50.91 | 16.37 | 21.68 | 13.60 | 20.11 |
| 7 | 24.24 | 25.36 | 13.10 | 10.29 | 7.02 | 7.77 | 17.13 | 9.12 | 28.85 | 41.71 | 13.68 | 17.92 | 13.69 | 19.33 |
| 8 | 23.17 | 22.18 | 11.80 | 8.11 | 6.51 | 8.20 | 17.84 | 9.27 | 28.02 | 40.19 | 11.50 | 18.58 | 10.49 | 18.66 |
| 9 | 21.38 | 18.44 | 10.35 | 6.99 | 4.48 | 7.98 | 15.22 | 5.48 | 25.24 | 34.61 | 10.32 | 15.17 | 10.56 | 12.77 |
| 10 | 20.00 | 19.18 | 10.92 | 7.33 | 5.34 | 7.28 | 13.28 | 6.05 | 24.03 | 33.13 | 10.39 | 14.64 | 8.36 | 16.35 |
| 11 | 16.77 | 15.79 | 8.42 | 5.99 | 3.93 | 5.64 | 12.75 | 4.90 | 19.84 | 26.46 | 9.11 | 11.35 | 7.96 | 12.52 |
| 12 | 20.32 | 17.45 | 9.80 | 6.57 | 4.91 | 7.42 | 12.92 | 6.68 | 23.56 | 29.77 | 9.81 | 13.15 | 10.47 | 12.51 |
| 13 | 27.14 | 23.93 | 13.13 | 9.44 | 6.15 | 10.30 | 15.76 | 9.42 | 27.42 | 40.95 | 11.94 | 16.91 | 10.35 | 17.48 |
| 14 | 24.94 | 23.28 | 12.68 | 8.86 | 6.01 | 9.68 | 15.81 | 9.38 | 31.01 | 41.82 | 13.94 | 17.80 | 10.54 | 15.49 |
| 15 | 27.52 | 25.81 | 14.21 | 10.37 | 7.10 | 9.92 | 17.67 | 9.51 | 31.83 | 43.30 | 14.62 | 20.14 | 13.69 | 19.63 |
| 16 | 23.66 | 20.81 | 11.62 | 9.11 | 6.23 | 8.47 | 16.77 | 8.47 | 26.50 | 40.91 | 10.56 | 17.78 | 10.48 | 14.73 |
| 17 | 44.60 | 49.53 | 25.20 | 19.89 | 15.14 | 18.06 | 30.34 | 15.95 | 54.93 | 81.61 | 34.00 | 35.82 | 30.62 | 34.05 |
| 18 | 45.20 | 53.79 | 25.06 | 20.50 | 15.27 | 19.50 | 33.89 | 16.14 | 56.77 | 84.90 | 34.50 | 39.39 | 30.55 | 40.30 |
| 19 | 40.64 | 43.34 | 21.27 | 17.23 | 11.59 | 16.81 | 28.60 | 15.09 | 48.98 | 79.02 | 29.63 | 34.58 | 27.68 | 36.76 |
| 20 | 42.42 | 46.74 | 22.59 | 16.59 | 13.32 | 18.97 | 26.78 | 13.91 | 50.46 | 78.25 | 28.12 | 35.78 | 25.65 | 36.41 |
| 21 | 51.58 | 58.78 | 26.22 | 21.42 | 17.31 | 22.66 | 37.68 | 18.12 | 65.10 | 96.17 | 34.18 | 45.00 | 36.54 | 44.04 |
| 22 | 38.17 | 39.75 | 20.05 | 15.88 | 10.29 | 15.28 | 25.90 | 10.12 | 44.08 | 66.10 | 22.94 | 27.95 | 22.06 | 32.39 |
| 23 | 34.76 | 37.53 | 18.10 | 13.97 | 10.89 | 15.13 | 23.74 | 11.75 | 41.62 | 66.18 | 20.96 | 28.61 | 20.39 | 26.74 |
| 24 | 36.78 | 37.82 | 17.84 | 14.04 | 10.77 | 14.30 | 26.86 | 11.93 | 41.91 | 63.20 | 22.23 | 27.54 | 21.27 | 25.77 |
| 25 | 36.51 | 40.93 | 18.97 | 16.64 | 10.72 | 15.36 | 23.52 | 12.70 | 43.21 | 64.42 | 23.45 | 29.42 | 21.71 | 27.53 |
| 26 | 49.58 | 60.76 | 25.04 | 21.78 | 16.64 | 22.47 | 37.56 | 17.54 | 61.62 | 95.31 | 36.45 | 43.11 | 38.85 | 44.34 |
| 27 | 42.02 | 49.49 | 23.09 | 18.02 | 13.20 | 17.35 | 30.63 | 13.48 | 52.49 | 79.40 | 32.43 | 35.76 | 26.16 | 32.82 |
| 28 | 23.54 | 27.92 | 10.94 | 8.58 | 5.86 | 10.18 | 14.78 | 6.09 | 27.63 | 37.27 | 13.83 | 16.02 | 11.28 | 16.80 |
| 29 | 60.89 | 79.20 | 33.20 | 27.42 | 24.13 | 31.29 | 46.97 | 20.24 | 79.02 | 115.47 | 42.38 | 51.01 | 42.84 | 57.49 |
| 30 | 56.68 | 73.26 | 31.20 | 27.93 | 19.28 | 23.97 | 44.16 | 19.00 | 68.40 | 107.36 | 46.08 | 48.33 | 43.28 | 50.80 |
| 31 | 47.67 | 59.11 | 24.20 | 20.27 | 15.16 | 19.84 | 35.16 | 15.41 | 59.18 | 89.41 | 36.73 | 40.56 | 35.34 | 43.97 |
| 32 | 46.63 | 58.18 | 25.66 | 20.33 | 15.03 | 20.62 | 33.43 | 15.44 | 58.95 | 91.28 | 32.93 | 38.35 | 33.52 | 40.88 |
| 33 | 54.24 | 55.59 | 26.95 | 20.53 | 18.14 | 24.80 | 35.51 | 14.94 | 63.39 | 90.07 | 30.54 | 36.67 | 32.50 | 44.43 |
| 34 | 27.91 | 25.84 | 13.50 | 9.13 | 7.89 | 12.10 | 17.15 | 7.73 | 33.58 | 40.98 | 13.61 | 18.57 | 12.80 | 17.69 |
| 35 | 36.03 | 35.04 | 17.55 | 13.41 | 11.45 | 15.53 | 26.32 | 10.01 | 42.05 | 56.81 | 20.13 | 27.62 | 16.86 | 27.72 |
| 36 | 34.77 | 34.67 | 17.18 | 12.45 | 9.63 | 15.32 | 21.17 | 10.76 | 39.26 | 53.02 | 17.16 | 22.47 | 16.47 | 22.07 |
| 37 | 26.61 | 27.43 | 13.16 | 9.82 | 6.44 | 10.09 | 17.69 | 8.23 | 28.59 | 41.80 | 11.92 | 19.24 | 9.53 | 15.90 |
| 38 | 25.94 | 24.99 | 12.74 | 9.46 | 7.47 | 10.52 | 16.49 | 7.29 | 30.42 | 40.45 | 12.00 | 17.59 | 10.88 | 19.64 |
| 39 | 23.99 | 25.78 | 10.71 | 8.85 | 7.25 | 9.32 | 16.86 | 7.09 | 29.28 | 39.78 | 13.91 | 17.49 | 12.25 | 17.89 |
| 40 | 24.05 | 22.09 | 11.10 | 7.97 | 5.58 | 9.95 | 16.36 | 6.09 | 26.68 | 35.21 | 11.43 | 16.24 | 9.60 | 13.51 |
| 41 | 21.48 | 20.68 | 10.37 | 7.98 | 5.00 | 7.89 | 12.07 | 5.26 | 23.41 | 32.11 | 9.87 | 14.02 | 9.13 | 13.11 |
| 42 | 24.90 | 22.51 | 12.06 | 8.26 | 6.37 | 9.91 | 15.53 | 5.70 | 27.63 | 37.78 | 11.73 | 17.15 | 10.55 | 16.42 |
| 43 | 20.21 | 19.83 | 10.44 | 7.15 | 4.66 | 7.66 | 13.10 | 5.69 | 23.41 | 30.78 | 11.27 | 13.27 | 10.15 | 15.47 |
| 44 | 24.22 | 23.05 | 11.48 | 8.35 | 5.96 | 9.72 | 13.59 | 5.39 | 26.14 | 34.92 | 9.92 | 15.79 | 9.86 | 12.64 |
| 45 | 23.83 | 24.82 | 12.99 | 8.65 | 6.33 | 8.86 | 17.27 | 5.97 | 28.98 | 39.60 | 13.30 | 17.15 | 11.59 | 17.17 |
| 46 | 22.03 | 20.30 | 10.51 | 7.53 | 5.04 | 9.37 | 13.27 | 5.42 | 25.09 | 33.10 | 9.10 | 15.10 | 8.94 | 13.10 |
| 47 | 18.23 | 16.72 | 8.66 | 5.77 | 4.29 | 7.38 | 11.75 | 3.93 | 19.41 | 26.23 | 7.85 | 13.03 | 6.82 | 10.04 |
| 48 | 18.38 | 17.71 | 9.11 | 6.31 | 4.44 | 6.68 | 12.45 | 4.99 | 22.43 | 27.97 | 8.71 | 13.34 | 8.10 | 12.73 |
| mean = | 31.61 | 33.44 | 16.26 | 12.46 | 9.22 | 12.96 | 21.75 | 10.02 | 37.65 | 54.27 | 18.99 | 24.08 | 17.77 | 24.47 |

| specimen | pelvic fin base length $X_{15}$ | pelvic fin length $X_{16}$ | pectoral-pelvic fin distance $X_{17}$ | pectoral fin base length $X_{18}$ | prectoral fin length $X_{19}$ | dorsal-caudal fin distance $X_{20}$ | caudal peduncle heigth $X_{21}$ | caudal peduncle width $X_{22}$ | caudal fin length $X_{23}$ | caudal fin heigth $X_{24}$ | total body lenght $X_{25}$ | pre-anal length $X_{27}$ | body heigth $X_{28}$ | body width $X_{29}$ |
|---|---|---|---|---|---|---|---|---|---|---|---|---|---|---|
| 1 | 5.44 | 18.75 | 5.80 | 6.02 | 17.57 | 7.05 | 10.46 | 3.38 | 21.09 | 11.54 | 91.24 | 45.95 | 33.92 | 11.07 |
| 2 | 9.92 | 33.87 | 11.19 | 9.95 | 32.41 | 11.18 | 18.42 | 4.31 | 36.40 | 20.90 | 151.45 | 76.46 | 55.55 | 19.22 |
| 3 | 7.11 | 21.86 | 6.83 | 6.76 | 17.18 | 7.63 | 12.29 | 3.61 | 22.47 | 14.46 | 98.35 | 52.50 | 35.11 | 14.04 |
| 4 | 7.00 | 23.39 | 7.84 | 5.59 | 22.31 | 8.45 | 13.70 | 3.86 | 22.57 | 10.19 | 101.19 | 53.10 | 39.92 | 18.61 |
| 5 | 3.03 | 11.09 | 4.63 | 3.02 | 11.60 | 4.30 | 6.15 | 1.45 | 12.22 | 5.62 | 47.71 | 23.64 | 17.41 | 6.09 |
| 6 | 7.47 | 22.57 | 8.97 | 6.10 | 22.95 | 10.80 | 15.53 | 5.59 | 27.17 | 14.36 | 106.26 | 47.98 | 39.76 | 15.36 |
| 7 | 5.88 | 20.63 | 6.50 | 5.83 | 18.65 | 7.08 | 12.06 | 3.28 | 22.40 | 13.92 | 92.05 | 46.51 | 36.18 | 11.32 |
| 8 | 5.62 | 18.28 | 7.77 | 6.04 | 19.20 | 7.16 | 11.64 | 3.17 | 20.79 | 10.41 | 85.66 | 42.94 | 32.76 | 10.89 |
| 9 | 4.66 | 18.00 | 5.18 | 4.68 | 18.03 | 5.61 | 9.99 | 2.93 | 17.89 | 9.56 | 74.70 | 35.79 | 24.97 | 11.03 |
| 10 | 4.70 | 14.01 | 5.83 | 4.58 | 15.78 | 5.13 | 9.78 | 2.71 | 16.28 | 10.50 | 72.32 | 38.78 | 25.62 | 10.90 |
| 11 | 4.07 | 13.25 | 4.77 | 4.21 | 13.61 | 5.43 | 9.25 | 1.94 | 13.79 | 7.93 | 59.04 | 31.40 | 21.43 | 7.74 |
| 12 | 4.60 | 16.52 | 4.85 | 3.89 | 14.62 | 5.54 | 8.50 | 2.06 | 14.56 | 9.97 | 67.34 | 35.95 | 23.49 | 9.08 |
| 13 | 5.98 | 17.58 | 6.40 | 5.61 | 17.78 | 7.37 | 10.85 | 2.71 | 19.94 | 11.57 | 91.62 | 47.54 | 31.14 | 11.86 |
| 14 | 5.56 | 19.26 | 5.28 | 5.66 | 20.19 | 7.89 | 11.70 | 3.02 | 21.41 | 10.88 | 89.31 | 43.51 | 33.24 | 11.31 |
| 15 | 6.55 | 20.24 | 8.01 | 6.13 | 21.92 | 8.20 | 13.25 | 3.47 | 23.28 | 15.47 | 97.10 | 48.96 | 33.38 | 14.95 |
| 16 | 5.39 | 18.19 | 6.39 | 5.09 | 18.87 | 6.49 | 10.95 | 2.89 | 20.80 | 9.76 | 84.48 | 41.25 | 28.70 | 13.06 |
| 17 | 12.17 | 36.18 | 15.11 | 10.57 | 37.41 | 13.50 | 22.31 | 6.44 | 41.71 | 28.92 | 172.00 | 84.57 | 68.24 | 24.19 |
| 18 | 11.66 | 35.30 | 14.53 | 10.29 | 35.57 | 14.32 | 25.48 | 7.47 | 42.74 | 29.80 | 178.00 | 89.73 | 72.52 | 23.96 |
| 19 | 10.96 | 28.97 | 12.96 | 9.50 | 30.82 | 11.83 | 21.20 | 6.21 | 38.04 | 31.36 | 161.00 | 77.46 | 60.78 | 20.24 |
| 20 | 11.02 | 31.38 | 12.28 | 9.92 | 31.28 | 14.26 | 21.90 | 7.54 | 37.64 | 30.96 | 170.00 | 80.97 | 61.01 | 23.40 |
| 21 | 13.95 | 48.86 | 14.42 | 11.48 | 42.46 | 13.19 | 26.72 | 8.44 | 49.98 | 28.27 | 204.00 | 97.64 | 78.03 | 28.90 |
| 22 | 9.14 | 31.56 | 9.91 | 7.99 | 28.22 | 10.18 | 18.19 | 5.16 | 33.60 | 25.09 | 141.83 | 69.19 | 51.45 | 20.44 |
| 23 | 9.00 | 28.66 | 10.56 | 8.12 | 26.58 | 10.35 | 17.46 | 5.63 | 28.99 | 19.89 | 134.02 | 63.63 | 52.95 | 18.29 |
| 24 | 8.74 | 32.06 | 9.21 | 7.11 | 29.62 | 9.63 | 16.99 | 5.76 | 33.20 | 28.03 | 131.46 | 61.97 | 48.57 | 18.21 |
| 25 | 9.63 | 31.15 | 9.96 | 8.13 | 27.06 | 10.44 | 17.83 | 5.24 | 30.81 | 24.35 | 134.80 | 65.17 | 53.48 | 18.46 |
| 26 | 13.80 | 42.53 | 15.35 | 12.12 | 38.51 | 14.46 | 26.02 | 8.70 | 47.86 | 30.39 | 195.00 | 94.27 | 78.38 | 29.61 |
| 27 | 11.05 | 34.70 | 12.67 | 9.73 | 33.27 | 13.76 | 23.20 | 6.51 | 36.05 | 33.38 | 164.00 | 78.93 | 64.24 | 21.63 |
| 28 | 5.46 | 18.12 | 7.06 | 5.05 | 17.59 | 4.48 | 10.60 | 2.70 | 19.28 | 13.62 | 81.43 | 39.54 | 32.08 | 10.06 |
| 29 | 16.50 | 52.85 | 19.09 | 13.94 | 45.41 | 19.23 | 31.88 | 8.45 | 56.13 | 38.77 | 245.00 | 117.32 | 92.98 | 33.00 |
| 30 | 15.20 | 53.31 | 14.63 | 13.14 | 43.74 | 12.73 | 28.38 | 8.80 | 51.48 | 37.13 | 217.00 | 100.19 | 85.99 | 29.37 |
| 31 | 12.31 | 40.91 | 14.31 | 10.27 | 38.22 | 14.42 | 24.79 | 7.32 | 45.39 | 35.47 | 191.00 | 90.33 | 74.20 | 28.80 |
| 32 | 13.21 | 39.56 | 14.32 | 11.05 | 37.60 | 14.74 | 25.77 | 8.31 | 44.29 | 39.09 | 190.00 | 88.54 | 71.10 | 26.06 |
| 33 | 12.25 | 42.76 | 12.49 | 10.22 | 41.94 | 13.66 | 23.40 | 7.79 | 46.30 | 30.90 | 203.00 | 99.98 | 63.03 | 26.72 |
| 34 | 5.49 | 18.77 | 7.30 | 5.23 | 19.97 | 7.89 | 12.44 | 4.21 | 22.61 | 10.65 | 99.42 | 50.26 | 31.54 | 12.95 |
| 35 | 9.12 | 27.45 | 7.05 | 8.10 | 27.86 | 9.42 | 16.04 | 5.35 | 28.34 | 16.75 | 126.33 | 62.67 | 41.62 | 16.62 |
| 36 | 6.93 | 23.42 | 8.43 | 7.32 | 24.99 | 9.58 | 15.41 | 4.47 | 29.36 | 17.49 | 122.67 | 60.86 | 40.23 | 16.48 |
| 37 | 5.79 | 19.31 | 6.97 | 5.59 | 17.92 | 7.08 | 11.68 | 3.53 | 17.64 | 12.06 | 88.68 | 46.85 | 31.90 | 13.80 |
| 38 | 5.65 | 17.60 | 6.04 | 5.63 | 17.86 | 6.57 | 11.36 | 2.93 | 17.53 | 10.38 | 86.46 | 46.18 | 31.05 | 13.03 |
| 39 | 5.79 | 17.80 | 6.52 | 5.13 | 19.60 | 6.28 | 11.11 | 3.60 | 20.03 | 9.89 | 86.50 | 43.50 | 31.68 | 12.76 |
| 40 | 4.82 | 17.23 | 5.65 | 5.22 | 18.00 | 5.86 | 9.79 | 3.20 | 19.39 | 9.21 | 81.57 | 39.56 | 27.04 | 11.29 |
| 41 | 4.68 | 15.51 | 5.39 | 4.34 | 15.89 | 5.43 | 9.25 | 2.65 | 15.73 | 9.45 | 73.17 | 37.55 | 25.68 | 10.03 |
| 42 | 4.92 | 18.00 | 5.35 | 4.64 | 18.45 | 7.07 | 9.57 | 3.16 | 19.92 | 8.54 | 86.28 | 42.60 | 27.19 | 10.64 |
| 43 | 4.68 | 15.81 | 4.61 | 4.35 | 16.05 | 5.10 | 8.93 | 2.52 | 17.00 | 9.04 | 72.31 | 38.40 | 24.25 | 10.45 |
| 44 | 5.31 | 17.13 | 5.38 | 5.06 | 16.83 | 5.16 | 9.89 | 2.77 | 18.42 | 8.16 | 78.16 | 38.97 | 26.71 | 10.98 |
| 45 | 5.87 | 18.84 | 6.40 | 4.99 | 19.94 | 6.58 | 10.74 | 3.38 | 20.62 | 9.54 | 87.68 | 43.06 | 29.99 | 12.87 |
| 46 | 4.54 | 15.65 | 4.65 | 4.65 | 16.42 | 5.19 | 8.88 | 2.71 | 16.53 | 9.60 | 74.94 | 36.73 | 23.42 | 10.56 |
| 47 | 3.86 | 12.93 | 3.88 | 3.61 | 13.50 | 3.44 | 7.72 | 1.80 | 13.50 | 7.96 | 61.96 | 31.82 | 20.68 | 7.15 |
| 48 | 4.17 | 13.47 | 3.95 | 3.74 | 13.45 | 4.93 | 7.94 | 2.14 | 15.23 | 6.87 | 64.03 | 31.86 | 21.74 | 7.82 |
| mean = | 7.72 | 25.11 | 8.62 | 6.99 | 24.29 | 8.88 | 15.13 | 4.48 | 27.24 | 17.87 | 116.95 | 57.53 | 42.82 | 16.11 |

Values (mm) were obtained by the measurement of the sample organisms.

Table 9 contains the quotients obtained by dividing each of the values of the variables contained in Table 8 by the average of its respective variable. This was done to unify the scale on which the Observed and Theoretical Systemic Phenotypic Spaces will be represented, and also to equalize the centroids of both spaces. If the data had not been normalized, the centroids would have different values and therefore both types of spaces could not be represented in the same graph, making their comparison difficult.

**Table 9. Matrix of the original normalized values for the morphometric variables *of V. fenestrata***

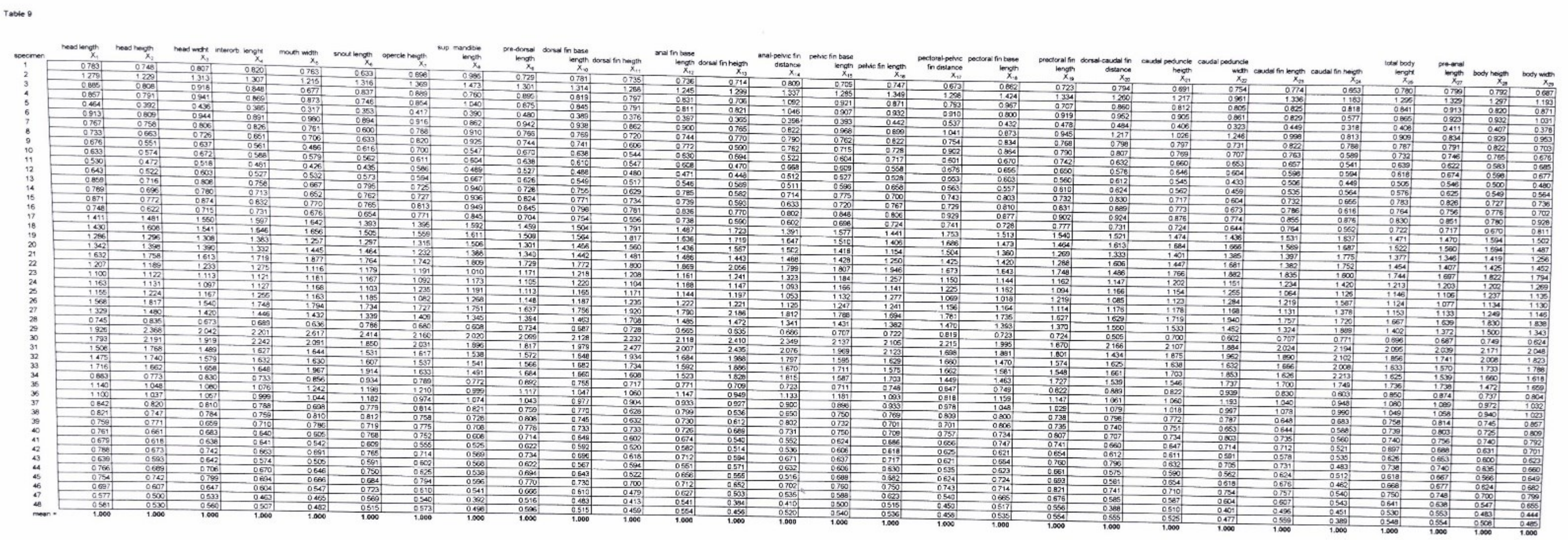

Table 9

| specimen | head length $X_1$ | head heigth $X_2$ | head widht $X_3$ | interorb. lenght $X_4$ | mouth width $X_5$ | snout length $X_6$ | opercle heigth $X_7$ | sup. mandible length $X_8$ | pre-dorsal length $X_9$ | dorsal fin base length $X_{10}$ | dorsal fin heigth $X_{11}$ | anal fin base length $X_{12}$ | dorsal fin heigth $X_{13}$ | anal-pelvic fin distance $X_{14}$ |
|---|---|---|---|---|---|---|---|---|---|---|---|---|---|---|
| 1 | 0.783 | 0.748 | 0.807 | 0.820 | 0.763 | 0.633 | 0.698 | 0.986 | 0.729 | 0.781 | 0.735 | 0.736 | 0.714 | 0.809 |
| 2 | 1.279 | 1.229 | 1.313 | 1.307 | 1.215 | 1.316 | 1.369 | 1.473 | 1.301 | 1.314 | 1.288 | 1.245 | 1.299 | 1.337 |
| 3 | 0.885 | 0.808 | 0.918 | 0.848 | 0.677 | 0.837 | 0.889 | 0.760 | 0.895 | 0.819 | 0.797 | 0.831 | 0.706 | 1.092 |
| 4 | 0.857 | 0.791 | 0.941 | 0.869 | 0.873 | 0.746 | 0.854 | 1.040 | 0.875 | 0.845 | 0.791 | 0.811 | 0.821 | 1.046 |
| 5 | 0.464 | 0.392 | 0.436 | 0.385 | 0.317 | 0.353 | 0.417 | 0.390 | 0.480 | 0.389 | 0.376 | 0.397 | 0.365 | 0.398 |
| 6 | 0.913 | 0.809 | 0.944 | 0.891 | 0.980 | 0.894 | 0.916 | 0.862 | 0.942 | 0.938 | 0.862 | 0.900 | 0.765 | 0.822 |
| 7 | 0.767 | 0.758 | 0.806 | 0.826 | 0.761 | 0.600 | 0.788 | 0.910 | 0.766 | 0.769 | 0.720 | 0.744 | 0.770 | 0.790 |
| 8 | 0.733 | 0.663 | 0.726 | 0.651 | 0.706 | 0.633 | 0.820 | 0.925 | 0.744 | 0.741 | 0.606 | 0.772 | 0.590 | 0.762 |
| 9 | 0.676 | 0.551 | 0.637 | 0.561 | 0.486 | 0.616 | 0.700 | 0.547 | 0.670 | 0.638 | 0.544 | 0.630 | 0.594 | 0.522 |
| 10 | 0.633 | 0.574 | 0.672 | 0.588 | 0.579 | 0.562 | 0.611 | 0.604 | 0.638 | 0.610 | 0.547 | 0.608 | 0.470 | 0.668 |
| 11 | 0.530 | 0.472 | 0.518 | 0.481 | 0.426 | 0.435 | 0.586 | 0.489 | 0.527 | 0.488 | 0.480 | 0.471 | 0.448 | 0.512 |
| 12 | 0.643 | 0.522 | 0.603 | 0.527 | 0.532 | 0.573 | 0.594 | 0.667 | 0.626 | 0.549 | 0.517 | 0.546 | 0.589 | 0.511 |
| 13 | 0.858 | 0.716 | 0.808 | 0.758 | 0.667 | 0.795 | 0.725 | 0.940 | 0.728 | 0.755 | 0.629 | 0.785 | 0.582 | 0.714 |
| 14 | 0.789 | 0.696 | 0.780 | 0.713 | 0.652 | 0.762 | 0.727 | 0.936 | 0.824 | 0.771 | 0.734 | 0.739 | 0.593 | 0.633 |
| 15 | 0.871 | 0.772 | 0.874 | 0.832 | 0.770 | 0.765 | 0.813 | 0.949 | 0.845 | 0.798 | 0.781 | 0.836 | 0.770 | 0.802 |
| 16 | 0.748 | 0.622 | 0.715 | 0.731 | 0.676 | 0.654 | 0.771 | 0.845 | 0.704 | 0.754 | 0.556 | 0.738 | 0.590 | 0.602 |
| 17 | 1.411 | 1.481 | 1.550 | 1.597 | 1.642 | 1.393 | 1.395 | 1.592 | 1.459 | 1.504 | 1.791 | 1.487 | 1.723 | 1.391 |
| 18 | 1.430 | 1.608 | 1.541 | 1.646 | 1.656 | 1.505 | 1.559 | 1.611 | 1.508 | 1.564 | 1.817 | 1.636 | 1.719 | 1.647 |
| 19 | 1.286 | 1.296 | 1.308 | 1.383 | 1.257 | 1.297 | 1.315 | 1.506 | 1.301 | 1.456 | 1.560 | 1.436 | 1.557 | 1.502 |
| 20 | 1.342 | 1.398 | 1.390 | 1.332 | 1.445 | 1.464 | 1.232 | 1.388 | 1.340 | 1.442 | 1.481 | 1.486 | 1.443 | 1.488 |
| 21 | 1.632 | 1.758 | 1.613 | 1.719 | 1.877 | 1.764 | 1.742 | 1.809 | 1.729 | 1.772 | 1.800 | 1.869 | 2.056 | 1.799 |
| 22 | 1.207 | 1.189 | 1.233 | 1.275 | 1.116 | 1.179 | 1.191 | 1.010 | 1.171 | 1.218 | 1.208 | 1.161 | 1.241 | 1.323 |
| 23 | 1.100 | 1.122 | 1.113 | 1.121 | 1.181 | 1.167 | 1.092 | 1.173 | 1.105 | 1.220 | 1.104 | 1.188 | 1.147 | 1.093 |
| 24 | 1.163 | 1.131 | 1.097 | 1.127 | 1.168 | 1.103 | 1.235 | 1.191 | 1.113 | 1.165 | 1.171 | 1.144 | 1.197 | 1.053 |
| 25 | 1.155 | 1.224 | 1.167 | 1.255 | 1.163 | 1.185 | 1.082 | 1.268 | 1.148 | 1.187 | 1.235 | 1.222 | 1.221 | 1.125 |
| 26 | 1.568 | 1.817 | 1.540 | 1.748 | 1.794 | 1.734 | 1.727 | 1.751 | 1.637 | 1.756 | 1.920 | 1.790 | 2.186 | 1.812 |
| 27 | 1.329 | 1.480 | 1.420 | 1.446 | 1.432 | 1.339 | 1.409 | 1.345 | 1.394 | 1.463 | 1.708 | 1.485 | 1.472 | 1.341 |
| 28 | 0.745 | 0.835 | 0.673 | 0.689 | 0.636 | 0.786 | 0.680 | 0.608 | 0.734 | 0.687 | 0.728 | 0.665 | 0.635 | 0.686 |
| 29 | 1.926 | 2.368 | 2.042 | 2.201 | 2.617 | 2.414 | 2.160 | 2.020 | 2.099 | 2.128 | 2.232 | 2.118 | 2.410 | 2.349 |
| 30 | 1.793 | 2.191 | 1.919 | 2.242 | 2.091 | 1.850 | 2.031 | 1.896 | 1.817 | 1.979 | 2.427 | 2.007 | 2.435 | 2.076 |
| 31 | 1.508 | 1.768 | 1.489 | 1.627 | 1.644 | 1.531 | 1.617 | 1.538 | 1.572 | 1.648 | 1.934 | 1.684 | 1.988 | 1.797 |
| 32 | 1.475 | 1.740 | 1.579 | 1.632 | 1.630 | 1.607 | 1.537 | 1.541 | 1.566 | 1.682 | 1.734 | 1.592 | 1.886 | 1.670 |
| 33 | 1.716 | 1.662 | 1.658 | 1.648 | 1.967 | 1.914 | 1.633 | 1.491 | 1.684 | 1.660 | 1.608 | 1.523 | 1.828 | 1.815 |
| 34 | 0.883 | 0.773 | 0.830 | 0.733 | 0.856 | 0.934 | 0.789 | 0.772 | 0.892 | 0.755 | 0.717 | 0.771 | 0.709 | 0.723 |
| 35 | 1.140 | 1.048 | 1.080 | 1.076 | 1.242 | 1.198 | 1.210 | 0.999 | 1.117 | 1.047 | 1.060 | 1.147 | 0.949 | 1.133 |
| 36 | 1.100 | 1.037 | 1.057 | 0.999 | 1.044 | 1.182 | 0.974 | 1.074 | 1.043 | 0.977 | 0.904 | 0.933 | 0.927 | 0.900 |
| 37 | 0.842 | 0.820 | 0.810 | 0.788 | 0.698 | 0.779 | 0.814 | 0.821 | 0.759 | 0.770 | 0.628 | 0.799 | 0.536 | 0.650 |
| 38 | 0.821 | 0.747 | 0.784 | 0.759 | 0.810 | 0.812 | 0.758 | 0.728 | 0.808 | 0.745 | 0.632 | 0.730 | 0.612 | 0.802 |
| 39 | 0.759 | 0.771 | 0.659 | 0.710 | 0.786 | 0.719 | 0.775 | 0.708 | 0.778 | 0.733 | 0.733 | 0.726 | 0.689 | 0.731 |
| 40 | 0.761 | 0.661 | 0.683 | 0.640 | 0.605 | 0.768 | 0.752 | 0.608 | 0.714 | 0.649 | 0.602 | 0.674 | 0.540 | 0.552 |
| 41 | 0.679 | 0.618 | 0.638 | 0.641 | 0.542 | 0.609 | 0.555 | 0.525 | 0.622 | 0.592 | 0.520 | 0.582 | 0.514 | 0.536 |
| 42 | 0.788 | 0.673 | 0.742 | 0.663 | 0.691 | 0.765 | 0.714 | 0.569 | 0.734 | 0.696 | 0.618 | 0.712 | 0.594 | 0.671 |
| 43 | 0.639 | 0.593 | 0.642 | 0.574 | 0.505 | 0.591 | 0.602 | 0.568 | 0.622 | 0.567 | 0.594 | 0.551 | 0.571 | 0.632 |
| 44 | 0.766 | 0.689 | 0.706 | 0.670 | 0.646 | 0.750 | 0.625 | 0.538 | 0.694 | 0.643 | 0.522 | 0.656 | 0.555 | 0.516 |
| 45 | 0.754 | 0.742 | 0.799 | 0.694 | 0.686 | 0.684 | 0.794 | 0.596 | 0.770 | 0.730 | 0.700 | 0.712 | 0.652 | 0.702 |
| 46 | 0.697 | 0.607 | 0.647 | 0.604 | 0.547 | 0.723 | 0.610 | 0.541 | 0.666 | 0.610 | 0.479 | 0.627 | 0.503 | 0.535 |
| 47 | 0.577 | 0.500 | 0.533 | 0.463 | 0.465 | 0.569 | 0.540 | 0.392 | 0.516 | 0.483 | 0.413 | 0.541 | 0.384 | 0.410 |
| 48 | 0.581 | 0.530 | 0.560 | 0.507 | 0.482 | 0.515 | 0.573 | 0.498 | 0.596 | 0.515 | 0.459 | 0.554 | 0.456 | 0.520 |
| mean = | 1.000 | 1.000 | 1.000 | 1.000 | 1.000 | 1.000 | 1.000 | 1.000 | 1.000 | 1.000 | 1.000 | 1.000 | 1.000 | 1.000 |

| specimen | pelvic fin base length $X_{15}$ | pelvic fin length $X_{16}$ | pectoral-pelvic fin distance $X_{17}$ | pectoral fin base length $X_{18}$ | prectoral fin length $X_{19}$ | dorsal-caudal fin distance $X_{20}$ | caudal peduncle heigth $X_{21}$ | caudal peduncle width $X_{22}$ | caudal fin length $X_{23}$ | caudal fin heigth $X_{24}$ | total body lenght $X_{25}$ | pre-anal length $X_{27}$ | body heigth $X_{28}$ | body width $X_{29}$ |
|---|---|---|---|---|---|---|---|---|---|---|---|---|---|---|
| 1 | 0.705 | 0.747 | 0.673 | 0.862 | 0.723 | 0.794 | 0.691 | 0.754 | 0.774 | 0.653 | 0.780 | 0.799 | 0.792 | 0.687 |
| 2 | 1.285 | 1.349 | 1.298 | 1.424 | 1.334 | 1.260 | 1.217 | 0.961 | 1.336 | 1.183 | 1.295 | 1.329 | 1.297 | 1.193 |
| 3 | 0.921 | 0.871 | 0.793 | 0.967 | 0.707 | 0.860 | 0.812 | 0.805 | 0.825 | 0.818 | 0.841 | 0.913 | 0.820 | 0.871 |
| 4 | 0.907 | 0.932 | 0.910 | 0.800 | 0.919 | 0.952 | 0.905 | 0.861 | 0.829 | 0.577 | 0.865 | 0.923 | 0.932 | 1.031 |
| 5 | 0.393 | 0.442 | 0.537 | 0.432 | 0.478 | 0.484 | 0.406 | 0.323 | 0.449 | 0.318 | 0.408 | 0.411 | 0.407 | 0.378 |
| 6 | 0.968 | 0.899 | 1.041 | 0.873 | 0.945 | 1.217 | 1.026 | 1.246 | 0.998 | 0.813 | 0.909 | 0.834 | 0.929 | 0.953 |
| 7 | 0.762 | 0.822 | 0.754 | 0.834 | 0.768 | 0.798 | 0.797 | 0.731 | 0.822 | 0.788 | 0.787 | 0.791 | 0.822 | 0.703 |
| 8 | 0.715 | 0.728 | 0.902 | 0.864 | 0.790 | 0.807 | 0.769 | 0.707 | 0.763 | 0.589 | 0.732 | 0.746 | 0.765 | 0.676 |
| 9 | 0.604 | 0.717 | 0.601 | 0.670 | 0.742 | 0.632 | 0.660 | 0.653 | 0.657 | 0.541 | 0.639 | 0.622 | 0.583 | 0.685 |
| 10 | 0.609 | 0.558 | 0.676 | 0.655 | 0.650 | 0.578 | 0.646 | 0.604 | 0.598 | 0.594 | 0.618 | 0.674 | 0.598 | 0.677 |
| 11 | 0.527 | 0.528 | 0.553 | 0.603 | 0.560 | 0.612 | 0.545 | 0.433 | 0.506 | 0.449 | 0.505 | 0.546 | 0.500 | 0.480 |
| 12 | 0.596 | 0.658 | 0.563 | 0.557 | 0.610 | 0.624 | 0.562 | 0.459 | 0.535 | 0.564 | 0.576 | 0.625 | 0.549 | 0.564 |
| 13 | 0.775 | 0.700 | 0.743 | 0.803 | 0.732 | 0.830 | 0.717 | 0.604 | 0.732 | 0.655 | 0.783 | 0.826 | 0.727 | 0.736 |
| 14 | 0.720 | 0.767 | 0.729 | 0.810 | 0.831 | 0.889 | 0.773 | 0.673 | 0.786 | 0.616 | 0.764 | 0.756 | 0.776 | 0.702 |
| 15 | 0.848 | 0.806 | 0.929 | 0.877 | 0.902 | 0.924 | 0.876 | 0.774 | 0.855 | 0.876 | 0.830 | 0.851 | 0.780 | 0.928 |
| 16 | 0.698 | 0.724 | 0.741 | 0.728 | 0.777 | 0.731 | 0.724 | 0.644 | 0.764 | 0.552 | 0.722 | 0.717 | 0.670 | 0.811 |
| 17 | 1.577 | 1.441 | 1.753 | 1.513 | 1.540 | 1.521 | 1.474 | 1.436 | 1.531 | 1.637 | 1.471 | 1.470 | 1.594 | 1.502 |
| 18 | 1.510 | 1.406 | 1.686 | 1.473 | 1.464 | 1.613 | 1.684 | 1.666 | 1.569 | 1.687 | 1.522 | 1.560 | 1.694 | 1.487 |
| 19 | 1.418 | 1.154 | 1.504 | 1.360 | 1.269 | 1.333 | 1.401 | 1.385 | 1.397 | 1.775 | 1.377 | 1.346 | 1.419 | 1.256 |
| 20 | 1.428 | 1.250 | 1.425 | 1.420 | 1.288 | 1.606 | 1.447 | 1.681 | 1.382 | 1.752 | 1.454 | 1.407 | 1.425 | 1.452 |
| 21 | 1.807 | 1.946 | 1.673 | 1.643 | 1.748 | 1.486 | 1.766 | 1.882 | 1.835 | 1.600 | 1.744 | 1.697 | 1.822 | 1.794 |
| 22 | 1.184 | 1.257 | 1.150 | 1.144 | 1.162 | 1.147 | 1.202 | 1.151 | 1.234 | 1.420 | 1.213 | 1.203 | 1.202 | 1.269 |
| 23 | 1.166 | 1.141 | 1.225 | 1.162 | 1.094 | 1.166 | 1.154 | 1.255 | 1.064 | 1.126 | 1.146 | 1.106 | 1.237 | 1.135 |
| 24 | 1.132 | 1.277 | 1.069 | 1.018 | 1.219 | 1.085 | 1.123 | 1.284 | 1.219 | 1.587 | 1.124 | 1.077 | 1.134 | 1.130 |
| 25 | 1.247 | 1.241 | 1.156 | 1.164 | 1.114 | 1.176 | 1.178 | 1.168 | 1.131 | 1.378 | 1.153 | 1.133 | 1.249 | 1.146 |
| 26 | 1.788 | 1.694 | 1.781 | 1.735 | 1.627 | 1.629 | 1.719 | 1.940 | 1.757 | 1.720 | 1.667 | 1.639 | 1.830 | 1.838 |
| 27 | 1.431 | 1.382 | 1.470 | 1.393 | 1.370 | 1.550 | 1.533 | 1.452 | 1.324 | 1.889 | 1.402 | 1.372 | 1.500 | 1.343 |
| 28 | 0.707 | 0.722 | 0.819 | 0.723 | 0.724 | 0.505 | 0.700 | 0.602 | 0.707 | 0.771 | 0.696 | 0.687 | 0.749 | 0.624 |
| 29 | 2.137 | 2.105 | 2.215 | 1.995 | 1.870 | 2.166 | 2.107 | 1.884 | 2.024 | 2.194 | 2.095 | 2.039 | 2.171 | 2.048 |
| 30 | 1.969 | 2.123 | 1.698 | 1.881 | 1.801 | 1.434 | 1.875 | 1.962 | 1.890 | 2.102 | 1.856 | 1.741 | 2.008 | 1.823 |
| 31 | 1.595 | 1.629 | 1.660 | 1.470 | 1.574 | 1.625 | 1.638 | 1.632 | 1.666 | 2.008 | 1.633 | 1.570 | 1.733 | 1.788 |
| 32 | 1.711 | 1.575 | 1.662 | 1.581 | 1.548 | 1.661 | 1.703 | 1.853 | 1.626 | 2.213 | 1.625 | 1.539 | 1.660 | 1.618 |
| 33 | 1.587 | 1.703 | 1.449 | 1.463 | 1.727 | 1.539 | 1.546 | 1.737 | 1.700 | 1.749 | 1.736 | 1.738 | 1.472 | 1.659 |
| 34 | 0.711 | 0.748 | 0.847 | 0.749 | 0.822 | 0.889 | 0.822 | 0.939 | 0.830 | 0.603 | 0.850 | 0.874 | 0.737 | 0.804 |
| 35 | 1.181 | 1.093 | 0.818 | 1.159 | 1.147 | 1.061 | 1.060 | 1.193 | 1.040 | 0.948 | 1.080 | 1.089 | 0.972 | 1.032 |
| 36 | 0.898 | 0.933 | 0.978 | 1.048 | 1.029 | 1.079 | 1.018 | 0.997 | 1.078 | 0.990 | 1.049 | 1.058 | 0.940 | 1.023 |
| 37 | 0.750 | 0.769 | 0.809 | 0.800 | 0.738 | 0.798 | 0.772 | 0.787 | 0.648 | 0.683 | 0.758 | 0.814 | 0.745 | 0.857 |
| 38 | 0.732 | 0.701 | 0.701 | 0.806 | 0.735 | 0.740 | 0.751 | 0.653 | 0.644 | 0.588 | 0.739 | 0.803 | 0.725 | 0.809 |
| 39 | 0.750 | 0.709 | 0.757 | 0.734 | 0.807 | 0.707 | 0.734 | 0.803 | 0.735 | 0.560 | 0.740 | 0.756 | 0.740 | 0.792 |
| 40 | 0.624 | 0.686 | 0.656 | 0.747 | 0.741 | 0.660 | 0.647 | 0.714 | 0.712 | 0.521 | 0.697 | 0.688 | 0.631 | 0.701 |
| 41 | 0.606 | 0.618 | 0.625 | 0.621 | 0.654 | 0.612 | 0.611 | 0.591 | 0.578 | 0.535 | 0.626 | 0.653 | 0.600 | 0.623 |
| 42 | 0.637 | 0.717 | 0.621 | 0.664 | 0.760 | 0.796 | 0.632 | 0.705 | 0.731 | 0.483 | 0.738 | 0.740 | 0.635 | 0.660 |
| 43 | 0.606 | 0.630 | 0.535 | 0.623 | 0.661 | 0.575 | 0.590 | 0.562 | 0.624 | 0.512 | 0.618 | 0.667 | 0.566 | 0.649 |
| 44 | 0.688 | 0.682 | 0.624 | 0.724 | 0.693 | 0.581 | 0.654 | 0.618 | 0.676 | 0.462 | 0.668 | 0.677 | 0.624 | 0.682 |
| 45 | 0.760 | 0.750 | 0.743 | 0.714 | 0.821 | 0.741 | 0.710 | 0.754 | 0.757 | 0.540 | 0.750 | 0.748 | 0.700 | 0.799 |
| 46 | 0.588 | 0.623 | 0.540 | 0.665 | 0.676 | 0.585 | 0.587 | 0.604 | 0.607 | 0.543 | 0.641 | 0.638 | 0.547 | 0.655 |
| 47 | 0.500 | 0.515 | 0.450 | 0.517 | 0.556 | 0.388 | 0.510 | 0.401 | 0.496 | 0.451 | 0.530 | 0.553 | 0.483 | 0.444 |
| 48 | 0.540 | 0.536 | 0.458 | 0.535 | 0.554 | 0.555 | 0.525 | 0.477 | 0.559 | 0.389 | 0.548 | 0.554 | 0.508 | 0.485 |
| mean = | 1.000 | 1.000 | 1.000 | 1.000 | 1.000 | 1.000 | 1.000 | 1.000 | 1.000 | 1.000 | 1.000 | 1.000 | 1.000 | 1.000 |

Values were obtained by dividing each variable value in Table 8 by its arithmetic mean.

The vector of the averages of the morphometric variables in Table 9 represents the centroid of the Observed SPS described based on SMD.

Table 10 presents the vectors of variables that represent the individuals of the theoretical SPS calculated using the SMM, which was resolved from untransformed values of the variable $X_{26}$, in the interval [(μ - 2.38σ), (μ + 3.00σ)]. In Table 7, values were calculated using (μ - 2.3σ) as the lower limit of $X_{26}$ and it was commented that after said limit, the value of the variable becomes negative when intervals of 0.2σ are taken. In the case of Table 10, the extreme values of the lower limit were explored with the intention of better representing the theoretical SPS, obtaining values greater than 0 in (μ - 2.38σ).

**Table 10. Matrix of values for 28 of the 29 morphometric values of *V. fenestrata***

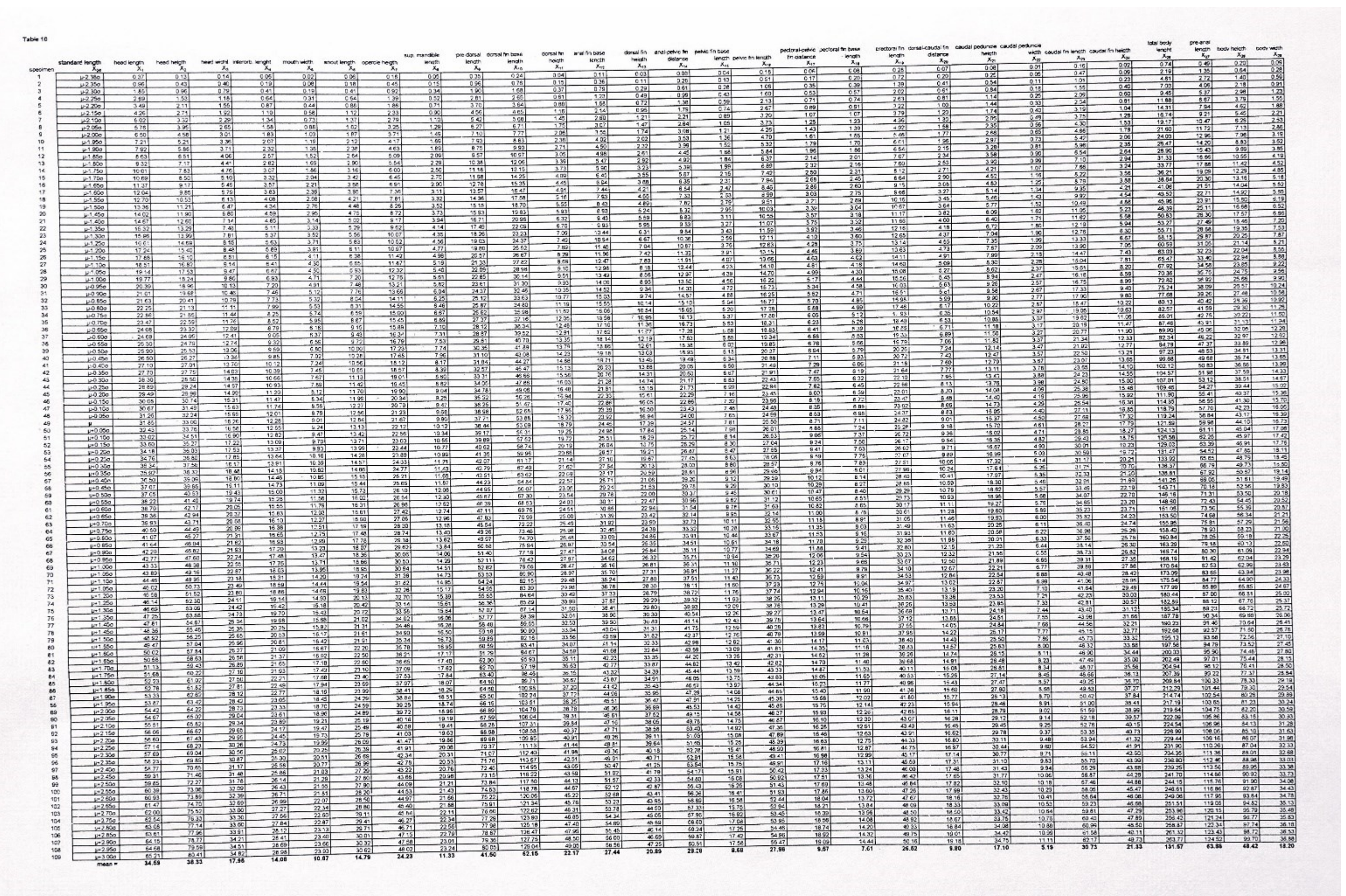

Values (mm) were calculated from the variable $X_{26}$ in the range [(μ-2.3σ), (μ+3.0σ)] of the untransformed values of that variable. With μ = 89.64 and σ = 37.4

Table 11 contains the normalized values of the values in Table 10. The vector of the averages of the morphometric variables in that table represents the centroid of the Theoretical SPS described based on the SMDs.

A generalized application of Eq. 1 to simultaneously calculate the distances of all individuals with respect to the centroid is as follows [25]:

$$Diag\ [(I_n - \frac{1}{n}J_n)XS^{-1}\,X^T\,(I_n - \frac{1}{n}\,J_n\,)] \qquad (2)$$

Where:

$I_n$ is the identity matrix of order n

$J_n$ is the matrix of 1 of order n

X is the data matrix that defines the n individuals

$S^{-1}$ is the inverse of the covariance matrix

$^T$ denotes the transpose of a matrix

**Table 11. Matrix of the normalized calculated values for the morphometric variables *of V. fenestrata***

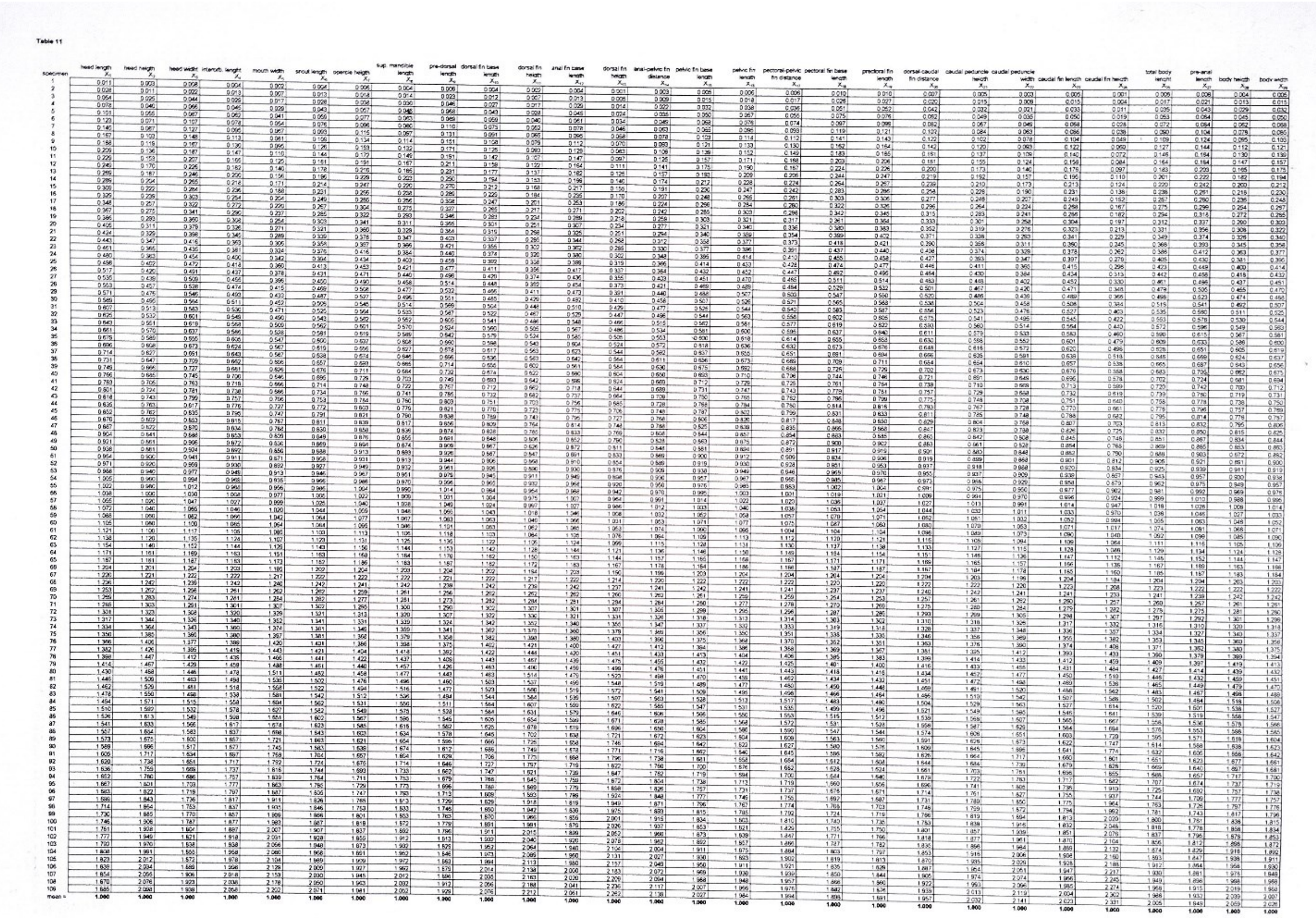

Values were obtained by dividing values in each column in Table 10 by its arithmetic mean.

Table 12 shows the values of the SMD for both the individuals in the Observed Phenotypic Space and for those in the Theoretical Space, calculated using equation Eq. 2.

**Table 12. Systemic morphometric distances, (SMD), of the individuals in the observed and theoretical SPS**

Table 12

| specimen | obs SMD norm. | theo SMD norm. |
|---|---|---|
| 1 | 25.50 | 75.73 |
| 2 | 28.30 | 47.43 |
| 3 | 35.88 | 38.62 |
| 4 | 35.88 | 29.72 |
| 5 | 19.62 | 29.15 |
| 6 | 33.99 | 26.98 |
| 7 | 24.90 | 41.99 |
| 8 | 26.73 | 28.19 |
| 9 | 15.37 | 42.28 |
| 10 | 19.27 | 32.36 |
| 11 | 19.01 | 24.95 |
| 12 | 24.11 | 30.20 |
| 13 | 23.14 | 27.21 |
| 14 | 25.20 | 30.13 |
| 15 | 22.09 | 25.38 |
| 16 | 27.56 | 21.80 |
| 17 | 39.56 | 22.14 |
| 18 | 36.24 | 28.86 |
| 19 | 36.12 | 33.70 |
| 20 | 29.48 | 36.35 |
| 21 | 37.95 | 18.08 |
| 22 | 23.91 | 28.86 |
| 23 | 23.72 | 23.21 |
| 24 | 37.45 | 32.31 |
| 25 | 20.34 | 24.95 |
| 26 | 33.76 | 29.94 |
| 27 | 29.16 | 24.04 |
| 28 | 31.14 | 26.13 |
| 29 | 42.25 | 29.00 |
| 30 | 43.16 | 24.76 |
| 31 | 34.93 | 25.55 |
| 32 | 34.22 | 21.90 |
| 33 | 36.00 | 33.50 |
| 34 | 25.60 | 25.83 |
| 35 | 35.52 | 32.75 |
| 36 | 24.60 | 17.87 |
| 37 | 29.59 | 29.09 |
| 38 | 26.73 | 33.50 |
| 39 | 22.28 | 22.05 |
| 40 | 19.27 | 23.53 |
| 41 | 15.13 | 23.46 |
| 42 | 22.66 | 14.77 |
| 43 | 17.56 | 31.96 |
| 44 | 22.85 | 27.14 |
| 45 | 24.30 | 31.47 |
| 46 | 17.98 | 18.19 |
| 47 | 20.52 | 25.58 |
| 48 | 15.21 | 34.57 |
| 49 |  | 19.78 |
| 50 |  | 35.44 |
| 51 |  | 25.83 |
| 52 |  | 28.57 |
| 53 |  | 15.90 |
| 54 |  | 26.12 |
| 55 |  | 28.67 |
| 56 |  | 25.09 |
| 57 |  | 23.36 |
| 58 |  | 34.70 |
| 59 |  | 28.76 |
| 60 |  | 18.95 |
| 61 |  | 26.58 |
| 62 |  | 22.99 |
| 63 |  | 31.47 |
| 64 |  | 18.45 |
| 65 |  | 24.69 |
| 66 |  | 23.97 |
| 67 |  | 16.46 |
| 68 |  | 26.94 |
| 69 |  | 28.06 |
| 70 |  | 22.70 |
| 71 |  | 24.50 |
| 72 |  | 28.43 |
| 73 |  | 23.52 |
| 74 |  | 26.88 |
| 75 |  | 25.99 |
| 76 |  | 23.49 |
| 77 |  | 29.14 |
| 78 |  | 22.84 |
| 79 |  | 30.10 |
| 80 |  | 24.67 |
| 81 |  | 38.44 |
| 82 |  | 23.25 |
| 83 |  | 25.72 |
| 84 |  | 22.99 |
| 85 |  | 22.08 |
| 86 |  | 19.32 |
| 87 |  | 27.00 |
| 88 |  | 23.59 |
| 89 |  | 24.59 |
| 90 |  | 20.89 |
| 91 |  | 26.13 |
| 92 |  | 19.83 |
| 93 |  | 32.46 |
| 94 |  | 25.70 |
| 95 |  | 25.64 |
| 96 |  | 26.42 |
| 97 |  | 28.25 |
| 98 |  | 39.15 |
| 99 |  | 20.02 |
| 100 |  | 25.08 |
| 101 |  | 27.93 |
| 102 |  | 45.75 |
| 103 |  | 28.15 |
| 104 |  | 26.98 |
| 105 |  | 22.99 |
| 106 |  | 24.79 |
| 107 |  | 29.99 |
| 108 |  | 39.11 |
| 109 |  | 41.54 |
| mean = | 27.41 | 27.74 |
| $\sigma$ = | 7.52 | 7.70 |
| $\sigma^2$ = | 56.58 | 59.31 |
| cv = | 27% | 28% |
| max = | 43.16 | 75.73 |
| min = | 15.13 | 14.77 |
| Range = | 28.03 | 60.96 |
| interval $\mu \pm 2\sigma$ = | 12.7 - 42.2 | 12.6 - 42.8 |

Stem and leave plots:

obs. SMD norm: unit = 1.0   1|2 represents 12.0

```
   9   1|555779999
  22   2|0022223334444
(10)   2|5556678999
  16   3|13344
  11   3|555666779
   2   4|23
```

theo. SMD norm: unit = 1.0   1|2 represents 12.0

```
   1   1|4
  11   1|5678888999
  41   2|001122222222333333333444444444
(41)   2|55555555555666666666777788888888888889999999
  27   3|000111222233344
  12   3|568899

       Hi |41.54 41.99 42.28 45.75 47.43 75.73
```

The distances are calculated with respect to their centroids and based on the values of Tables 9 and 11. We also show statistics such as the mean, standard deviation, variance, etc. and a stem and leaf diagram.

An advantage of analyzing the SPS using the MQD is that it allows analytical comparisons to be made between the observed and theoretical SPS. For example, the mean and standard deviation, and therefore the variance, are very similar between both SPSs. According to these values, the variability within each of the spaces is similar, however when we consider other statistics such as the Range, we realize that the theoretical SPS has extreme values in its upper limit that far exceed those corresponding to the upper limit of the SPS observed.

The extreme values of the stem-leaf diagram for the SMDs of the theoretical SPS show what was commented about the SMD values very distant from its centroid which have no parallel in the observed SPS.

Based once again on the descriptive statistics of the SMD values, we can observe that 95%, the interval $\mu \pm 2\sigma$ of the SMD, of the individuals in both spaces are found in a region of the SMD field whose limits are very similar.

The box and whiskers plots and the density plots , (Fig. 6), for the SMD values of the observed and theoretical SPS show that, even though the second contains a number of organisms greater than twice that of the first, the distribution of both groups is very similar. The density plot shows the degree of overlap between both SPSs.

**Fig. 6. Box and whiskers, and density plots of the observed and theoretical SPS**

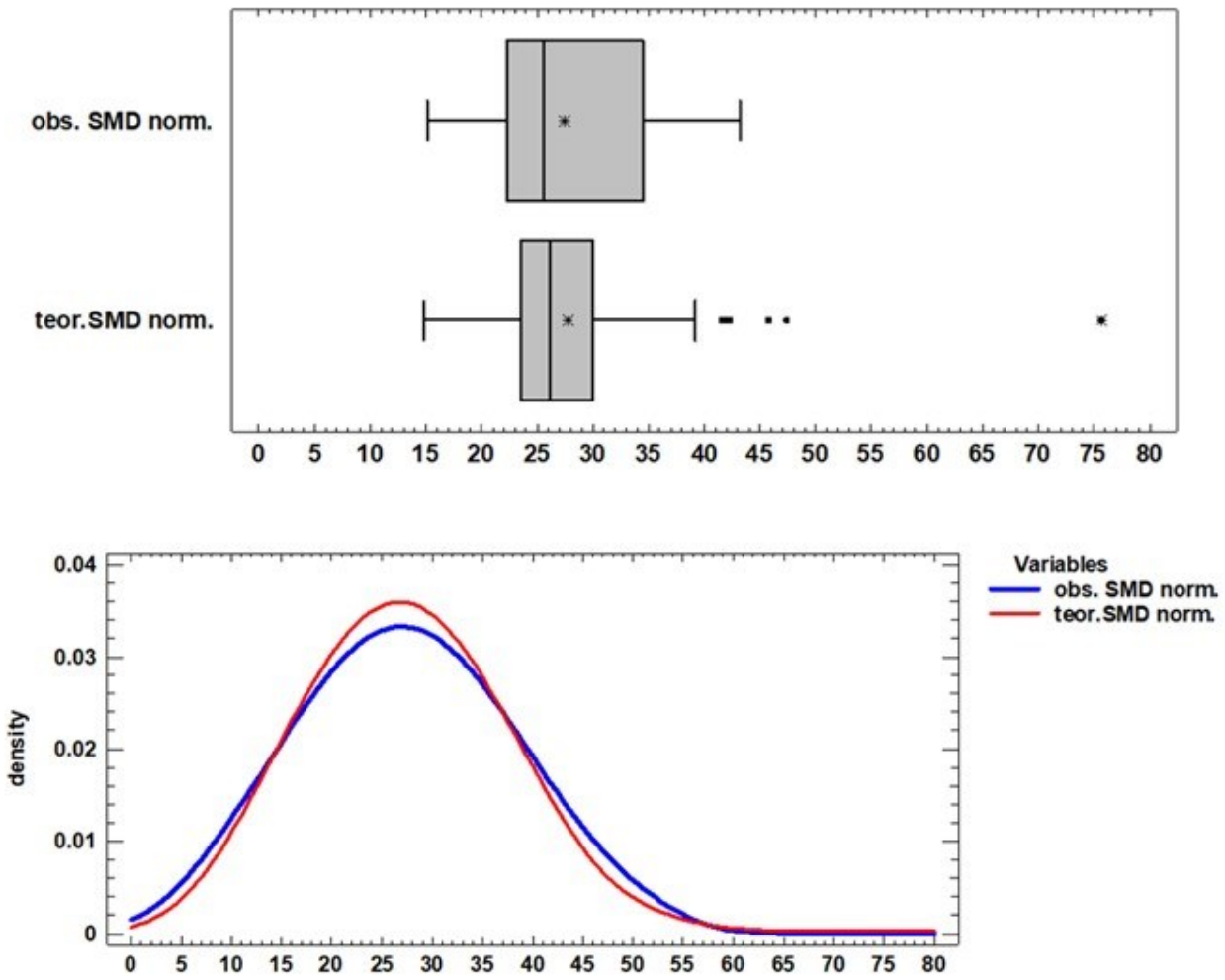

Representation of the SMDs values of real organisms, measured in the laboratory, and the theoretical values, inferred using the SMM. The box-and-whisker plot allows for determining the quartile values and locating the mean and median of these data sets; the density plot is an estimate of the distribution shape of the two data sets.

Figure 7 shows the individuals of the observed SPS, blue boxes, and of the theoretical SPS, red triangles. Both sets of figures are found in a rectangle at the base of which is the scale of the SMD values, which range from 0 to 80. The centroids of both SPS are found at SMD = 0 and the individuals in both spaces are placed at different distances from their respective centroids. The intervals of SMD values form bands or strata starting from the SMD = 0 line; organisms of both types found in these strata have similar SDMs.

**Fig. 7. Observed and theoretical SPS of Vieja fenestrata using SMDs**

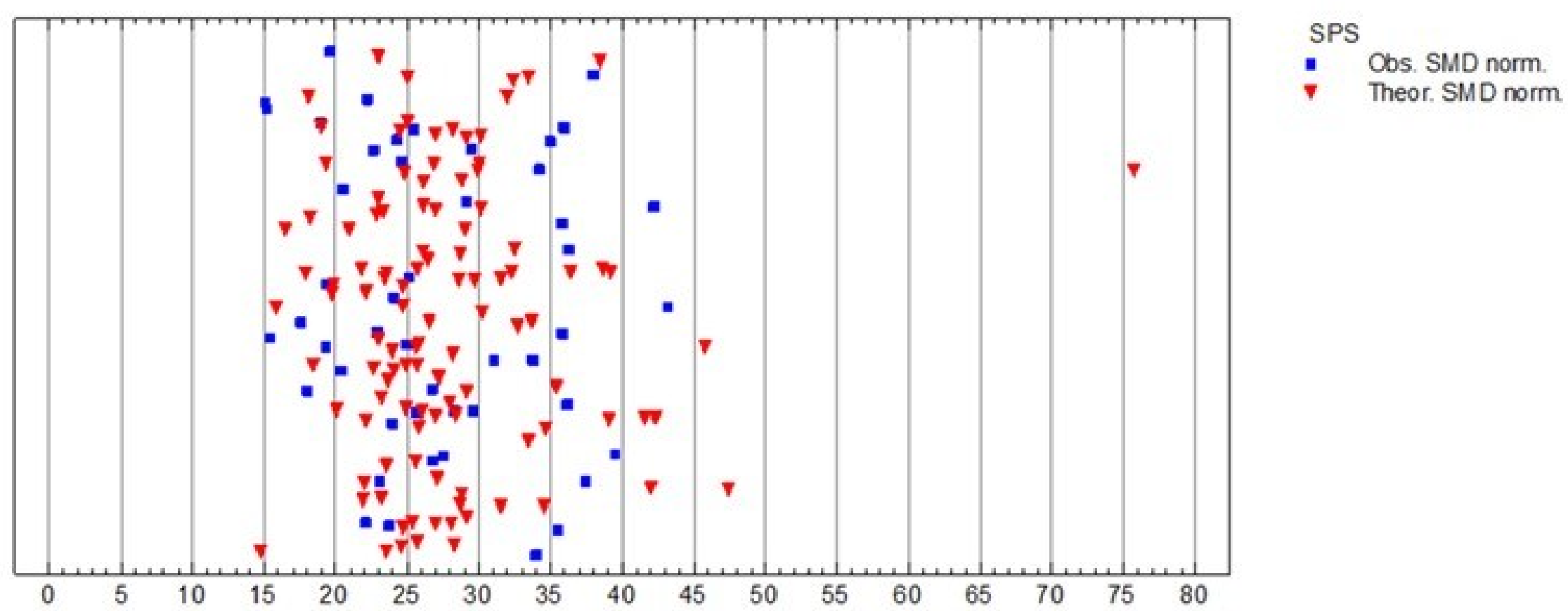

A representation of observed and theoretical phenotypic spaces based on the concept of systemic morphometric distance. It should be interpreted as a line of observed and theoretical individuals located at a certain distance from their respective centroid.

This representation of the SPS is similar to a grid, however, they are different because the individuals in this graph are not in a two-dimensional space, but rather on one, the axis of SMD values. It was presented in this way so that it was possible to observe the strata of the SMD in which observed and theoretical individuals coexist or the strata in which only some type of these individuals are found.

The fact that two or more individuals, observed or theoretical, have a similar SMD does not necessarily indicate that they have a similar body structure. In stratum 40-45 of Figure 7, it can be seen that 3 theoretical and 2 observed individuals coexist, the average SMD for this group is 42; However, their standard lengths ($X_{26}$) are: 11.1, 14.8 and 201.9 for the firsts and 166 and 191 for the seconds. This seems to indicate that individuals in the early phase of development present differences in body structure with respect to the centroid of the same magnitude as mature ones. To explain this, it can be considered that even though the structural differences between the mentioned stages and the centroid are mathematically similar, the biological causes of these differences are not the same; in the case of individuals in the early phase of development, such differences are evident, however, mature individuals have gone through structural changes that move them away from the centroid, an argument for this is presented in the Discussion section.

The empty spaces between the blue boxes and the red triangles have no biological meaning; they are a random result of the representation of both types of figures in the strata. They should not be interpreted as empty spaces in which there are no individuals.
Considering the above, it can be said that the observed and theoretical SPS overlap between the values of SMD 15 and 45, as is easy to see in Figure 7. Beyond this last value, the theoretical SPS shows the possibility of the existence of observable organisms markedly different from their centroid.

It can be noted that no individual, observed or theoretical, has a SMD < 12.6, this implies that in this set of observed and theoretical organisms there is no one that has a body structure very similar to the average body structure. At the other extreme, systemic morphometry predicts that organisms could exist that are very different structurally from those observed and from 95% of the theoretical ones. In this zone of the Figure 7, five empty strata are observed before reaching the stratum with SMD = 75 in which the last theoretical individual is found. These gaps must be interpreted from a biological perspective.

An alternative way to jointly visualize the SPS, observed and theoretical, using SDM is to construct a double histogram in which both spaces are represented. Said double histogram is presented in Figure 8. The information it provides with respect to the relationship between the observed and theoretical phenotypic spaces is similar to what we obtained when analyzing the graph in Figure 7. The advantage of this representation is that it allows us to compare the extent and abundance, the frequency, of organisms in each bar of the histogram of both SPSs and, therefore, compare their sizes. Once again, the lower limit of both spaces becomes evident and the fact that there is a gap between the values 50 and 70 of the SMD, an aspect that will be discussed later. Another advantage of this simultaneous representation of the phenotypic spaces is that it eliminates the perception of empty spaces within each stratum of SDM values, a problem that is observed in the graph of Figure 7 and that can generate interpretation errors.

**Fig. 8. Alternative representation of observed and theoretical SPSs**

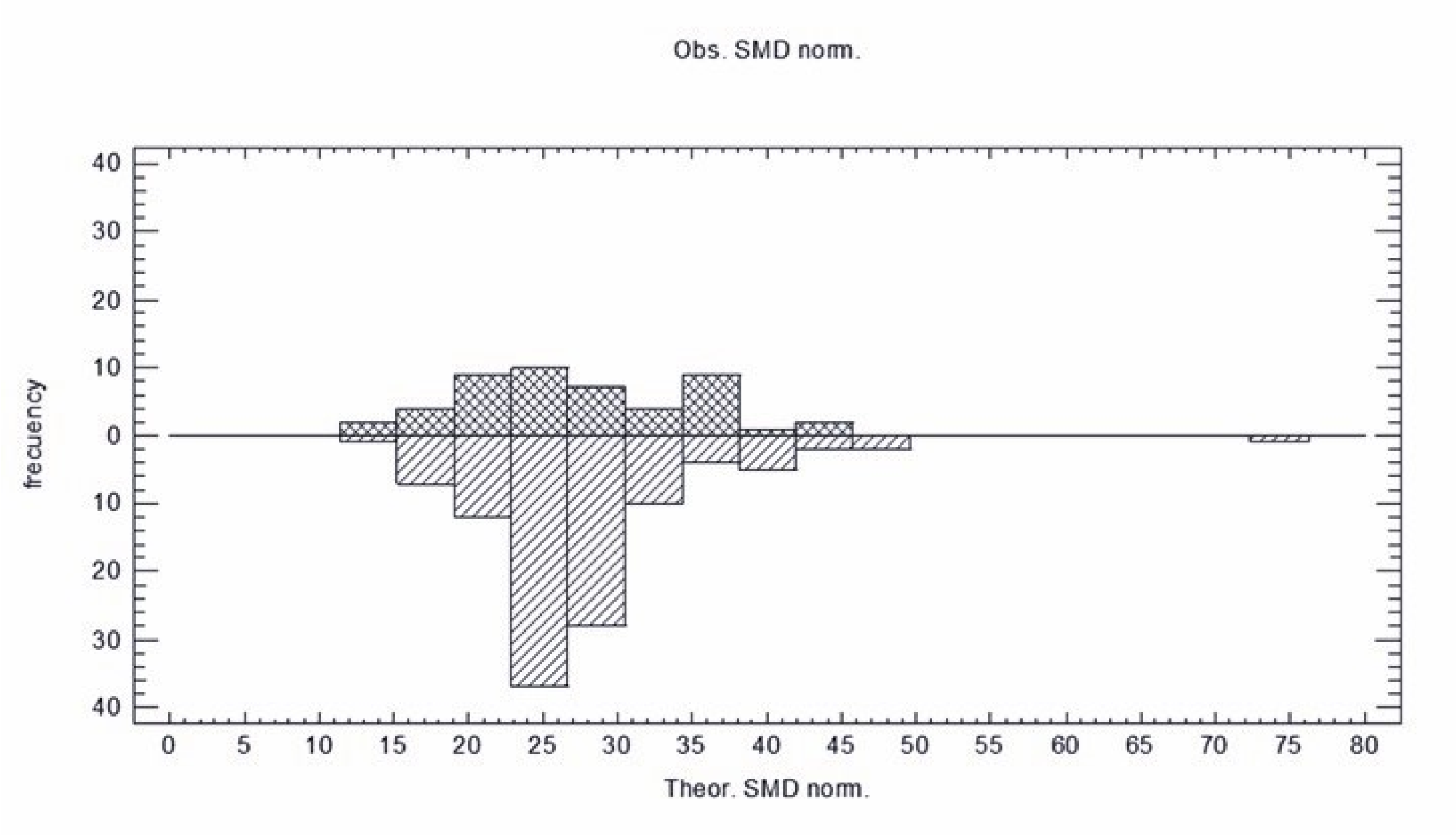

Another option for representing and comparing SPSs is to use a double histogram. This image allows you to visualize the shape of the SPSs as well as the frequency distribution of the SMDs in each of them.

Recapitulating, when comparing the observed and theoretical SPS of *V. fenestrata* based on Figures 7 and 8, the following is observed: a) the first is contained in the second, b) both have the same lower limit, c) both contain subgroups, or strata, consisting of individuals with similar SMD, and d) there is an empty space between the strata of SMD 50 to 70 which is necessary to explain biologically as well as the existence of a common lower limit.

Using SMD statistics, the exploration of the properties of the SPS presented in Figures 7 and 8 can be further refined. For example, studying differences in body structure between individuals from different strata within the same SPS or between homologous strata of different SPSs. Table 13 presents the values for the basic descriptive statistics of the subspaces, (ss), of the observed SPS. These subspaces were defined by taking as reference the grouping of the SMDs in the stem and leaf diagram of Table 12. Table 14 presents the same information as Table 13, but the SMDs corresponds to the individuals of the theoretical SPS.

**Table 13. Descriptive statistics of the SMDs of the observed SPS strati**

Table 13

| specimen | ss1-obs | ss2-obs | ss3-obs | ss4-obs | ss5-obs | ss6-obs |
|---|---|---|---|---|---|---|
| 1 | 15.13 | 20.34 | 25.20 | 31.14 | 35.52 | 42.25 |
| 2 | 15.21 | 20.52 | 25.50 | 33.76 | 35.88 | 43.16 |
| 3 | 15.37 | 22.09 | 25.60 | 33.99 | 35.88 | |
| 4 | 17.56 | 22.28 | 26.73 | 34.22 | 36.00 | |
| 5 | 17.98 | 22.66 | 26.73 | 34.93 | 36.12 | |
| 6 | 19.01 | 22.85 | 27.56 | | 36.24 | |
| 7 | 19.27 | 23.14 | 28.30 | | 37.45 | |
| 8 | 19.27 | 23.72 | 29.16 | | 37.95 | |
| 9 | 19.62 | 23.91 | 29.48 | | 39.56 | |
| 10 | | 24.11 | 29.59 | | | |
| 11 | | 24.30 | | | | |
| 12 | | 24.60 | | | | |
| 13 | | 24.90 | | | | |
| n = | 9 | 13 | 10 | 5 | 9 | 2 |
| mean = | 17.60 | 23.03 | 27.39 | 33.61 | 36.73 | 42.71 |
| median = | 17.98 | 23.14 | 27.15 | 33.99 | 36.12 | 42.71 |
| std desv = | 1.89 | 1.45 | 1.69 | 1.45 | 1.33 | 0.64 |
| variance = | 3.57 | 2.10 | 2.85 | 2.10 | 1.76 | 0.41 |

**Table 14. Descriptive statistics of the SMDs of the theoretical SPS strati**

Table 14

| specimen | ss1-teo | ss2-teo | ss3-teo | ss4-teo | ss5-teo | ss6-teo | ss7-teo |
|---|---|---|---|---|---|---|---|
| 1 | 14.77 | 20.02 | 25.08 | 30.10 | 35.44 | 41.54 | 75.73 |
| 2 | 15.90 | 20.89 | 25.09 | 30.13 | 36.35 | 41.99 | |
| 3 | 16.46 | 21.80 | 25.38 | 30.20 | 38.44 | 42.28 | |
| 4 | 17.87 | 21.90 | 25.55 | 31.47 | 38.62 | 45.75 | |
| 5 | 18.08 | 22.05 | 25.58 | 31.47 | 39.11 | 47.43 | |
| 6 | 18.19 | 22.08 | 25.64 | 31.96 | 39.15 | | |
| 7 | 18.45 | 22.14 | 25.70 | 32.31 | | | |
| 8 | 18.95 | 22.70 | 25.72 | 32.36 | | | |
| 9 | 19.32 | 22.84 | 25.83 | 32.46 | | | |
| 10 | 19.78 | 22.99 | 25.83 | 32.75 | | | |
| 11 | 19.83 | 22.99 | 25.99 | 33.50 | | | |
| 12 | | 22.99 | 26.12 | 33.50 | | | |
| 13 | | 23.21 | 26.13 | 33.70 | | | |
| 14 | | 23.25 | 26.13 | 34.57 | | | |
| 15 | | 23.36 | 26.42 | 34.70 | | | |
| 16 | | 23.46 | 26.58 | | | | |
| 17 | | 23.49 | 26.88 | | | | |
| 18 | | 23.52 | 26.94 | | | | |
| 19 | | 23.53 | 26.98 | | | | |
| 20 | | 23.59 | 26.98 | | | | |
| 21 | | 23.97 | 27.00 | | | | |
| 22 | | 24.04 | 27.14 | | | | |
| 23 | | 24.50 | 27.21 | | | | |
| 24 | | 24.59 | 27.93 | | | | |
| 25 | | 24.67 | 28.06 | | | | |
| 26 | | 24.69 | 28.15 | | | | |
| 27 | | 24.76 | 28.19 | | | | |
| 28 | | 24.79 | 28.25 | | | | |
| 29 | | 24.95 | 28.43 | | | | |
| 30 | | 24.95 | 28.57 | | | | |
| 31 | | | 28.67 | | | | |
| 32 | | | 28.76 | | | | |
| 33 | | | 28.86 | | | | |
| 34 | | | 28.86 | | | | |
| 35 | | | 29.00 | | | | |
| 36 | | | 29.09 | | | | |
| 37 | | | 29.14 | | | | |
| 38 | | | 29.15 | | | | |
| 39 | | | 29.72 | | | | |
| 40 | | | 29.94 | | | | |
| 41 | | | 29.99 | | | | |
| n = | 11 | 30 | 41 | 15 | 6 | 5 | 1 |
| mean = | 17.96 | 23.29 | 27.33 | 32.35 | 37.85 | 43.80 | --- |
| median = | 18.19 | 23.41 | 27.00 | 32.36 | 38.53 | 42.28 | --- |
| std desv = | 1.63 | 1.23 | 1.48 | 1.50 | 1.57 | 2.63 | --- |
| variance = | 2.66 | 1.52 | 2.19 | 2.24 | 2.46 | 6.92 | --- |

Observing Tables 13 and 14, it is possible to notice that the number of individuals varies between the strata of each of the SPS, however the variability of body structures within these strata, measured based on the standard deviation of the SMD, tends to be homogeneous between the first 5 of the 6 subspaces of the observed SPS, the same happening for the theoretical SPS.

It is interesting to investigate whether the standard deviations, and therefore the variances, are the same or different between the strata of the observed and theoretical SPS of *V. fenestrata*. Table 15 presents a matrix of comparisons between the standard deviations of pairs of observed and theoretical subspaces. The results of these comparisons indicate that there are no significant differences between the standard deviations of the compared strata or subspaces, with 95% confidence.

**Table 15. Comparison of the standard deviations among pairs of strati of the observed and theoretical SPS**

**Table 15**

| comparison | Sigma1 | Sigma2 | F-Ratio | p-valor |
|---|---|---|---|---|
| ss1-obs / ss1-teo | 1.89 | 1.63 | 1.34 | 0.65 |
| ss2-obs / ss2-teo | 1.45 | 1.23 | 1.38 | 0.46 |
| ss3-obs / ss3-teo | 1.69 | 1.48 | 1.30 | 0.53 |
| ss4-obs / ss4-teo | 1.45 | 1.50 | 0.93 | 0.94 |
| ss5-obs / ss5-teo | 1.33 | 1.57 | 0.72 | 0.64 |
| ss6-obs / ss6-teo | 0.64 | 2.63 | 0.06 | 0.36 |

With $p \geq 0.05$ there are not statistically significant differences between pairs of standard deviations with 95% confidence.

# Discussion

The application of the systemic approach to the study of the theoretical morphology of *V. fenestrata* allowed obtaining a morphometric model with different characteristics from those models developed with geometric and traditional morphometry for the genera *Amphilophus* and *Vieja* [7,8,26]. The central idea is that the systemic model reflects the relationships established between specific morphometric properties of the body plan of the taxon studied, avoiding the implicit reductionism by analyzing only the relationship between two features, as allometric models do [27–30].

The SMM foundation is the idea that all elements of the body plan are structurally intertwined and influence one each other. Evidently, a single morphometrical characteristic is distinctly influenced by every other variable within the system (by some more than by others), which is reflected in the correlation coefficient between them. Therefore, the body plan capacity of acquiring various configurations (phenotypes) depends on the intensity with which its elements influence each other.

Since all the variables used to build the MMS are expressed in the same linear units (millimeters), the expected allometric exponent is 1. But, as shown in the MMS (Table 5), there are some exponents less than 1 and others greater to this value, which means that some pairs of body plan features exhibit negative allometry (exponent < 1), while others exhibit positive allometry (exponent > 1). This mixed growth model (isometric and allometric) has been described in some vertebrates such as amphibians and reptiles [31,32].

Table 16 contains a stem-and-leaf plot that shows the distribution of the exponents of the equations contained in the MMS; 12 exponents have values <1, 11>1 and 6 ≈ 1. Since some of these exponents differ little from 1, while others differ widely, a geometric similarity or isometry test was applied [33], to determine analytically which of the equations can be identified as allometric or linear. This test consists of determining the exponent a in the allometric equation $y = bX^{a}$, and its Standard Error (SEa), for a pair of morphometric variables, to then estimate the probability (p), under a null hypothesis of isometrics, of randomly obtaining a high (a>1) or low (a<1) value, such as those resulting from the relationship between the variables studied. The p value is approximated by a t-distribution with n-2 degrees of freedom, where n is the number of individuals measured: $t_{n-2} = \frac{a-1}{SE_a}$ ; if $p < 0.05$, allometry can be considered to exist with 95% confidence. The results presented in Table 17 show that 6 relationships between morphometric variables are isometric with exponents between [0.99-1.03], 11 show negative allometry with exponents between [0.85-0.96] and 12 show positive allometry with exponents between [1.04- 1.24]. Strictly, according to this last analysis, the SMM of *V. fenestrata* contains 6 linear and 23 allometric equations. The exponents identified as linear or allometric with this test are not exactly the same as those identified in the stem-and-leaf plot, but both analyses agreed with the number of linear and allometric equations in the SMM.

**Table 16. Stem and leaf diagram for the allometric exponents of the SMM**

**Table 16**

unit = 0.01    1|2 represents 0.12

| | |
|---|---|
| 1 | 8\|4 |
| 3 | 8\|89 |
| 9 | 9\|022334 |
| 12 | 9\|558 |
| (6) | 10\|011234 |
| 11 | 10\|79 |
| 9 | 11\|0123 |
| 5 | 11\|68 |
| 3 | 12\|134 |

**Table 17. Results of the geometric similitude or isometry test**

**Table 17**

| equation | variables | variables name | | exponents | SEa | $t_{48-2=46}$ | p value | type of allometry |
|---|---|---|---|---|---|---|---|---|
| Eq.1 | $X_1 - X_{26}$ | Head length | Standard length | 0.883 | 0.0123 | -9.570 | <0.005 | ( - ) |
| Eq.2 | $X_{10} - X_{26}$ | Dorsal fin base length | Standard length | 1.076 | 0.0143 | 5.264 | <0.005 | ( + ) |
| Eq.3 | $X_{25} - X_{10}$ | Total length | Dorsal fin base length | 0.933 | 0.0112 | -5.997 | <0.005 | ( - ) |
| Eq.4 | $X_{27} - X_{25}$ | Preanal length | Total length | 0.944 | 0.0129 | -4.325 | <0.005 | ( - ) |
| Eq.5 | $X_{12} - X_{10}$ | Anal fin base length | Dorsal fin base length | 0.986 | 0.0139 | -1.041 | >0.100 | isometry |
| Eq.6 | $X_2 - X_1$ | Head height | Head length | 1.243 | 0.0285 | 8.504 | <0.005 | ( + ) |
| Eq.7 | $X_{21} - X_{25}$ | Caudal peduncle height | Total length | 1.027 | 0.0198 | 1.376 | >0.05 | isometry |
| Eq.8 | $X_3 - X_{21}$ | Head width | Caudal peduncle height | 0.910 | 0.0210 | -4.309 | <0.005 | ( - ) |
| Eq.9 | $X_{15} - X_3$ | Pelvic fin base length | Head width | 1.094 | 0.0244 | 3.842 | <0.005 | ( + ) |
| Eq.10 | $X_4 - X_{15}$ | Interorbital length | Pelvic fin base length | 1.031 | 0.0216 | 1.451 | >0.05 | isometry |
| Eq.11 | $X_{26} - X_{15}$ | Standard length | Pelvic fin base length | 0.928 | 0.0198 | -3.630 | <0.005 | ( - ) |
| Eq.12 | $X_7 - X_{12}$ | Opercle height | Anal fin base length | 0.924 | 0.0220 | -3.443 | <0.005 | ( - ) |
| Eq.13 | $X_6 - X_1$ | Snout length | Head length | 1.212 | 0.0143 | 14.796 | <0.005 | ( + ) |
| Eq.14 | $X_5 - X_{10}$ | Mouth width | Dorsal fin base length | 1.118 | 0.0287 | 4.105 | <0.005 | ( + ) |
| Eq.15 | $X_9 - X_{21}$ | Predorsal length | Caudal peduncle height | 0.899 | 0.0188 | -5.357 | <0.005 | ( - ) |
| Eq.16 | $X_{11} - X_{15}$ | Dorsal fin height | Pelvic fin base length | 1.183 | 0.0297 | 6.153 | <0.005 | ( + ) |
| Eq.17 | $X_{28} - X_3$ | Body height | Head width | 1.129 | 0.0290 | 4.432 | <0.005 | ( + ) |
| Eq.18 | $X_{23} - X_{10}$ | Caudal fin length | Dorsal fin base length | 0.950 | 0.0199 | -2.497 | <0.010 | ( - ) |
| Eq.19 | $X_{13} - X_{11}$ | Anal fin height | Dorsal fin height | 1.041 | 0.0234 | 1.770 | <0.050 | ( + ) |
| Eq.20 | $X_{16} - X_7$ | Pelvic fin length | Opercle height | 1.002 | 0.0143 | 0.160 | >0.100 | isometry |
| Eq.21 | $X_{18} - X_4$ | Pectoral fin base length | Interorbital length | 0.847 | 0.0215 | -7.117 | <0.005 | ( - ) |
| Eq.22 | $X_{19} - X_9$ | Pectoral fin length | Predorsal length | 0.959 | 0.0223 | -1.839 | <0.050 | ( - ) |
| Eq.23 | $X_8 - X_{21}$ | Superior mandible length | Caudal peduncle height | 1.020 | 0.0458 | 0.426 | >0.100 | isometry |
| Eq.24 | $X_{14} - X_{25}$ | Anal-pelvic fin distance | Total length | 1.135 | 0.0380 | 3.564 | <0.005 | ( + ) |
| Eq.25 | $X_{17} - X_{28}$ | Pectoral-pelvic fin distance | Body height | 0.938 | 0.0284 | -2.181 | <0.050 | ( - ) |
| Eq.26 | $X_{20} - X_3$ | Dorsal-caudal fin distance | Head width | 1.019 | 0.0429 | 0.436 | >0.100 | isometry |
| Eq.27 | $X_{22} - X_{21}$ | Caudal peduncle width | Caudal peduncle height | 1.107 | 0.0368 | 2.912 | <0.005 | ( + ) |
| Eq.28 | $X_{24} - X_{10}$ | Caudal fin height | Dorsal fin base length | 1.239 | 0.0495 | 4.836 | <0.005 | ( + ) |
| Eq.29 | $X_{29} - X_1$ | Body width | Head length | 1.161 | 0.0350 | 4.603 | <0.005 | ( + ) |

Applied to the exponents of the SMM equations to find which are really allometric or linear. Critical value with 46 d. f.= 1.638; with $\alpha = 0.05$

The first derivative of the allometric equation $y = bX^a$, can be interpreted as the Infinitesimal Change Rate (IChR) at which a morphological characteristic change in relation to another (Eq.3). In other words, it is an instantaneous speed and can be expressed as follows:

Given: $X_m = bX_n^a$

$$\frac{d(X_m)}{d(X_n)} = IChR = X'_m = abX_n^{(a-1)} \qquad (3)$$

Where $X_m$ y $X_n$ are morphometric variables measured in the same units, while a and b are constants.

Therefore:

if a < 1 then: $IChR = X'_m = \frac{ab}{X_n^{1-a}}$ (4)

Which implies that: $X'_m \rightarrow 0$, when $X_n \rightarrow \infty$

The biological interpretation of this argument is that, when two elements of the body plan (measured in the same units) show negative allometry, the IChR is never zero, since for this to be possible it is necessary that $X_n$ be infinite. The consequence of this fact is that the pairs of properties of the elements of the body plan that exhibit negative allometry do not suspend their growth during the life of the organism.

When characteristics show positive allometric relationships:

a > 1 therefore $IChR = X'_m = abX_n^{a-1}$ (5)

$X'_m \rightarrow 0$, when $X_n \rightarrow 0$

IChR is never zero unless $X_n$ is zero.

The corollary of the two previous statements is that, in allometry, the body plan never stops growing and, consequently, the shape of organisms changes throughout their life. This change in the body plan is clearly reflected in the differences shown in the shape of organisms during the first stages of their development, however in the adult stage they may not be so evident.

In isometric growth $X_m = bX_n$, and $X'_m = b$ (6)

Therefore, the IChR is a constant ≠ 0.

Table 18 shows the IChR derived from the SMM equations, which have been integrated into a system of simultaneous equations. This implies that, based on the IChR of a certain pair of variables in the system, it is possible to estimate the IChR corresponding to the other pairs of variables in that system. But it is important to clarify that obtained IChR values show the change rate between a particular pair of variables. For example, the equation: $X'_1 = 0.5314X_{26}^{-0.1113}$ allows to calculate IChR between $X_1$ y $X_{26}$, while: $X'_{15} = 0.3960X_3^{0.0938}$ will do the same for $X_{15}$ and $X_3$.

**Table 18. System of equations that represent the IChR of the equations contained in the SMM.**

**Table 18**

| equation | |
|---|---|
| Eq.1 | $X'_{1} = 0.5314 * X_{26}^{-0.1113}$ |
| Eq.2 | $X'_{10} = 0.4604 * X_{26}^{0.0755}$ |
| Eq.3 | $X'_{25} = 2.6423 * X_{10}^{-0.0671}$ |
| Eq.4 | $X'_{27} = 0.6081 * X_{25}^{-0.0557}$ |
| Eq.5 | $X'_{12} = 0.4635 * X_{10}^{-0.0145}$ |
| Eq.6 | $X'_{2} = 0.556 * X_{1}^{0.2426}$ |
| Eq.7 | $X'_{21} = 0.1163 * X_{25}^{0.0272}$ |
| Eq.8 | $X'_{3} = 1.2563 * X_{21}^{-0.0905}$ |
| Eq.9 | $X'_{15} = 0.3960 * X_{3}^{0.0938}$ |
| Eq.10 | $X'_{4} = 1.6155 * X_{15}^{0.0313}$ |
| Eq.11 | $X'_{26} = 12.5421 * X_{15}^{-0.0718}$ |
| Eq.12 | $X'_{7} = 1.0648 * X_{12}^{-0.0756}$ |
| Eq.13 | $X'_{6} = 0.2346 * X_{1}^{0.2122}$ |
| Eq.14 | $X'_{5} = 0.1168 * X_{10}^{0.1180}$ |
| Eq.15 | $X'_{9} = 2.9606 * X_{21}^{-0.1006}$ |
| Eq.16 | $X'_{11} = 1.9532 * X_{15}^{0.1829}$ |
| Eq.17 | $X'_{28} = 2.0479 * X_{3}^{0.1285}$ |
| Eq.18 | $X'_{23} = 0.5818 * X_{10}^{-0.0498}$ |
| Eq.19 | $X'_{13} = 0.8538 * X_{11}^{0.0414}$ |
| Eq.20 | $X'_{16} = 1.1475 * X_{7}^{0.0023}$ |
| Eq.21 | $X'_{18} = 0.7067 * X_{4}^{-0.1531}$ |
| Eq.22 | $X'_{19} = 0.7195 * X_{9}^{-0.0411}$ |
| Eq.23 | $X'_{8} = 0.6363 * X_{21}^{0.0195}$ |
| Eq.24 | $X'_{14} = 0.1225 * X_{25}^{0.1353}$ |
| Eq.25 | $X'_{17} = 0.2388 * X_{28}^{-0.0619}$ |
| Eq.26 | $X'_{20} = 0.5251 * X_{3}^{0.0187}$ |
| Eq.27 | $X'_{22} = 0.2420 * X_{21}^{0.1072}$ |
| Eq.28 | $X'_{24} = 0.1493 * X_{10}^{0.2392}$ |
| Eq.29 | $X'_{29} = 0.3351 * X_{1}^{0.1610}$ |

Table 19 contains the IChR values calculated based on their respective equation and the values of morphometric variables, determined as mentioned when explaining Table 7. To locate the moment in which these IChR values occur during the life of the *V. fenestrata*, values of $X_{26}$, (standard length), are used as a frame of reference during its development; without forgetting that IChR values are not a function of this variable, except when it is part of the derived equation, such as $X'_{10} = 0.4604 X_{26}^{0.0755}$.

**Table 19. Values of IChR for each pair of variables contained in the SMM.**

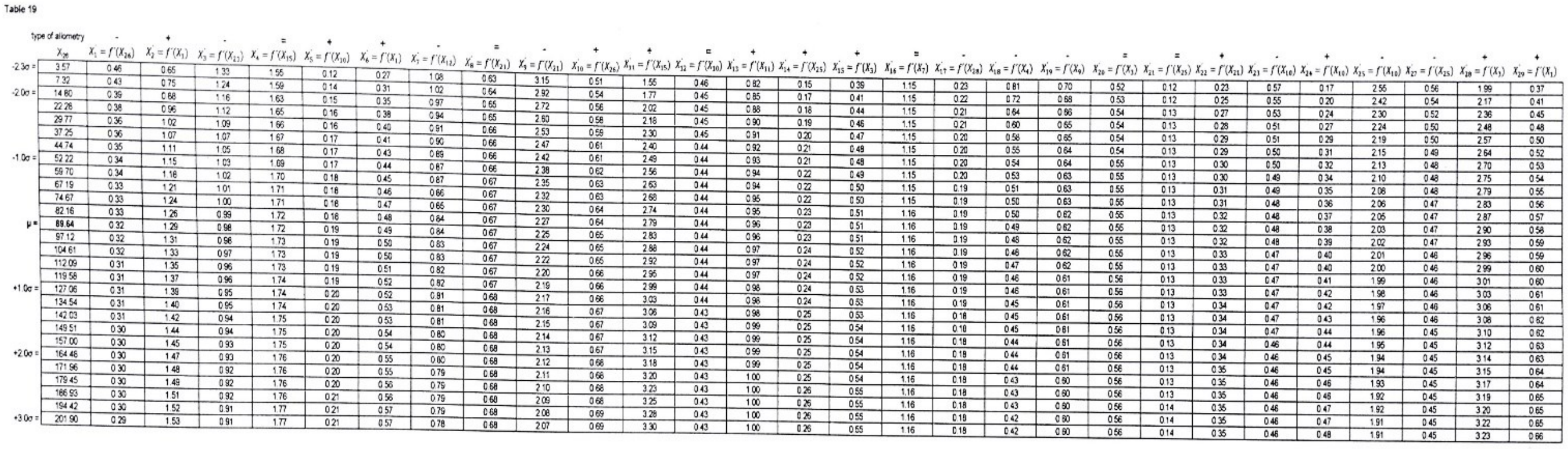

Table 19

| | $X_{26}$ | $X'_1 = f'(X_{26})$ | $X'_2 = f'(X_1)$ | $X'_3 = f'(X_{23})$ | $X'_4 = f'(X_{15})$ | $X'_5 = f'(X_{10})$ | $X'_6 = f'(X_1)$ | $X'_7 = f'(X_{12})$ | $X'_8 = f'(X_{21})$ | $X'_9 = f'(X_{21})$ | $X'_{10} = f'(X_{26})$ | $X'_{11} = f'(X_{15})$ | $X'_{12} = f'(X_{10})$ | $X'_{13} = f'(X_{11})$ | $X'_{14} = f'(X_{25})$ |
|---|---|---|---|---|---|---|---|---|---|---|---|---|---|---|---|
| type of allometry | | - | + | - | = | + | + | - | = | - | + | + | = | + | + |
| -2.3σ = | 3.57 | 0.46 | 0.65 | 1.33 | 1.55 | 0.12 | 0.27 | 1.08 | 0.63 | 3.15 | 0.51 | 1.55 | 0.46 | 0.82 | 0.15 |
| | 7.32 | 0.43 | 0.75 | 1.24 | 1.59 | 0.14 | 0.31 | 1.02 | 0.64 | 2.92 | 0.54 | 1.77 | 0.45 | 0.85 | 0.17 |
| -2.0σ = | 14.80 | 0.39 | 0.88 | 1.16 | 1.63 | 0.15 | 0.35 | 0.97 | 0.65 | 2.72 | 0.56 | 2.02 | 0.45 | 0.88 | 0.18 |
| | 22.28 | 0.38 | 0.96 | 1.12 | 1.65 | 0.16 | 0.38 | 0.94 | 0.65 | 2.60 | 0.58 | 2.18 | 0.45 | 0.90 | 0.19 |
| | 29.77 | 0.36 | 1.02 | 1.09 | 1.66 | 0.16 | 0.40 | 0.91 | 0.66 | 2.53 | 0.59 | 2.30 | 0.45 | 0.91 | 0.20 |
| | 37.25 | 0.36 | 1.07 | 1.07 | 1.67 | 0.17 | 0.41 | 0.90 | 0.66 | 2.47 | 0.61 | 2.40 | 0.44 | 0.92 | 0.21 |
| | 44.74 | 0.35 | 1.11 | 1.05 | 1.68 | 0.17 | 0.43 | 0.89 | 0.66 | 2.42 | 0.61 | 2.49 | 0.44 | 0.93 | 0.21 |
| -1.0σ = | 52.22 | 0.34 | 1.15 | 1.03 | 1.69 | 0.17 | 0.44 | 0.87 | 0.66 | 2.38 | 0.62 | 2.56 | 0.44 | 0.94 | 0.22 |
| | 59.70 | 0.34 | 1.18 | 1.02 | 1.70 | 0.18 | 0.45 | 0.87 | 0.67 | 2.35 | 0.63 | 2.63 | 0.44 | 0.94 | 0.22 |
| | 67.19 | 0.33 | 1.21 | 1.01 | 1.71 | 0.18 | 0.46 | 0.86 | 0.67 | 2.32 | 0.63 | 2.68 | 0.44 | 0.95 | 0.22 |
| | 74.67 | 0.33 | 1.24 | 1.00 | 1.71 | 0.18 | 0.47 | 0.85 | 0.67 | 2.30 | 0.64 | 2.74 | 0.44 | 0.95 | 0.23 |
| | 82.16 | 0.33 | 1.26 | 0.99 | 1.72 | 0.18 | 0.48 | 0.84 | 0.67 | 2.27 | 0.64 | 2.79 | 0.44 | 0.96 | 0.23 |
| μ = | 89.64 | 0.32 | 1.29 | 0.98 | 1.72 | 0.19 | 0.49 | 0.84 | 0.67 | 2.25 | 0.65 | 2.83 | 0.44 | 0.96 | 0.23 |
| | 97.12 | 0.32 | 1.31 | 0.98 | 1.73 | 0.19 | 0.50 | 0.83 | 0.67 | 2.24 | 0.65 | 2.88 | 0.44 | 0.97 | 0.24 |
| | 104.61 | 0.32 | 1.33 | 0.97 | 1.73 | 0.19 | 0.50 | 0.83 | 0.67 | 2.22 | 0.65 | 2.92 | 0.44 | 0.97 | 0.24 |
| | 112.09 | 0.31 | 1.35 | 0.96 | 1.73 | 0.19 | 0.51 | 0.82 | 0.67 | 2.20 | 0.66 | 2.95 | 0.44 | 0.97 | 0.24 |
| | 119.58 | 0.31 | 1.37 | 0.96 | 1.74 | 0.19 | 0.52 | 0.82 | 0.67 | 2.19 | 0.66 | 2.99 | 0.44 | 0.98 | 0.24 |
| +1.0σ = | 127.06 | 0.31 | 1.39 | 0.95 | 1.74 | 0.20 | 0.52 | 0.81 | 0.68 | 2.17 | 0.66 | 3.03 | 0.44 | 0.98 | 0.24 |
| | 134.54 | 0.31 | 1.40 | 0.95 | 1.74 | 0.20 | 0.53 | 0.81 | 0.68 | 2.16 | 0.67 | 3.06 | 0.43 | 0.98 | 0.25 |
| | 142.03 | 0.31 | 1.42 | 0.94 | 1.75 | 0.20 | 0.53 | 0.81 | 0.68 | 2.15 | 0.67 | 3.09 | 0.43 | 0.99 | 0.25 |
| | 149.51 | 0.30 | 1.44 | 0.94 | 1.75 | 0.20 | 0.54 | 0.80 | 0.68 | 2.14 | 0.67 | 3.12 | 0.43 | 0.99 | 0.25 |
| | 157.00 | 0.30 | 1.45 | 0.93 | 1.75 | 0.20 | 0.54 | 0.80 | 0.68 | 2.13 | 0.67 | 3.15 | 0.43 | 0.99 | 0.25 |
| +2.0σ = | 164.48 | 0.30 | 1.47 | 0.93 | 1.76 | 0.20 | 0.55 | 0.80 | 0.68 | 2.12 | 0.68 | 3.18 | 0.43 | 0.99 | 0.25 |
| | 171.96 | 0.30 | 1.48 | 0.92 | 1.76 | 0.20 | 0.55 | 0.79 | 0.68 | 2.11 | 0.68 | 3.20 | 0.43 | 1.00 | 0.25 |
| | 179.45 | 0.30 | 1.49 | 0.92 | 1.76 | 0.20 | 0.56 | 0.79 | 0.68 | 2.10 | 0.68 | 3.23 | 0.43 | 1.00 | 0.26 |
| | 186.93 | 0.30 | 1.51 | 0.92 | 1.76 | 0.21 | 0.56 | 0.79 | 0.68 | 2.09 | 0.68 | 3.25 | 0.43 | 1.00 | 0.26 |
| | 194.42 | 0.30 | 1.52 | 0.91 | 1.77 | 0.21 | 0.57 | 0.79 | 0.68 | 2.08 | 0.69 | 3.28 | 0.43 | 1.00 | 0.26 |
| +3.0σ = | 201.90 | 0.29 | 1.53 | 0.91 | 1.77 | 0.21 | 0.57 | 0.78 | 0.68 | 2.07 | 0.69 | 3.30 | 0.43 | 1.00 | 0.26 |

| | $X_{26}$ | $X'_{15} = f'(X_3)$ | $X'_{16} = f'(X_7)$ | $X'_{17} = f'(X_{28})$ | $X'_{18} = f'(X_4)$ | $X'_{19} = f'(X_9)$ | $X'_{20} = f'(X_3)$ | $X'_{21} = f'(X_{25})$ | $X'_{22} = f'(X_{21})$ | $X'_{23} = f'(X_{10})$ | $X'_{24} = f'(X_{10})$ | $X'_{25} = f'(X_{10})$ | $X'_{27} = f'(X_{25})$ | $X'_{28} = f'(X_3)$ | $X'_{29} = f'(X_1)$ |
|---|---|---|---|---|---|---|---|---|---|---|---|---|---|---|---|
| type of allometry | | + | = | - | - | - | = | = | + | - | + | - | - | + | + |
| -2.3σ = | 3.57 | 0.39 | 1.15 | 0.23 | 0.81 | 0.70 | 0.52 | 0.12 | 0.23 | 0.57 | 0.17 | 2.56 | 0.56 | 1.99 | 0.37 |
| | 7.32 | 0.41 | 1.15 | 0.22 | 0.72 | 0.68 | 0.53 | 0.12 | 0.25 | 0.55 | 0.20 | 2.42 | 0.54 | 2.17 | 0.41 |
| -2.0σ = | 14.80 | 0.44 | 1.15 | 0.21 | 0.64 | 0.66 | 0.54 | 0.13 | 0.27 | 0.53 | 0.24 | 2.30 | 0.52 | 2.36 | 0.45 |
| | 22.28 | 0.46 | 1.15 | 0.21 | 0.60 | 0.65 | 0.54 | 0.13 | 0.28 | 0.51 | 0.27 | 2.24 | 0.50 | 2.48 | 0.48 |
| | 29.77 | 0.47 | 1.15 | 0.20 | 0.58 | 0.65 | 0.54 | 0.13 | 0.29 | 0.51 | 0.29 | 2.19 | 0.50 | 2.57 | 0.50 |
| | 37.25 | 0.48 | 1.15 | 0.20 | 0.55 | 0.64 | 0.54 | 0.13 | 0.29 | 0.50 | 0.31 | 2.15 | 0.49 | 2.64 | 0.52 |
| | 44.74 | 0.48 | 1.15 | 0.20 | 0.54 | 0.64 | 0.55 | 0.13 | 0.30 | 0.50 | 0.32 | 2.13 | 0.48 | 2.70 | 0.53 |
| -1.0σ = | 52.22 | 0.49 | 1.15 | 0.20 | 0.53 | 0.63 | 0.55 | 0.13 | 0.30 | 0.49 | 0.34 | 2.10 | 0.48 | 2.75 | 0.54 |
| | 59.70 | 0.50 | 1.15 | 0.19 | 0.51 | 0.63 | 0.55 | 0.13 | 0.31 | 0.49 | 0.35 | 2.08 | 0.48 | 2.79 | 0.56 |
| | 67.19 | 0.50 | 1.15 | 0.19 | 0.50 | 0.63 | 0.55 | 0.13 | 0.31 | 0.48 | 0.36 | 2.06 | 0.47 | 2.83 | 0.56 |
| | 74.67 | 0.51 | 1.16 | 0.19 | 0.50 | 0.62 | 0.55 | 0.13 | 0.32 | 0.48 | 0.37 | 2.05 | 0.47 | 2.87 | 0.57 |
| | 82.16 | 0.51 | 1.16 | 0.19 | 0.49 | 0.62 | 0.55 | 0.13 | 0.32 | 0.48 | 0.38 | 2.03 | 0.47 | 2.90 | 0.58 |
| μ = | 89.64 | 0.51 | 1.16 | 0.19 | 0.48 | 0.62 | 0.55 | 0.13 | 0.32 | 0.48 | 0.39 | 2.02 | 0.47 | 2.93 | 0.59 |
| | 97.12 | 0.52 | 1.16 | 0.19 | 0.48 | 0.62 | 0.55 | 0.13 | 0.33 | 0.47 | 0.40 | 2.01 | 0.46 | 2.96 | 0.59 |
| | 104.61 | 0.52 | 1.16 | 0.19 | 0.47 | 0.62 | 0.55 | 0.13 | 0.33 | 0.47 | 0.40 | 2.00 | 0.46 | 2.99 | 0.60 |
| | 112.09 | 0.52 | 1.16 | 0.19 | 0.46 | 0.61 | 0.56 | 0.13 | 0.33 | 0.47 | 0.41 | 1.99 | 0.46 | 3.01 | 0.60 |
| | 119.58 | 0.53 | 1.16 | 0.19 | 0.46 | 0.61 | 0.56 | 0.13 | 0.33 | 0.47 | 0.42 | 1.98 | 0.46 | 3.03 | 0.61 |
| +1.0σ = | 127.06 | 0.53 | 1.16 | 0.19 | 0.45 | 0.61 | 0.56 | 0.13 | 0.34 | 0.47 | 0.42 | 1.97 | 0.46 | 3.06 | 0.61 |
| | 134.54 | 0.53 | 1.16 | 0.18 | 0.45 | 0.61 | 0.56 | 0.13 | 0.34 | 0.47 | 0.43 | 1.96 | 0.46 | 3.08 | 0.62 |
| | 142.03 | 0.54 | 1.16 | 0.18 | 0.45 | 0.61 | 0.56 | 0.13 | 0.34 | 0.47 | 0.44 | 1.96 | 0.45 | 3.10 | 0.62 |
| | 149.51 | 0.54 | 1.16 | 0.18 | 0.44 | 0.61 | 0.56 | 0.13 | 0.34 | 0.46 | 0.44 | 1.95 | 0.45 | 3.12 | 0.63 |
| | 157.00 | 0.54 | 1.16 | 0.18 | 0.44 | 0.61 | 0.56 | 0.13 | 0.34 | 0.46 | 0.45 | 1.94 | 0.45 | 3.14 | 0.63 |
| +2.0σ = | 164.48 | 0.54 | 1.16 | 0.18 | 0.44 | 0.61 | 0.56 | 0.13 | 0.35 | 0.46 | 0.45 | 1.94 | 0.45 | 3.15 | 0.64 |
| | 171.96 | 0.54 | 1.16 | 0.18 | 0.43 | 0.60 | 0.56 | 0.13 | 0.35 | 0.46 | 0.46 | 1.93 | 0.45 | 3.17 | 0.64 |
| | 179.45 | 0.55 | 1.16 | 0.18 | 0.43 | 0.60 | 0.56 | 0.13 | 0.35 | 0.46 | 0.46 | 1.92 | 0.45 | 3.19 | 0.65 |
| | 186.93 | 0.55 | 1.16 | 0.18 | 0.43 | 0.60 | 0.56 | 0.14 | 0.35 | 0.46 | 0.47 | 1.92 | 0.45 | 3.20 | 0.65 |
| | 194.42 | 0.55 | 1.16 | 0.18 | 0.42 | 0.60 | 0.56 | 0.14 | 0.35 | 0.46 | 0.47 | 1.91 | 0.45 | 3.22 | 0.65 |
| +3.0σ = | 201.90 | 0.55 | 1.16 | 0.18 | 0.42 | 0.60 | 0.56 | 0.14 | 0.35 | 0.46 | 0.48 | 1.91 | 0.45 | 3.23 | 0.66 |

Note that the IChR values are not constant whether the relationship between the variables is isometric, (in which case the variation is minimal), for example between $X_{21}$ (height of the caudal peduncle) and $X_{25}$ (total length); or allometric as is the case between $X_{28}$ (body height) and $X_3$ (head width). In the case of the latter, it is reasonable that the rate of change does not remain constant since this would indicate that the organisms grow at the same speed throughout their life given that, as has been demonstrated previously in this work, under the growth model allometric IChR $\neq 0$. It is also notable that while some IChR change rapidly, as in the case of the one between $X_{28}$ and $X_3$, others do so very slowly, for example the one between $X_{10}$ (dorsal fin length) and $X_{26}$ (standard length), however they all change throughout the life of the organism.

In Table 19, the IChR for variables with positive allometry increase with the age of the individual, the negative ones decrease, and the isometric ones remain reasonably constant. Possibly the SMD of adult individuals increases due to the behavior of the IChRs of positive allometric relationships, which constitute 41% of those with which the MMS was built. These IChRs are the engine that drive the morphological differentiation of older fish.

In the case of the observed individuals, the fact that they have the same age, but different SMD could be explained by considering the effect of internal biological factors such as nutrition, developmental anomalies, mutations, etc., or external factors such as food availability, competition, predation, etc.

If the MMS of a taxon is solved for a specific variable, for example $X_{26}$, the standard length in fishes, and the rest of the MMS variables are plotted against this variable, the trajectories of relative growth are obtained. These allow us to simulate what the development of the body plan would “look like” from the perspective of a particular variable; For example, from the perspective of the development of the standard length of a fish we would “see” how, as it increases, structures and organs such as eyes, opercles, fins, etc. appear and develop; Figure 9 presents an example with the variables $X_{26}$ vs $X_9$, $X_{10}$, $X_{12}$, $X_{21}$, $X_{27}$ and $X_{28}$. This subset of variables was chosen because they are among those that establish a greater number of relationships with the rest of the variables contained in the MMS such that $0.99 < r < 1.00$, see Figure 3A, and because they describe the main block of the body, the cephalic region and the trunk, of the *V. fenestrata* as can be seen in Figure 2.

**Figure 9. Trajectories of relative growth of some elements of the body plan of *V. fenestrata***

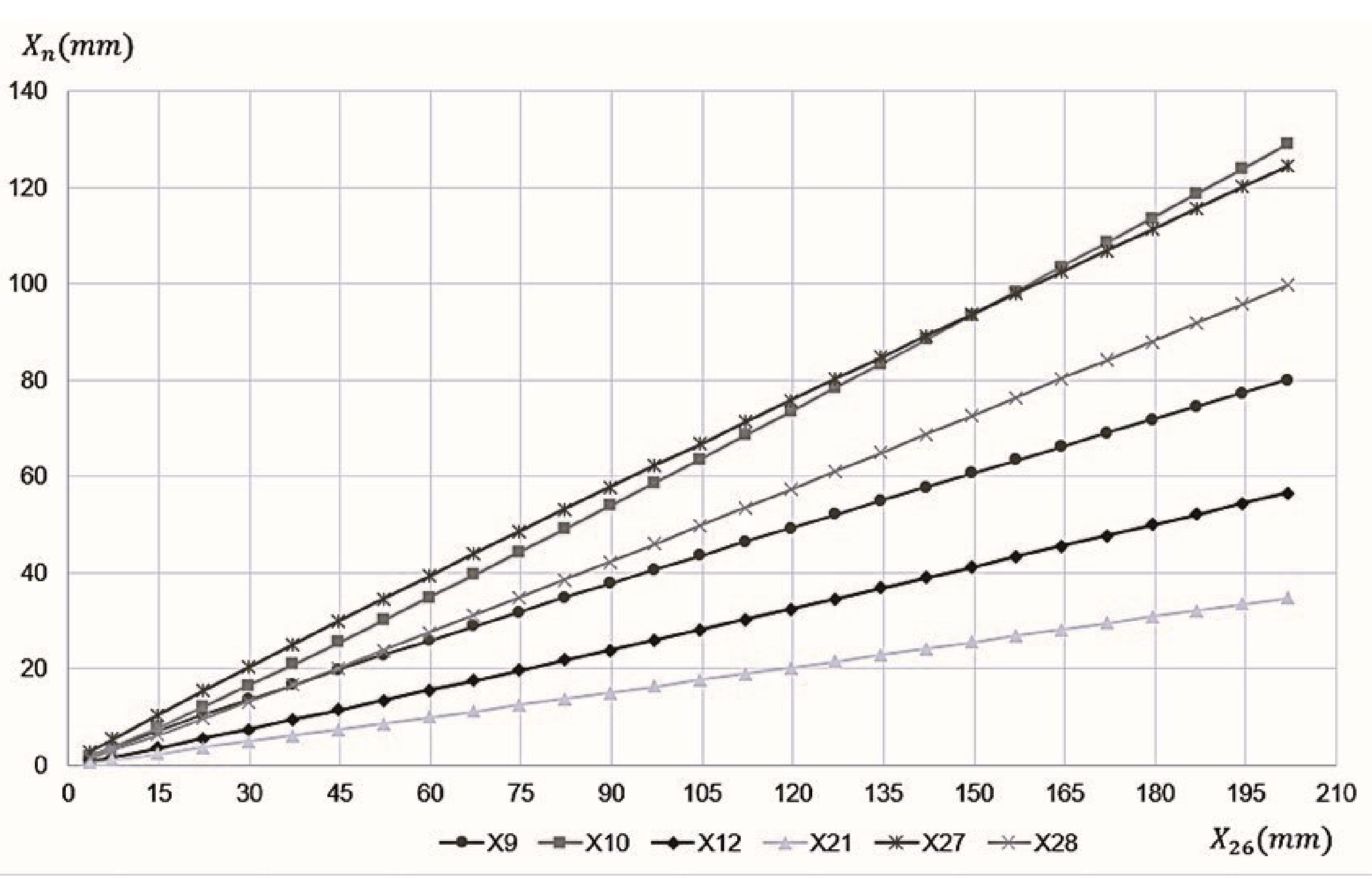

The graph shows the relative growth of some elements of the body structure of *Vieja fenestrata* during the growth of the complete individual. By observing the increase in values on the axis of the variable X26, the standard length, it is possible to observe how the other body elements considered develop.

On the other hand, when a graph is constructed with the IChR of the pairs of variables of which the variables graphed in Figure 9 are part, (see Table 19), taking again the variable $X_{26}$ (standard length) as a temporal reference, the trajectories of the Figure 10 are obtained, these show what was mentioned above in relation to the variation of IChR values for pairs of allometric or isometrically related variables.

**Figure 10. Behavior of the IChR between elements of the body plan of *Vieja fenestrata***

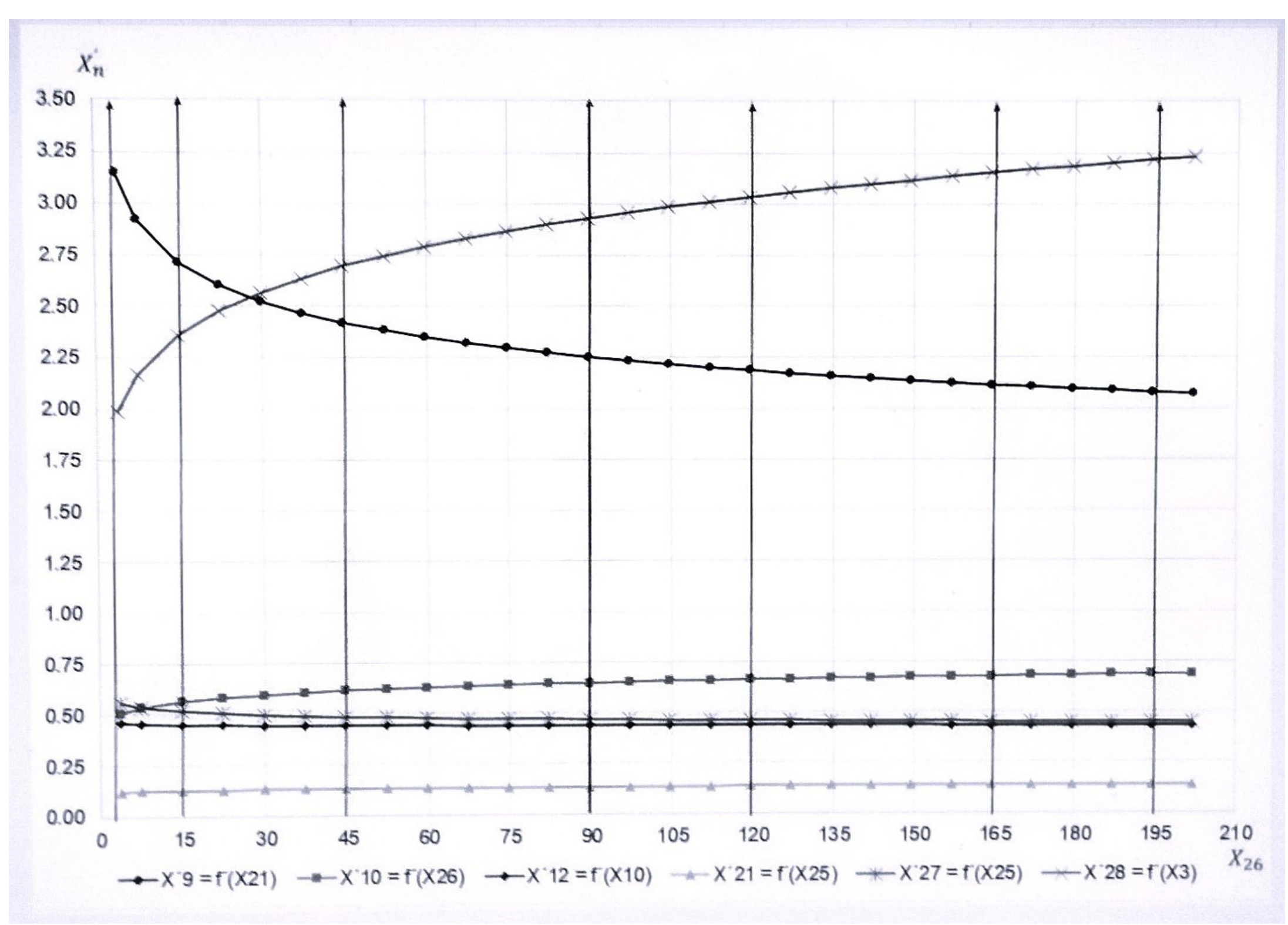

Graph of the behavior of the infinitesimal change rates, IChR, between the variables in Fig. 9 and their pairs considered in Table 19, during the ontogeny of *Vieja fenestrata*

By using the graphs in Figures 9 and 10, it is possible to simultaneously determine the value and IChR of a morphometric variable, in relation to another variable when $X_{26}$ (standard length) takes a specific value. For example, in Figure 9, it is easily observed that when $X_{26}$ takes a value ≈ μ-2.0σ, close to the inferior limit of the theoretical SPS of *V. fenestrata* (equivalent to 14.8 mm), the value of $X_{12}$ (anal fin length) is 3.55 mm.

On the other hand, Figure 10 shows that, for the same value of $X_{26}$, $X_{12}$ increases isometrically with an IChR = 0.45 in relation to $X_{10}$ (dorsal fin length), which increases with positive allometry with respect to $X_{26}$ with IChR=0.56, as can also be seen in Table 19.

Systemic morphometric models allow determining the limits of the phenotypic space; however, it must be considered that this is valid for a certain time and place. If the population develops a structural change due to an evolutionary force, such as a mutation, the variance, ($\sigma^2$), of the possible phenotypes will increase and the boundaries of the SPS, the systemic phenotypic space, will be modified. Therefore, these are dynamic in time and space, and it is not possible to infer the SPS properties of a taxon in the distant past from information on the current taxon.

The SMM of *V. fenestrata* is predictive, since it allows to know the magnitude of a variable with respect to other components of the system and vice versa: it allows to know the system's answer with respect to changes in one of its components. This property might be essential to study the effect of the regulation exercised by selective pressure from certain ecological changes; it has been observed in phylogenetically related species to *V. fenestrata*, where trophic structure influence has been documented to influence traits, such as tooth types, snout length and pharyngeal bones, apparently in an independent way from the body plan [34]. This cichlid belongs to a phenotypically very plastic group of organisms, with a wide range of chewing apparatus diversification [35], giving it a potentially complex phenotypical space, and a high body plan variability.

At this point, it is convenient to mention that the use of SMM for a given body plan offers the possibility of inferring when an element of the plan disappears; for example, the limbs of a lizard or the fins of a fish, but the model cannot predict when a limb could appear.

By using the SMM it has been possible to infer some of the Blackstripe Cichlid body plan properties and SPS, but it must be acknowledged that even though the species is endemic to Mexico and the samples come from a representative area of its distribution, the sample size is far from representative of its total phenotypical variability. Hence, it is necessary to continue to register species morphometrical data to refine the calculation of the SMM parameters and to define the body plan and phenotypical space properties more precisely. As mentioned above, the species is endemic to Mexico, in Veracruz and Oaxaca states, which makes it easier the collection of a sufficiently representative sample of its entire distribution range.

Despite the latter, the simulation capacity of the SMM allows us to propose that the body structure of the Blackstripe Cichlid is plastic enough to generate phenotypic diversity based on the dimensional variations of its constituent elements, which propagate through the body plane due to its systemic nature. This may help explain what was reported by [35]. From this perspective, quantitative variation and variation in the structural, or spatial, arrangement of the elements of the body plan contribute to the phenotypic variance of a taxon because they generate structural variance.

Based on this, we propose adding the structural variance to the classic phenotypic variance equation of Quantitative Genetics [36]. Rewriting the equation:

$$\sigma_f^2 = \sigma_g^2 + \sigma_e^2 + \sigma_{g-e}^2 \qquad (7)$$

Where: $\sigma_f^2$: phenotypic variance, $\sigma_g^2$: genetic variance, $\sigma_e^2$: environmental variance and $\sigma_{g-e}^2$ : variance due to the genotype - environment interaction); as:

$$\sigma_{fs}^2 = \sigma_g^2 + \sigma_e^2 + \sigma_{g-e}^2 + \sigma_s^2 \qquad (8)$$

The last term is the structural variance, generated by the random variation of the assembly of the elements of the body plane and its propagation due to its systemic nature.

This could be the most representative equation of the Systemic Morphometry approach because it reflects the potential for phenotypic variation of a taxon, not only due to its genetics or the selective pressures of the environment, but also due to random metric variation and the spatial arrangement of elements of the body plane.

A consequence of this is that it can be hypothesized that, under conditions of genetic and environmental homogeneity, body structure could vary due to random changes in the dimensions and spatial arrangement of its constituent elements. Furthermore, it can be proposed that: $\sigma_{phenotypic}^2 = SPS_{theoretical}$ , (the phenotypic variance is equal to the theoretical systemic phenotypic space), which represents a point of convergence between Quantitative Genetics and Systemic Morphometry.

With the values obtained from the measured *V. fenestrata* individuals, a representation of the observed SPS was constructed and, using the SMM as a tool for the simulation of biologically possible phenotypes, we obtained basic information to construct the theoretical SPS. This characteristic of the SPS differentiates it from phenotypic spaces constructed through mathematical models that do not consider relationships or interactions between the constituent elements of the body structure.

However, it must be recognized that, in a way, the characteristics of the theoretical SPS depend on the representativeness of the samples of the organisms and the morphometric variables selected to describe the body structure of the taxon and their observed SPS; which implies that the theoretical SPS depends on the quality of the description of the observed SPS. The representation of both SPS in the same graph allows us to compare them in their shape, extension, limits, etc., or consider population aspects such as the homogeneity in the distribution of the individuals they contain (population density).

The representation of both observed and theoretical SPS of *V. fenestrata*, shown in Figure 5, was constructed following the guidelines and considerations that [1] used to represent the SPS of a lizard species. This representation has an inconvenience: it does not allow the entire SPS to be represented in a single image because it would require a 29-dimensional coordinate system (in the case of *V. fenestrata*). Because of this, Figure 5 only represents one facet of the entire space in a three-axis cartesian system, and in it, the observed SPS is contained within the limits of the theoretical SPS, but without focusing on a specific zone, practically both spaces occupy the same area. On the axes it is possible to read the limits of the observed and theoretical SPS for the variables used in its construction; though it was also demonstrated that it is possible to analytically determine the limit values for each of the variables. The same type of analysis could be performed by constructing other 3D images of both SPS using other variable triads.

Regarding the limits of the theoretical SPS, it was said that the lower limit is clearly identified because it is the value at which the SMM solutions are above zero for all the contained variables. However, it was also mentioned that the SMM is not capable of detecting the upper limit of the theoretical SPS because it presents solutions greater than zero for all variables when the variable $X_{26}$ (standard length) takes values greater than $\mu+3\sigma$. This fact was also observed when the theoretical SPS of the lizard *U. stansburiana* was built [1], where they argued the possibility that the body structure of the species allowed sizes much larger than those observed, but non-existent due to internal (i.e., metabolic, or other types of physiological functions) or external (predation, competition, etc.) constraints. These same considerations could apply to the Blackstripe Cichlid, but it is evident that the explanation is not entirely satisfactory and further research is needed to answer in more detail the reasons why there is structurally no upper limit to the size of fishes and lizards.
It is important to mention that, from the perspective of Systemic Morphometry, the properties of the basic body structure of a taxon, at the individual level, determine the properties or characteristics of the SPS at the population level. For this reason, it is proposed that the SPS is an emergent property of the complex system called bauplan of a taxon.

Alternatively, to represent the SPS in a 3D cartesian system, we explore the possibility of using a parameter to concentrate all the variability information of the elements of the body structure, as well as the relationships established between them since embryogenesis. The MQD [21] considers the variance and covariance of the variables that define an individual in a multivariate space, precisely what is needed, given that Systemic Morphometry seeks to conceptualize the properties and diversity of the baupläne considering the variance and covariance of the elements that constitute them.

The construction of the theoretical and observed SPS using the MQD, referred to as Systemic Morphometrics Distance (SMD) in the context of the Systemics Morphometrics, is totally different to what has been analyzed in the former paragraphs. First, the representation of the SPS is a distribution of the frequencies of the values of the SMD, that can be represented with a graph as in Figure 7 (resembling a grid) or by a double histogram, as in Figure 8, that represents and allows the comparing of the theoretical and observed SPS simultaneously. Other options to represent such spaces are a stem and leaf diagram, Table 12, or a box and whisker plot Figure 6.

In the SPS shown in Figures 7 and 8, the origin of the axis of SMD values is a theoretical individual or centroid, described based on the means of the morphometric variables used, 28 for the black striped cichlid *V. fenestrata*. Another characteristic of the SPS, observed and theoretical, based on SMD is that they require a common scale for their representation in the same plot, allowing them to be compared under the same conditions. To achieve this, we normalize the values of each variable by dividing them by their respective arithmetic mean. This unifies the scale and allows both centroids to have the same SMD=0 value, in order to measure the SMD of each individual, theoretical or observed, with respect to their respective centroid. If the SMDs are not calculated with normalized values, the centroids will have different sizes and require different scales to be represented, making it impossible to compare the SPSs.

The theoretical value of the SMD is that it represents the degree of structural deviation that an individual has with respect to its centroid, which can be considered as the basic body structure or bauplan of a taxon if the sample of the organisms is, without a doubt, representative. Another situation in which the centroid can be equated to the bauplan is when it is obtained as the average of several observed SPS centroids of the same taxon, calculated based on samples from different sites throughout its distribution range.

The joint representation of the observed and theoretical SPS in Figure 7 shows how individuals form subsets with very similar SMD, which can be considered as strata, levels or layers that have theoretical and observed centroids as a nucleus or origin. In this image there is no distribution of individuals in two dimensions, there is only one, the SMD value axis, it was represented in this way to visualize how the observed organisms coexist in the same stratum with the theoretical ones. Therefore, the empty spaces observed within the strata should not be interpreted as an absence of individuals. This is why it was previously mentioned that perhaps it would be better to represent the SPS as a double histogram, as in Figure 8.

Both SPS plots show that they practically have the same lower limit, near SMD = 15, and that the upper limit of the theoretical SPS far exceeds that of the observed SPS.

The use of the SMDs to construct the SPS allows its analytical comparison through statistical and/or algebraic methods, as shown in Table 12. The values of the descriptive statistics indicate that the observed and theoretical SPS have similarities and differences; Perhaps the most important of the differences is the extension of the theoretical SPS, evaluated based on the frequency distribution range of the SMDs. A fact that supports the similarity of the SPSs is that 95% of the individuals (from both spaces) are practically in the same range of SMD values: 12.6-42.8. When these data were presented, it was noted that none of the organisms, observed or theoretical, had an SMD value <12.6 (Tab. 12). This study did not generate enough information to provide a convincing answer to this observation, although it could be considered as a restriction due to the architecture, the spatial arrangement, of the body structure.

On the other extreme, the superior boundary of the theoretical SPS, only one individual with SMD = 75.73 was found, and the nearest individual had SMD = 47.43; with a gap in between where no individuals, nor theoretical or observed, was found. Considering that the individual with SMD = 75.73 is in $\mu - 2.38\sigma$ of the $X_{26}$ variable, meaning the superior boundary in Table 10, we thought that, since the cichlid sample does not include such small individuals, it could possibly explain the superior limit of SMD for the observed SPS, 43.16. However, that does not explain the jump of the theoretical SMD values from 47.43 to 75.73, corresponding to the last and penultimate individuals of the inferior limit of the variable $X_{26}$ (standard length) of *V. fenestrata*.

When representing the observed and theoretical SPS by means of a double histogram, this situation of the unexplained gap between the last and the penultimate theoretical value of the SMD is also evident.

To explain such a large value for the SMD, it could be considered that, although the SMM generates values greater than zero for all the descriptive variables of the body structure of *V. fenestrata*, when $X_{26} = \mu - 2.38\sigma$, some of them reach only a few hundredths of a millimeter, so it is very likely that the spatial configuration of this theoretical individual is significantly different from the theoretical centroid. The individual with $X_{26} = \mu - 2.34\sigma$, and SMD = 47.43 is defined by values of the order of tenths of a millimeter for almost all morphometric variables, and most likely with greater structural similarity to the centroid, this may explain the fact that the 5 strata between SMD values 50 and 70 are empty.

[37] studied the ontogenetic development of *Vieja melanura*, reporting that newly hatched larvae have a standard length $X_{26}$ = 3.8 mm, while pre-juvenile and juvenile individuals measure 15.7 and 23.6 mm.

These results reinforce the conclusion that *V. fenestrata* individuals close to the lower limit of the theoretical SPS, with a standard length $X_{26} < 3.57$ mm (Tab. S19), do not yet have the typical structure of their species and therefore have a large MSD.

Considering the above, it is possible that the individuals with $X_{26} = \mu - 1.8\sigma = 22.8$ mm, (Tab. S19), already present the typical body structure of the species *V. fenestrata*.

Not considering the individual with an extreme value of SMD in the theoretical SPS, the observed SPS occupies practically the entire theoretical SPS, which implies that the phenotypic variability observed, at the time and place of sampling, coincides with the expected phenotypic variability for *V. fenestrata*, according to the approach of Systemic Morphometry. In other words, $observed\ \sigma^2_{phenotipic} = theoretical\ \sigma^2_{phenotipic}$ . However, this does not mean that equality is valid at the species level because, as mentioned above, the sample of organisms for this study was not as representative, as can be seen in Table 1 with the coefficient of variation values for all morphometric variables used to describe the body structure of *V. fenestrata*. In all cases, the value of the coefficient is greater than 35%, so the centroid of the observed SPS could not be equivalent to the bauplan of this species.

Other useful information comes from the stem-and-leaf plots of both SPSs, which indicate that the frequency distribution of the SMDs is very similar in the area where the two overlap. In this area, six groups or strata of MDS are observed; the usefulness of this numerical representation of the strata is that it allows each body structure, observed or theoretical, to be included in a particular stratum and subsequently perform the descriptive statistics of each stratum in order to make comparisons within each SPS or between SPSs (Tabs. S13 and S14). The results of the comparisons of the standard deviations of the homologous strata in both SPS allow us to conclude that both spaces present the same distribution of body structures with respect to their centroid (Tab. 15).

Therefore, the observed SPS of *V. fenestrata* can be considered to have the expected structure, according to the properties of its body plan, since it covers six of the seven subspaces of the theoretical SPS. Furthermore, the variability between the six observed SPS strata is also as expected, according to the systemic morphometry approach, given that it does not present significant differences with respect to the variability existing within the first six strata of the theoretical SPS. The sixth subspace of both SPS has few individuals and its standard deviation is different with respect to the other strata of its SPS. Stratum 7 is exclusive of the theoretical SPS and has only one individual.

We conclude that the observed SPS analysis, using SMD, is made up of a set of organisms that can be grouped according to their degree of similarity in body structure. Its limits, in terms of SMD, are 15.13 and 43.16; 95% of these organisms are located within the zone from 12.7 to 42.2 SMD (Tab. 12).

The theoretical SPS of this species has its lower and upper limits at 14.77 and 75.73 SMD, respectively. The upper limit is a unique case and differs greatly from the set of values that constitute that space. In the box-and-whisker plot, this value appears as an extreme case as it does in the stem and leaf plot. 95% of the theoretical SPS individuals fall within the range of 12.6 to 42.8 SMD. It is evident that the observed SPS is within the limits of the theoretical SPS (Tab. 12).

In SPSs based on the MQD the boundaries of the space are expressed in terms of SMD, unlike the partial graphing method, in which the boundaries are established in terms of the values of the morphometric variables used to construct the 3D graph of the SPS or are determined analytically.

We consider that this study achieved its objectives, since it managed to analyze the body structure of the *V. fenestrata* species by applying the Systemic Morphometry approach, and obtain a Systemic Morphometric Model which served as a basis for the study of some structural properties of the bauplan of this species as well as for the simulation of the expected phenotypic variation according to these properties (the theoretical Systemic Phenotypic Space).

By using the SMM we obtained some useful concepts in the morphometric analysis of a taxon such as: Infinitesimal Change Rates (IChR), continuous growth throughout ontogeny, developmental trajectories and their relationship with the IChR, and the concept of changing scenario throughout embryonic development.

The representation of the systemic phenotypic space, SPS, observed and theoretical, using the conventional model, which consists of a three-dimensional graph generated by the random selection of three morphometric variables, intertwined with the rest of the variables of the set that was used to describe the body structure of the species, was compared with the SPS obtained through the use of the MQD to determine the SMD, at which an individual is located with respect to a centroid, which is interpreted as an individual whose body structure has the average values of all the morphometric variables with which it was described. This new method of analyzing phenotypic spaces allowed us to go beyond simple visual exploration to include a statistical analysis to compare them in a more objective and efficient way.

As previously mentioned, this work precedes others in which the methodology described here will be applied to build systemic morphometric models for other vertebrate taxa (birds and mammals), with the aim of integrating a Compendium of the Systemic Morphometry of Vertebrates.

This study, like those carried out previously, aims to support the proposal to develop a Systemic Biological Morphometry, whose philosophy, working methods, and scopes are more biological than other existing methodologies to date.

Systemic Morphometry has 3 properties or characteristics that define it:

*Descriptive capacity*: As already mentioned, by applying it is possible mathematically and graphically describe the relationships between the morphometric variables; obtain a Systemic Mathematical Model, based on these relationships; characterize and represent the observed Systemic Phenotypic Space, among other concepts.

*Inferential capacity*: Using the SMM it is possible to infer the total variability expected for a taxon based on the properties of its body structure; represent the theoretical SPS; determine the IChR, between pairs of variables; demonstrate that IChR $\neq 0$ during ontogeny; more efficiently compare the observed and theoretical SPS using techniques such as the MQD based; establish equality between Phenotypic Variance and theoretical SPS; describe the behavior of IChR depending on the type of allometry, etc.

*Integration capacity*: The results obtained when using Systemic Morphometry are capable of being integrated in a coherent manner. For example, to explain the coexistence of juvenile and adult individuals in a stratum with a given SMD (Fig. 7), concepts such as IChR, the behavior of IChR in positive allometric relationships, the relationship between SMDs and the size of the individual, the idea that IChRs are never equal to zero during ontogeny, etc. This integration of results and inferences based on the principles of Systemic Morphometry leads to a better understanding of the Theoretical Morphology of a taxon.

# Acknowledgments

To Jonnathan D. Rivera Ruíz, for his valuable contributions to the mathematical and computational analysis of data; and to Sebastián Huacuja Barraza, for his valuable suggestions and help in shaping the final manuscript.

We dedicate this work to Dr. Héctor Espinosa Pérez, our friend and teacher, for lending us the time and specimens to be measured, as well as for his guidance and support in the taxonomy and anatomy of fish.